\documentclass[fleqn,usenatbib]{mnras}

\usepackage[T1]{fontenc}

\DeclareRobustCommand{\VRIES}[3]{#2}
\let\VRIESthebibliography\thebibliography
\def\thebibliography{\DeclareRobustCommand{\VRIES}[3]{##3}\VRIESthebibliography}

\usepackage{subcaption}
\usepackage{float}

\usepackage{graphicx}	% Including figure files
\usepackage{amsmath}	% Advanced maths commands
\usepackage{amssymb}	% Extra maths symbols
\usepackage{multicol}        % Multi-column entries in tables
\usepackage{bm}		% Bold maths symbols, including upright Greek
\usepackage{pdflscape}	% Landscape pages
\usepackage[table,xcdraw]{xcolor}

\usepackage{xcolor}

\usepackage{newtxtext,newtxmath}

\title[Interactions of precession and convection]{Tidal dissipation and spin-orbit alignment due to the precessional instability in convection zones in rotating giant planets and stars}

\author[N. B. de Vries et al.]{
Nils B. de Vries,$^{1,2}$\thanks{E-mail: N.B.de-Vries@exeter.ac.uk (NBV)}
Adrian J. Barker,$^{2}$\thanks{E-mail: A.J.Barker@leeds.ac.uk (AJB)}
Rainer Hollerbach$^{2}$\thanks{E-mail: R.Hollerbach@leeds.ac.uk (RH)}
\\
$^{1}$ Department of Physics and Astronomy,  University of Exeter, Exeter EX4 4PY, UK\\
$^{2}$ School of Mathematics, University of Leeds, Leeds LS2 9JT, UK
}

\date{Accepted Nov X XXX. Received Nov Y YYY; in original form Nov Z ZZZ}

\pubyear{2025}

\begin{document}
\label{firstpage}
\pagerange{\pageref{firstpage}--\pageref{lastpage}}
\maketitle

% Abstract of the paper
\begin{abstract}
Tidal dissipation in star-planet systems occurs through various mechanisms, including the precessional instability. This is an instability of laminar flows (``Poincar\'{e} flows") forced by axial precession of a rotating, oblate, spin-orbit misaligned fluid planet or star, which excites inertial waves in convective regions if the dimensionless precession rate (``Poincar\'{e} number" $\mathrm{Po}$) is sufficiently large. We constrain the contribution of the precessional instability to tidal dissipation and heat transport, using Cartesian hydrodynamical simulations in a small patch of a planet, and study its interaction with turbulent convection, modelled as rotating Rayleigh-B\'{e}nard convection. The precessional instability without convection results in laminar flow at low values and turbulent flow at sufficiently high values of $\mathrm{Po}$. The associated tidal dissipation rate scales as $\mathrm{Po}^2$ and $\mathrm{Po}^3$ in each regime, respectively. 
With convection, the Poincar\'e number at which turbulent flow is achieved shifts to lower values for stronger convective driving. Convective motions also act on large-scale tidal flows like an effective viscosity, resulting in continuous tidal dissipation (scaling as $\mathrm{Po}^2$), which obfuscates or suppresses tidal dissipation due to precessional instability. The effective viscosities obtained agree with scaling laws previously derived using (rotating) mixing-length theory. By evaluating our scaling laws using interior models of Hot Jupiters, we find that the precessional instability is significantly more efficient than the effective viscosity of convection. The former drives alignment in 1 Gyr for a Jupiter-like planet orbiting within 23 days. Linearly excited inertial waves can be even more effective for wider orbits, aligning spins for orbits within 53-142 days.

\end{abstract} 

% Select between one and six entries from the list of approved keywords.
% Don't make up new ones.

%%%%%%%%%%%%%%%%%%%%%%%%%%%%%%%%%%%%%%%%%%%%%%%%%%

%%%%%%%%%%%%%%%%% BODY OF PAPER %%%%%%%%%%%%%%%%%%

\section{Introduction}

Gaseous planets are expected to have an oblate shape due to centrifugal deformations caused by their axial rotations (or an approximately triaxial ellipsoidal shape if there are also strong tidal deformations). If the spin and orbital rotation axes of these planets are also misaligned, i.e., if they possess a non-zero planetary obliquity, the gravitational tidal forces -- due to, for example, the host star acting on a Hot Jupiter -- cause its spin axis to rotate around a secondary rotation axis (approximately aligned with the orbit normal vector): they cause the Hot Jupiter to precess. In recent years, the excitation of obliquities of exoplanets has received a fair share of attention. Numerous mechanisms to explain the origin of this excitation have been explored, among which are planet-disk interactions \citep[][]{Millholland2019,Su2020,Martin2021}, mergers of proto-planets \citep[][]{Li2020} and planet-planet scattering \citep[][]{Li2021}. However, the close proximities of Hot Jupiters, and other short-period exoplanets, to their host stars yield strong tidal effects. This produces axial precession on relatively short timescales (of order a year or shorter). The resulting tidal deformations and tidal flows, and dissipation of the latter, additionally lead to transfers of angular momentum and energy from one body to its companion. This can result in many long-term effects in exoplanetary and close binary systems, such as tidal circularisation of orbits \citep[e.g.][]{binarycirc,Barker2022}, spin-orbit synchronisation \citep[e.g.][]{tidalcircularizationDobbs, Binarysync}, tidal heating \citep[potentially leading to radius inflation, e.g.][]{Bodenheimer_2001} and orbital decay, which has potentially been observed for WASP-12b \citep[e.g.][]{M2016,Orbitaldecaysummary,wasp12bdecay}. Crucially, the transfer of angular momentum due to tidal interactions may result in the spin and orbital axes of these Hot Jupiters tending toward alignment ($0^\circ$), anti-alignment ($180^\circ$), or perpendicularity ($90^\circ$), depending on the tidal mechanism involved \citep[though the ultimate tidal evolution, if allowed to proceed, will be towards alignment in isolated two-body systems, e.g.~][]{Lai2012,Ogilvie2014, Barker2016precession}.

The planetary obliquities of four extrasolar wide-orbiting Jupiter-like objects have been constrained by observations \citep[][]{Bryan2020,Bryan2021,Palma-Bifani2023,Poon2024}, and with ever-improving detection techniques \citep[][]{Barnes2003,Carter2010,Biersteker2017,Akinsanmi2020} it may be possible to measure obliquities due to differences (and time variability) in the transit signals caused by the oblateness of transiting Hot Jupiters. Hence, while awaiting further observations constraining their oblatenesses and obliquities, we can attempt to estimate theoretically the orbital period out to which Hot Jupiters can be tidally (re)aligned. Since the precession periods of these objects are expected to be relatively short and therefore the effects of precession to be strong, we study the efficiency with which a fluid instability of the flow in precessing objects, known as the precessional instability \citep[][]{Kerswell1993}, can align these objects. In particular, we will study whether and how this mechanism is impacted by 
convection.

The precessional instability is an instability associated with the $l=2,\ m=1,\ n=0$ component of the equilibrium tide, where $l$ denotes the spherical harmonic degree, $m$ the azimuthal wavenumber and $n$ the harmonic of the orbital frequency ($n=0$ implies that this component is stationary in an inertial frame). The equilibrium or non-wavelike tide is one of the two components that the tidal response in a star or planet is usually split into, the other being the dynamical or wave-like tide \citep[e.g.][]{Zahn1977tidesplit,OgilvieIW, Ogilvie2014,B2025}. The equilibrium tide is the quasi-hydrostatic fluid bulge, and associated flow, rotating around the body \citep[e.g.][]{Zahn1977tidesplit}, while the dynamical tide consists of waves generated by resonant tidal forcing (such as inertial waves in convection zones or internal gravity -- or gravito-inertial -- waves in radiation zones). Tidal dissipation of the equilibrium tide is thought to result from its interaction with turbulence, usually of a convective nature \citep[][]{Zahn1966,GoldreichNicholson1977,Zahn1989Turbvisceqtide,Goodmaneffvisc,Penev2007,Penev2009a,Penev2009b,Ogilvieeffvisc,Braviner_thesis,Craig2019effvisc,Craig2020effvisc,Vidal_Barker_2019,Vidal_Barker_2020,deVries2023b}, or by instabilities of the equilibrium tide itself, such as the precessional or elliptical instabilities, which can result in the excitation of waves \citep[e.g.][]{Cebron2010,librationellip,CebronellipHJ,Barker2013,BBO2016,Barker2016,deVries2023}. The $l=2,\ m=1,\ n=0$ component of the tidal potential, among others, produces what is known as obliquity tides \citep[][]{Ogilvie2014}. This component has a tidal frequency $\omega=-\Omega$ in the frame rotating with the body (using the sign convention where each tidal component is proportional to $\mathrm{e}^{\mathrm{i}(m\phi - \omega t)}$), with $\Omega$ its axial rotation rate or spin. It is static in the inertial frame and causes bulk precessional motion of the star or planet. These precessional flows within solid boundaries are described by the laminar flow solution known as the Poincar\'{e} solution \citep[][]{Poincare1910}{}{}. It is this laminar solution that is unstable to the precessional instability, which typically grows with a growth rate proportional to the precession rate, and may allow turbulence to arise \citep{Kerswell1993}. 

The properties of the precessional instability are very similar to the related elliptical instability \citep{Ellipticalinstability, LeBars_prec_2015}. The precessing flow excites pairs of inertial waves through a parametric resonance, and these inertial waves initially grow exponentially. The linear instability subsequently saturates, producing small-scale inertial waves which finally collapse to rotating turbulence, after which the cycle starts anew. This effect has been observed in simulations of the precessional instability executed in a Cartesian shearing box model with periodic boundaries \citep{Barker2016precession} and in spheroidal shells with stress-free boundaries \citep{LORENZANI2003}. Precession, and the associated precessional instability, produces sustained turbulence in realistic spherical \citep{TILGNER2001} or spheroidal geometries \citep{LORENZANI2001,LORENZANI2003}. As such, it is thought that the precessional instability could power a dynamo in, for example, Earth's liquid outer core \citep{Malkus1968,Kerswell1996,Tilgner2005,Wu2008, LeBars_prec_2015}.

The precessional instability can be studied in isolation by considering a local box approach in which the streamlines are circular, but vertically sheared. The initial linear instability analysis in this setup was performed in \citet[][]{Kerswell1993}. The fluid in the box rotates around the spin axis and the rotation of precession is represented by a secondary rotation perpendicular to the spin axis, with rotation rate $\Omega_p=\mathrm{Po}\Omega$. The parameter $\mathrm{Po}$ is the Poincar\'{e} number and indicates how fast precession occurs compared to the spin rotation; by analogy to the elliptical instability, this parameter is sometimes also denoted by $\epsilon$ and is usually small. The growth rate of the precessional instability is proportional to $\mathrm{Po}$ \citep{Kerswell1993}. This local approach to study the precessional instability has been taken in various papers \citep[e.g.][]{Barker2016precession,Khlifi2018,Salhi2019,Pizzi2022, kumar2023local}. The non-linear energy transfers due to the evolution of the instability were examined numerically in \citet{Pizzi2022}. They identified two different regimes, one at small $\mathrm{Po}$, in which large-scale columnar vortices are dominant, and one at higher $\mathrm{Po}$ where inertial waves contribute more significantly. Furthermore, the precessional instability was studied in tandem with weak magnetic fields as well, which show that the precessional flow and its instabilities can act as a dynamo \citep{Barker2016precession,kumar2023local}. Finally, the evolution of the precessional instability in isolation, together with walls in the vertical direction, has been studied in  \citet{masonkerswell} and \citet{Wu2008}. These works found that the introduction of walls imposes a mode selection constraint on the precessional instability. This constraint forces a distinct growth rate on each set of inertial waves that are able to resonate and grow, and reduces the maximum growth rate (though it typically remains $O(\mathrm{Po})$) compared to expectations from the unbounded analysis in \citet{Kerswell1993}.

The interactions of the precessional instability and convection have so far been studied using a linear stability analysis in cylindrical geometry in \citet{Benkacem2022}, and using simulations in spherical geometry in \citet{WeiTilgner2013}. The former found that the instability is slightly enhanced for weak convection, and is suppressed for strong convection. The latter also found that the precessional instability is enhanced for weak convection, allowing it to onset for smaller Po than it could in the absence of convection. Furthermore, it was found that convection can onset at weaker convective driving (smaller Rayleigh numbers) when weak precession is present. With strong precession, they also found that stronger convective driving was required for it to operate. Finally, in a similar fashion to the elliptical instability \citep[][]{Lavorelexperimentalellip,LE_BARS_elliptical, Cebron2010,Cebron_terrestrial, deVries2023}, it was found that precession can cause additional heat transport compared to convection in isolation.

In this paper, we will study the interactions of the precessional instability and convection in a local Cartesian box with walls in the vertical direction. This can be thought to model a small patch of the convection zone of a star or planet. Imposing walls is beneficial for controlling the properties of convection \citep[as explained in][who studied interactions of convection and oscillatory tidal flows]{Craig2019effvisc}, even if they may not be expected to be present in real stars or giant planets. We will examine whether the precessional instability is suppressed by its interaction with convectively-driven large-scale vortices, as we have found for the related elliptical instability \citep[][]{deVries2023}. Furthermore, we will examine whether a turbulent effective viscosity due to convection interacting with the equilibrium tide arises, and whether it can be used to interpret this interaction. Next, we will perform a large parameter sweep to determine scaling laws for the resulting tidal dissipation, examining the effects of varying the Poincar\'{e} number, Rayleigh number (which measures the strength of convective driving), and Ekman number (which measures the ratio of viscous to Coriolis forces). Finally, we will apply these scaling laws to compute the tidal dissipation due to the precessional instability and turbulent effective viscosity in interior models of Hot Jupiters obtained using the Modules for Experiments in Stellar Astrophysics (\textsc{mesa}) code. This allows us to make predictions for planetary (modified) tidal quality factors $Q'$, and the resulting timescales for significant spin-orbit evolution in hot and warm Jupiter systems.

In Section \ref{sec:model_setup} we will describe the model used in this work and discuss some theoretical predictions of the scaling laws for various quantities of interest. In Section \ref{sec:numerical_setup} we detail the numerical setup we employ and verify that we accurately capture both the precessional and convective instabilities with our simulations. Next, in Section \ref{sec:illustrative_simulations} we examine snapshots and time series of the flow to study the different behaviours that emerge within this system as the control parameters are varied. Then, in Section \ref{sec:scaling_laws} we fit the derived scaling laws to the time-averaged values of the quantities of interest. In Section \ref{sec: astro app} we outline the astrophysical implications of our results by generating interior profiles of a Jupiter-like and a Hot Jupiter planet using the \textsc{mesa} code, which we use to evaluate the dissipation of the equilibrium tide and that due to inertial waves. Finally, we conclude in Section \ref{sec:discussion}.

\section{Model setup}
\label{sec:model_setup}
\subsection{Governing equations and setup of the problem}

\begin{figure}
    \centering
    \includegraphics[width=\linewidth]{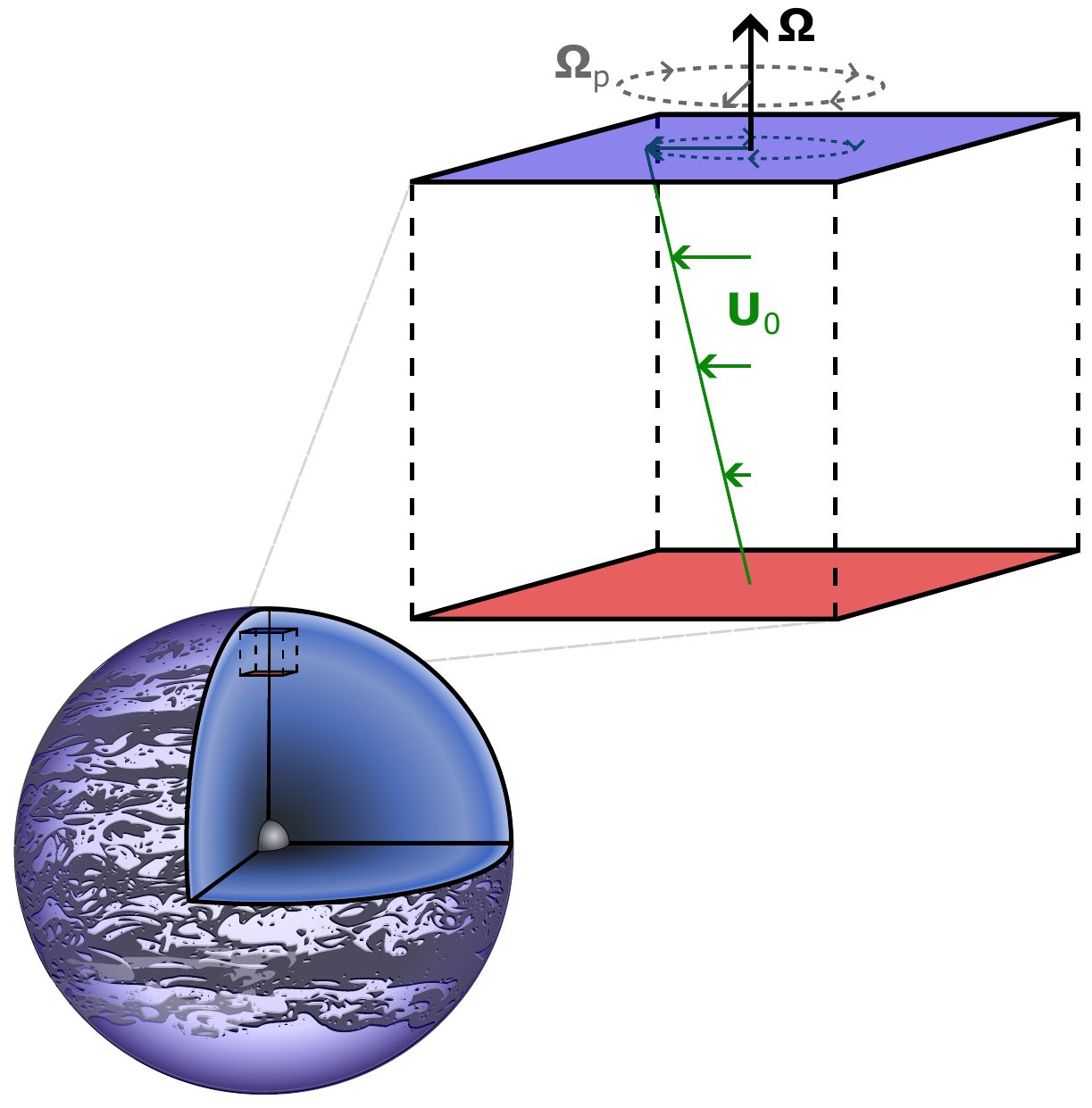}
    \caption{The location of the local box in the convection zone of a Hot Jupiter. We indicate the rotation vector $\bm{\Omega}$ in black, the precession vector $\bm{\Omega}_p$ in dark grey and the background flow $\bm{U}_0$ in green, all in the boundary frame and not to scale relative to each other. The background flow takes the form of a shear in $z$, the direction of which rotates as a function of time. The precession vector also rotates as a function of time, always lying in the $x,y$-plane. The time dependence of the direction of both $\bm{\Omega}_p$ and $\bm{U}_{0}$ is illustrated by dark and green dashed circles respectively. The local temperature gradient is represented by the red (hot) and blue (cold) sides of the box.}
    \label{fig:Precession_reference_image}
\end{figure}

We use a local Cartesian model, shown in Fig.~\ref{fig:Precession_reference_image}, building upon the previous works of
\citet{Kerswell1993,masonkerswell,Wu2008,Barker2016precession,Pizzi2022,kumar2023local}{}{} to study the interaction of the precessional instability with convection, with a focus on studying the resulting tidal dissipation. To study the precessional instability we choose to work in the ``boundary frame". In this frame, the flow inside the body appears to precess according to an observer located on the boundary of the body, and is therefore sometimes also referred to as the ``mantle frame". The spin axis points in the $z$-direction, and the fluid rotates in this frame at the rate $\Omega$. The component of the precessional vector that is parallel to the body's axial rotation is unimportant for the precessional instability, so it is set to zero. The precession vector is given by $\bm{\Omega}_p=\mathrm{Po}\,\Omega (\cos(\Omega t), -\sin(\Omega t),0)^T$, with $\mathrm{Po}=\Omega_p/\Omega$ the Poincar\'{e} number. The precession vector thus rotates around the body in the equatorial plane in this frame. The Poincar\'{e} number is a small parameter in most astrophysical systems. On Earth, for example, $\mathrm{Po}\approx10^{-7}$ due to lunisolar precession. Even so, \citet[][]{Kerswell1996}{}{} estimated that the maximum possible energy injected by the precessional instability could be sufficient in principle to power the geodynamo. In Hot Jupiters, with both larger oblateness and closer proximity to the tide-raising body, and therefore stronger torques, these values are more likely to be of order $\mathrm{Po}=\mathcal{O}(10^{-4}-10^{-3})$. These values are small, but still sufficiently large that the precessional instability could be important for aligning the spins and orbits of such Hot Jupiters \citep[][]{Barker2016precession}{}{}. Note that the frequency of the precessional flow and the frequency of the precession itself differ. In the boundary frame, ignoring the effects of viscosity and (stable or unstable) stratification, and considering just the shearing effects of the precessional flow, the precessional flow that satisfies the incompressible equations of motion can be written as \citep[][]{Kerswell1993}: 
\begin{equation}
\bm{U}_0=\mathrm{A}\bm{x}=-2\mathrm{Po}\Omega\begin{pmatrix}
0 & 0 & \sin(\Omega t)\\
0 & 0 & \cos(\Omega t)\\
0 & 0 & 0
\end{pmatrix}\bm{x},
\label{eq:backgroundflow}
\end{equation}
where $\bm{x}$ represents the position vector from the centre of the body. The precessing flow in this frame takes the form of an oscillatory shear flow in $z$, whose direction rotates in the $(x,y)$-plane. Alternatively, this can be interpreted as a rotating flow which is sheared in $z$.\\

We use stress-free impermeable walls in the $z$-direction to study the precessional instability, following ~\citet[][]{masonkerswell}{}{}, which allows us to study rotating Rayleigh-B\'{e}nard convection. This is the best-studied model of rotating convection, and it allows convection to be well controlled compared to models with periodic boundaries in $z$ (so-called ``homogeneous'' convection). We adopt a box with dimensions $[d,d,d]$, i.e., a box where $L_x=L_y=L_z$. We will defer studying different aspect ratios to later work. Gravity points in the vertical direction, $\bm{g}=g \bm{\hat{z}}$, and as a result it is (anti-)aligned with the rotation axis, i.e. the box is situated near the poles of the body. The background temperature profile is given by the conduction-state profile that depends only on $z$:
\begin{equation}
    T(z)=T_0-\frac{\Delta T}{d}z,
\end{equation}
where $T_0$ is a uniform background temperature that is set to zero without loss of generality, and $\Delta T$ is the temperature drop across the vertical extent of the box. The precession vector is often rewritten as $\bm{\epsilon}(t)=\mathrm{Po}\Omega(\cos(\Omega t), -\sin(\Omega t),0)^T$. This precession vector adds a second Coriolis-like term to the momentum equation. We decompose the total velocity, temperature and pressure profile into a background state and a perturbation, as follows: $\bm{U} = \bm{U}_0+\bm{u}$, $T_{\mathrm{tot}}= T(z) + \theta$ and $P_{\mathrm{tot}}=P(z) + p$. Applying these decompositions and non-dimensionalising the governing equations by scaling lengths with the vertical domain size $d$, times with $d^2/\kappa$, velocities with $\kappa/d$, temperatures with $\Delta T$, and pressures with $\rho_0 \kappa^2/d^2$, yields: 
\begin{equation}
     \frac{\mathrm{D} \bm{u}}{\mathrm{D} t} + \bm{u} \cdot \nabla \bm{U}_0 +  \frac{\mathrm{Pr}}{\mathrm{Ek}}(\bm{\hat{z}} + \bm{\epsilon} (t))\times \bm{u} = -\nabla p + \mathrm{RaPr}\theta \bm{\hat{z}} + \mathrm{Pr} \nabla^2 \bm{u},
     \label{eq:govern_momentum}
\end{equation}
\begin{equation}
    \nabla \cdot \bm{u} = 0,
    \label{eq:govern_incompress}
\end{equation}
\begin{equation}
    \frac{\mathrm{D} \theta}{\mathrm{D} t} - u_z = \nabla^2 \theta,
    \label{eq:govern_heat}
\end{equation}
\noindent where 
\begin{equation}
    \frac{\mathrm{D}}{\mathrm{D}t}\equiv \frac{\partial}{\partial t} + \bm{U}_0\cdot\nabla + \bm{u}\cdot\nabla.
\end{equation}

\noindent The non-dimensional parameters describing the system are the Ekman number, the Rayleigh number and the Prandtl number: 
\begin{equation}
\mathrm{Ek} = \frac{\nu}{2\Omega d^2}, \quad \mathrm{Ra} = \frac{\alpha g \Delta T d^3}{\nu\kappa}, \quad \mathrm{Pr}=\nu/\kappa,
\end{equation}
in addition to the Poincar\'{e} number Po.

The dimensional buoyancy frequency $N^2$ is related to the Rayleigh number according to $N^2 = -\mathrm{Ra\,Pr}\,\kappa^2/(\alpha g d^4)$, so that when $\mathrm{Pr}=1$, the dimensionless value (in thermal time units, where $d^2/\kappa=1$) is $N^2=-\mathrm{Ra}$. We set $\mathrm{Pr}=1$ for consistency with the previous works we will build upon e.g. \citet{Cebron2010,WeiTilgner2013,Barker2014RMLT,CelineLSV}. The boundary conditions in the horizontal directions are periodic, while in the vertical direction we impose walls that are impermeable, $u_z(z=0)=u_z(z=d)=0$, stress-free (vanishing tangential viscous stresses), $\partial_z u_x(z=0)=\partial_z u_x(z=d)=\partial_z u_y(z=0)=\partial_z u_y(z=d)=0$, and perfectly (thermally) conducting, $\theta(z=0)=\theta(z=d)=0$.

\subsection{Known properties of the precessional instability}
\label{sec:properties_prec}
A linear stability analysis of the precessional instability, in the absence of convection, but with walls in the vertical direction, has been performed in \citet{masonkerswell}. We summarise their most important results here. The most unstable modes take the form of inertial wave modes \citep[][]{masonkerswell}, which arise in a rotating local box with two stress-free and impermeable walls \citep{greenspan1968}. These two modes, denoted by $A$ and $B$, must satisfy $\lambda_B=\lambda_A\pm\Omega$, since the precessing flow has tidal frequency magnitude $\Omega$. Here, $\lambda_A$ and $\lambda_B$ are the frequencies of the inertial waves, which must each satisfy the inertial wave dispersion relation:
\begin{equation}
\lambda=\pm\frac{2\Omega k_z}{k},
\end{equation}
where $k_z$ is the vertical wavenumber, $k$ is the wavevector magnitude, and $\lambda$ is the frequency.

Furthermore, due to the introduction of stress-free impermeable walls at the top and bottom, solutions must be proportional to $\cos (n\pi z)$ (for $u_x,u_y,p$) or $\sin(n\pi z)$ (for $u_z,\theta$), with integer $n$. Hence, the vertical wavenumbers of the two waves, given by $k_z = n\pi$, must satisfy $(n_B-n_A)\ \mathrm{mod}\ 2 = 1$, i.e. the difference between $n_A$ and $n_B$ must be odd. This, in combination with the requirement that $k_{\perp,B}=k_{\perp,A}$, where $k_\perp$ is the horizontal wavenumber, implies that there is a value of $k_\perp$ that satisfies the resonance conditions on the frequency for every combination of $n_A$ and $n_B$. The precessional instability, with stress-free walls, thus has a characteristic size for each mode pair. The particular value of $k_\perp$ for each pair of modes also depends on $\mathrm{Po}$, as illustrated in Fig.~3 of \citet[][]{masonkerswell}{}{}. Some examples of the inviscid growth rates and associated values of $k_\perp$ are found in Table 1 of \citet[][]{masonkerswell}{}{}. The modes with the largest growth rates in the simulations in this work are those with $n_A=1$, $n_B=2$, due to the introduction of viscosity, in which case the velocity field -- in the absence of stratification and viscosity -- grows according to \citep[][]{masonkerswell}{}{}:
\begin{equation}
    \bm{u} = \mathrm{Re}\{(A_0\bm{u}_A\mathrm{e}^{-\mathrm{i}\lambda_At}+B_0\bm{u}_B\mathrm{e}^{-\mathrm{i}\lambda_Bt})\mathrm{e}^{0.3547\mathrm{Po}\Omega t}\},
    \label{eq:prec_growthn1n2}
\end{equation}
where $A_0$ and $B_0$ are the complex amplitudes at $t=0$ of the two modes and $\bm{u}_A$ and $\bm{u}_B$ are the velocity eigenvectors of the two waves.

The growth rate of the instability is thus $0.3547\mathrm{Po}\Omega$, which can be compared to the theoretical maximum value of $\approx0.385\mathrm{Po}\Omega$, which is a factor of $2/\pi$ smaller than in the triply periodic case \citep[][]{masonkerswell}{}{}. The introduction of walls therefore decreases the growth rate of the precessional instability. Finally, the growth rate vanishes identically when either of $n_A$ or $n_B$ are equal to zero, such that these modes should be unable to grow due to this instability mechanism.

\subsection{Energetic analysis of simulations}
\label{sec:diagnostics}

To analyse the flow we derive the kinetic energy equation by taking the scalar product with $\bm{u}$ of Eq.~\eqref{eq:govern_momentum} and then averaging over the box. We define the averaging operation on a quantity $X$ as $\langle X\rangle=\frac{1}{d^3}\int_V X\!\ \mathrm{d}V$. We obtain:
\begin{equation}
    \frac{\mathrm{d}}{\mathrm{d}t} K= I + W_B - D_{\nu},
    \label{eq:kin_energy}
\end{equation}

We have defined the mean kinetic energy $K$, the work done by buoyancy forces $W_B$, and the mean viscous dissipation rate $D_{\nu}$:
\begin{equation}
    K\equiv\frac{1}{2}\langle|\bm{u}|^2\rangle, \qquad W_B\equiv \mathrm{PrRa}\langle\theta u_z\rangle, \qquad D_{\nu}\equiv-\mathrm{Pr} \langle \bm{u} \cdot \nabla^2 \bm{u} \rangle,
    \label{eq:Kdef}
\end{equation}
as well as the energy transfer rate from the background precessional flow $I$:
\begin{equation}
    I\equiv-\langle\bm{u}\mathrm{A}\bm{u}\rangle=-\langle \bm{u}\cdot(\bm{u}\cdot \nabla \bm{U}_0)\rangle.
    \label{eq:Idef}
\end{equation}

To obtain an equation for the thermal (potential) energy when the stratification is convectively unstable, i.e. $\textrm{Ra}>0$, we multiply Eq.~\eqref{eq:govern_heat} by $\textrm{PrRa}\theta$ and average over the box to obtain, in a similar manner:
\begin{equation}
    \frac{\mathrm{d}}{\mathrm{d}t}P=W_B-D_{\kappa},
    \label{eq:therm_energy}
\end{equation}
where we have defined the mean thermal energy $P$ and the mean thermal dissipation rate $D_\kappa$ as:
\begin{equation}
    P\equiv\text{PrRa}\frac{1}{2}\langle\theta^2\rangle, \quad D_\kappa\equiv-\text{PrRa}\langle\theta\nabla^2\theta\rangle.
    \label{eq:Pdef_ch2}
\end{equation}     
\noindent The total energy is $E=K+P$, which thus obeys:
\begin{equation}
    \frac{\mathrm{d}}{\mathrm{d}t}E=I+2W_B-D_\nu-D_\kappa=I+2W_B-D,
    \label{eq:Energ_equation}
\end{equation}
\noindent where $D=D_\nu + D_\kappa$ is the total dissipation rate. In a steady state, i.e. no change in time of the total energy, it is expected that the (time-averaged value of the) energy injected together with the buoyancy work balances the total dissipation. Because of the presence of the buoyancy work term, the total dissipation is not equivalent to the tidal dissipation rate. However, the energy injected by the tide must be dissipated if a steady state is to be maintained. Therefore, to interpret the tidal energy dissipation rate, we examine the tidal energy injection rate $I$, which equals $D-2W_B$ in a steady state.

A scaling law for the dissipation due to the precessional instability was proposed in \citet[][]{Barker2016precession}. They considered a single most unstable mode whose amplitude saturates when its growth rate balances its non-linear cascade rate, i.e. $\sigma\sim ku\sim 1/t_{\mathrm{damp}}$, where $k$ is the wavenumber magnitude and $u$ is the velocity amplitude. This implies that the mode attains a velocity amplitude $u\sim\mathrm{Po}\Omega/k$, so the resulting tidal dissipation is therefore expected to scale as $D\sim u^2/t_{\mathrm{damp}}\sim\mathrm{Po}^3\Omega^3k^{-2}$. Or more generally:
\begin{equation}
    D=I\propto \mathrm{Po}^3.
\end{equation}
\noindent This scaling is consistent with some local simulations \citep{Barker2016precession,Pizzi2022} at sufficiently high (but still $\mathrm{Po}<1$) values of the Poincar\'{e} number. An energy injection rate that is consistent with a scaling closer to $\mathrm{Po}^2$ at low values of the Poincar\'{e} number was also observed by \citet[][]{Barker2016precession}. A theoretical basis for the latter scaling was not identified, but we will examine if it also arises in this work. 

Based on our studies in \citet[][]{deVries2023} of the elliptical instability and rotating convection, we expect to observe the non-linear creation of large-scale geostrophic flows from both the convective instability as well as the precessional instability. In this work, we also decompose the flow into a 2D $z$-invariant flow and a 3D $z$-dependent flow, with the energy of the $z$-invariant flow $K_{2D}$ defined as:
\begin{equation}
    K_{2D} = \frac{1}{2V}\iint\left(\int u_x \!\ \mathrm{d}z\right)^2\!\ \mathrm{d}y \!\ \mathrm{d}x + \frac{1}{2V}\iint\left(\int u_y \!\ \mathrm{d}z\right)^2\!\ \mathrm{d}y \!\ \mathrm{d}x.
    \label{eq:K2D_split}
\end{equation}
\noindent We have chosen not to include $u_z$ in the definition of $K_{2D}$ because the $z$-invariant vertical velocity must be zero due to the impermeability boundary condition, such that the $k_z=0$ mode cannot contribute to $K_{2D}$. The definition of $K_{3D}$ then follows as:
\begin{equation}
    K_{3D}= K - K_{2D}.
\end{equation}
\noindent Furthermore, we can expand the energy injection using the definition of the background flow:
\begin{equation}
    I = 2\mathrm{Po}\Omega\left(\langle u_xu_z\rangle\sin(\Omega t)+\langle u_yu_z\rangle\cos(\Omega t)\right).
    \label{eq:I_def_explicity}
\end{equation}
\noindent We can see that, by definition, $I_{2D}$ defined similarly to Eq.~\eqref{eq:K2D_split} must be zero, because every term in Eq.~\eqref{eq:I_def_explicity} contains the vertical velocity. Thus, we do not decompose $I$ into $I_{2D}$ and $I_{3D}$ \citep[unlike in][for the elliptical instability]{deVries2023}.

Finally, we can use the same technique as in \citet[][]{Goodmaneffvisc,Ogilvieeffvisc,Craig2019effvisc,deVries2023b} to obtain an estimate of the effective viscosity due to the convection acting on the precessional flow, which we have strong reason to believe will arise in these simulations as well. The rate at which energy of the flow $\boldsymbol{U}_0$ is viscously dissipated with viscosity $\nu_{\mathrm{eff}}$, is given by:
\begin{equation}
    \frac{2\nu_{\mathrm{eff}}}{V}\int_V e_{ij}^0e_{ij}^0 \ \!\mathrm{d}V=4\nu_{\mathrm{eff}}\mathrm{Po}^2\Omega^2,
\end{equation}
where $e_{ij}^0 \equiv \frac{1}{2}(\partial_iU_{0,j}+\partial_jU_{0,i})$ is the strain rate tensor. Upon equating this expression to $I$, we can \textit{define} the effective viscosity according to:
\begin{equation}
    \nu_{\mathrm{eff}}=I/(4\mathrm{Po}^2\Omega^2).
    \label{eq: I_and_nueff}
\end{equation}
The effective viscosity might be expected to be independent of $\mathrm{Po}$; this would imply that, if the interaction of turbulence acting on the precessional flow can be parametrised as a turbulent effective viscosity, it should scale as:
\begin{equation}
    D=I\propto \mathrm{Po}^2.
    \label{eq:I_effviscscaling}
\end{equation}

Furthermore, the effective viscosity is expected to depend on the convective velocity ($u_c$), lengthscale ($l_c$) and frequency ($\omega_c$) according to \citep[][]{Craig2020effvisc}:
\begin{equation}
  \nu_{\mathrm{eff}} =
    \begin{cases}
      5u_{c}l_{c} & \frac{| \omega|}{\omega_c}\lesssim 10^{-2},\\
      \frac{1}{2}u_{c}l_{c}\Big(\frac{\omega_c}{\omega}\Big)^{\frac{1}{2}} & \frac{| \omega|}{\omega_c}\in[10^{-2},5],\\
      \frac{25}{\sqrt{20}}u_{c}l_{c}\Big(\frac{\omega_c}{\omega}\Big)^2 & \frac{| \omega|}{\omega_c}\gtrsim 5.
    \end{cases} 
    \label{eq:effviscCraig_Ch4}
\end{equation}

Using rotating mixing length theory (RMLT) these can be expressed in terms of the relevant non-dimensional parameters in our temperature-based Rayleigh-B\'{e}nard setup (where we fix $\Delta T$ rather than the heat flux), omitting prefactors \citep[][]{deVries2023b}:
\begin{equation}
  \nu_{\mathrm{eff}} \propto
    \begin{cases}
      \mathrm{Ra}^{3/2}\mathrm{Ek}^2\mathrm{Pr}^{-1/2}\kappa& \mathrm{low}\ \mathrm{frequency},\\
\mathrm{Ra}^{7/4}\mathrm{Ek}^{2}\mathrm{Pr}^{-1/4}\kappa^{3/2}d^{-1}\omega^{-1/2} & \mathrm{intermediate}\ \mathrm{frequency},\\
    \mathrm{Ra}^{5/2}\mathrm{Ek}^{2}\mathrm{Pr}^{1/2}\kappa^{3}d^{-4}\omega^{-2}& \mathrm{high}\ \mathrm{frequency}.
    \end{cases} 
    \label{eq:effviscscalings}
\end{equation}

Finally, we will also study the heat transport in these simulations, to see if we observe modification of the heat transport by precession, as was found by \citet[][]{WeiTilgner2013}. To this end, we study the Nusselt number in the simulations, defined as:
\begin{equation}
    \mathrm{Nu} = 1+ \mathrm{RaPr}\langle\theta u_z\rangle,
\end{equation}
which represents the ratio of the total heat flux to the conductive heat flux, which is identically one in the absence of any flows.

\section{Numerical setup}
\label{sec:numerical_setup}

We will use two codes in tandem to study the precessional instability and its interactions with convection, namely \textsc{dedalus} \citep[][]{Dedalus2020} and \textsc{nek5000} \citep[][]{nek5000-web-page}. Obtaining agreement between the two codes partly ensures that the dynamics in the system are being captured accurately, as they are based on different numerical methods. \textsc{dedalus} implements a pseudo-spectral method. We use Chebsyhev polynomials in $z$ and the ``real" Fourier basis, i.e. sines and cosines instead of complex exponentials, in the $x$ and $y$ directions. We set the number of grid points in the $x$, $y$ and $z$ directions to $N_x=N_y=N_z=96$, unless otherwise specified. The choice of Chebyshev polynomials is responsible for our definition of $K_{2D}$ in real space in Eq.~\eqref{eq:K2D_split}. Pseudo-spectral methods converge exponentially with increasing resolution for smooth solutions \citep[]{boyd2001chebyshev}{}{}. The precessional instability in a triply periodic domain has previously been studied by employing so-called shearing waves, with time-dependent wavevectors, but the introduction of impermeable walls at the top and bottom has made these shearing waves redundant. We employ a two-stage, second-order, Runge-Kutta timestepping scheme, identified in \textsc{dedalus} as RK222 \citep[described in Sec.~2.6 of][]{Ascher1997}{}{}. In this scheme the non-linear terms, advection terms between the background flow and the velocity perturbations, and the Coriolis-like term that arises due to the precession are explicitly calculated, while the other terms -- including the usual Coriolis term -- are calculated implicitly. We employ a CFL safety factor set to $0.5$ to ensure that timesteps are small enough to accurately capture the dynamics. For de-aliasing purposes the resolution is expanded with a factor $3/2$ when calculating the non-linear terms, so as to satisfy the standard $2/3$ rule \citep{boyd2001chebyshev}. 

We also employ the \textsc{nek5000} code, which utilises a spectral element method with an equal number of regularly-spaced elements $\mathcal{E}$ in each direction, and thus $\mathcal{E}^3$ elements in total. Inside each element, the velocity, temperature, and pressure fields are then expanded using Legendre polynomials. The associated grid points are the Gauss-Legendre-Lobatto quadrature points, such that the edges of the elements are included in the grid. The number of grid points inside each element is given by $(\mathcal{N}+1)^3$. Due to overlapping grid points on the boundaries of the elements, the total number of grid points in a given simulation is $\mathcal{E}^3\mathcal{N}^3$. We work in a Cartesian box with sides of equal length, with $\mathcal{E}=10$. We set $\mathcal{N}=9$ unless otherwise specified. Due to the grid points adopted within each element in \textsc{nek5000}, the grid is irregular in all three directions even if the elements themselves are uniformly spaced. We have therefore opted to interpolate to a regular grid to calculate $K_{2D}$, so this quantity is susceptible to interpolation errors. We will interpolate to a $100^3$ regular grid for such integrations, unless otherwise specified. We timestep using an implicit second-order backward difference scheme for the pressure and diffusion terms, and a second-order characteristics-based method to evolve the non-linear, buoyancy and Coriolis terms, as well as the Coriolis-like term due to the precession and the advection terms of the background flow and the velocity perturbation. We employ a CFL safety factor set to $0.3$ to ensure sufficiently small timesteps, although note that characteristics-based timesteppers are allowed to exceed this value while still guaranteeing accuracy when integrating the advective non-linear terms. De-aliasing is dealt with by increasing the order of the polynomials inside each element when calculating non-linear terms, with a factor that approximately satisfies the 2/3 rule \citep[][]{boyd2001chebyshev}{}{}. 

We use \textsc{nek5000} in addition to \textsc{dedalus} because results obtained by both codes independently are necessarily more robust, particularly as the methods implemented are different. This allows for an additional check on our results.

\subsection{Benchmarking}

\begin{figure*}
    \centering\begin{minipage}[b]{0.49\textwidth}
         \centering
         \includegraphics[width=\textwidth]{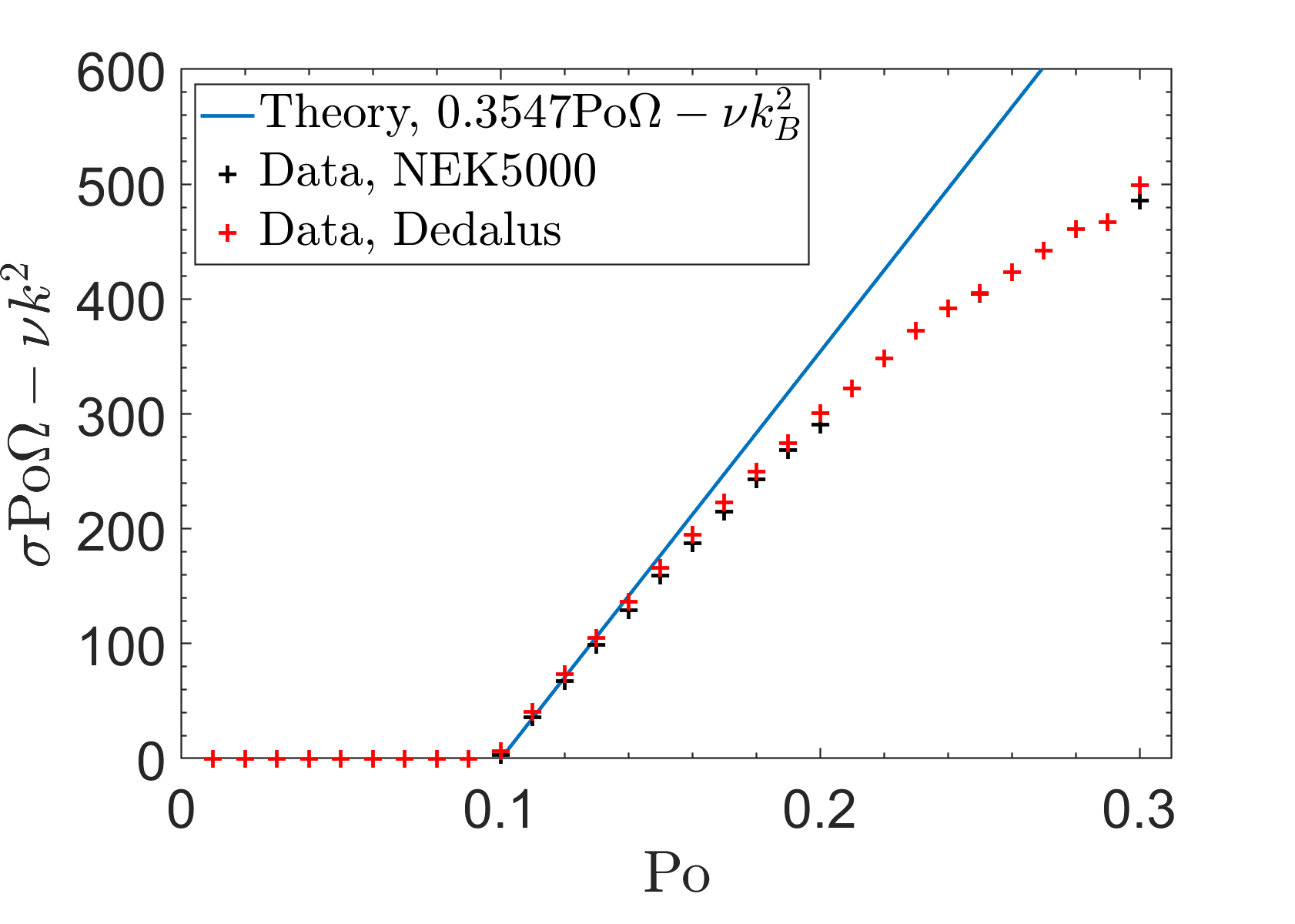}
     \end{minipage}
     \hfill
     \begin{minipage}[b]{0.49\textwidth}
         \centering
         \includegraphics[width=\textwidth]{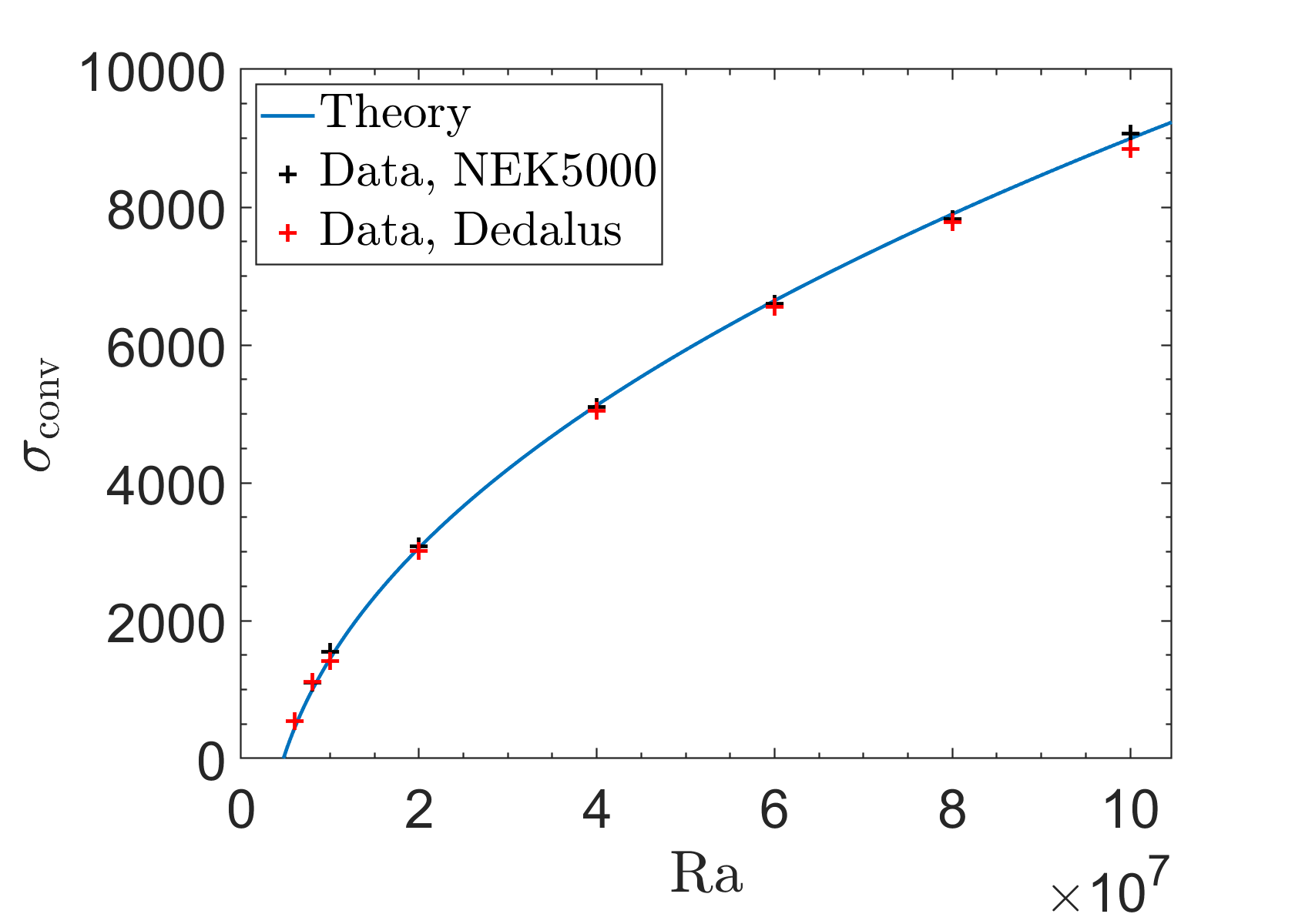}
     \end{minipage}
    \caption[Benchmark simulations of the linear growth rate of the precessional instability and convective instability, each in isolation.]{Benchmark simulations of the linear growth rate of the precessional instability and convective instability, each in isolation at $\mathrm{Ek}=5\cdot 10^{-5}$ in a 1-by-1-by-1 box using both \textsc{nek5000} (black) and \textsc{dedalus} (red). Left: growth rate of the precessional instability. The most unstable mode pair, after taking the viscous reduction into account, is $n_A=1,\ n_B=2$. The theoretical prediction for small $\mathrm{Po}$ for this mode is given in blue. The simulation data obtained using both codes match each other and the theoretical prediction well in the interval $\mathrm{Po}\in[0.1,0.15]$, and accurately predict the critical value of $\mathrm{Po}$ for inhibition of the instability by viscosity. Deviation from the predicted maximum growth at higher $\mathrm{Po}$ partly arises from effects that are of higher order in $\mathrm{Po}$ not captured by the linear analysis. In addition, the most unstable horizontal wavenumber changes as $\mathrm{Po}$ is increased, see Fig. 3a in \citet[][]{masonkerswell}{}{}, and therefore the mode becomes increasingly more de-tuned. Right: Benchmark simulations of the linear growth rate of convection in isolation at $\mathrm{Ek}=5\cdot 10^{-5}$ in a 1-by-1-by-1 box. The growth rates in the simulations match each other and the theoretical growth rate of convection well.}
    \label{fig:lin_growth_conv_prec}
\end{figure*}

We first perform benchmarking tests to ensure that the precessional and convective instabilities in isolation are each properly computed in the simulations. To this end, we performed numerous simulations with the perturbation non-linearities switched off in both \textsc{dedalus} and \textsc{nek5000}. We have fitted the growth rate of the kinetic energy in the simulations $K=\frac{1}{2}\langle|\bm{u}|^2\rangle$, where the brackets represent a volume-average. The growth rate of the velocity can then be found by taking half of the kinetic energy growth rate. We compare the growth rates from simulations to those from the theoretical predictions given in Eq.~\eqref{eq:prec_growthn1n2}, and obtained by numerically solving the relevant cubic dispersion relation, for the precessional and convective instabilities, respectively. The results of this benchmarking of both instabilities in isolation, along with the theoretical predictions, are plotted in Fig.~\ref{fig:lin_growth_conv_prec}. The \textsc{dedalus} simulations portrayed in this figure are executed in a 1-by-1-by-1 box, with a resolution of $32^3$. The \textsc{nek5000} simulations are executed in a 1-by-1-by-1 box with a resolution of $10^3$ elements, each with a polynomial of order $4$. Thus, the total number of grid points is $40^3$. 

The most unstable modes of the precessional instability are predicted to be those with the smallest horizontal wavenumber, according to \citet[][]{masonkerswell}{}{}, which are the modes with $n_A=1,\ n_B=2$ and $k_\perp=\sqrt{k_x^2+k_y^2}\approx18.059$. In the 1-by-1-by-1 box the modes with $n_x=n_y=2\Rightarrow k_x=k_y=4\pi$ have $k_\perp\approx17.77$, so the modes with these horizontal wavenumbers almost satisfy the perpendicular wavenumber requirement of the most unstable modes. Thus we expect these modes to match the theoretically predicted growth rate well, except for a small discrepancy, which can be attributed to de-tuning. The unstable horizontal wavenumber is found to change with $\mathrm{Po}$, however, and for this particular mode decreases as $\mathrm{Po}$ increases \citep[see Fig.~3a in][]{masonkerswell}. According to this figure, the unstable horizontal wavenumber of these modes satisfies $k_\perp\approx17.77$ around $\mathrm{Po}=0.1$. The secondary effects of viscosity in shifting the resonances are ignored, which are expected to be much weaker than the damping we do consider. The growth rate of the mode with $n_A=1,\ n_B=2$ is given by: $0.3547 \mathrm{Po} \Omega - \nu k^2$. This theoretically predicted growth rate, where we have chosen $k^2=k_\perp^2+n_B^2\pi^2$ (which probably slightly overestimates the viscous damping), is plotted in the solid-blue line in the left-hand panel of Fig.~\ref{fig:lin_growth_conv_prec} along with the simulation data plotted in black (\textsc{nek5000}) and red (\textsc{dedalus}) markers. The growth rates obtained from simulations executed with both \textsc{dedalus} and \textsc{nek5000} match each other very well. They also match the theoretically predicted growth rate very well in the interval $\mathrm{Po}\in[0.1, 0.15]$, where the most unstable wavenumber very accurately matches the one available within the box. Additionally, the theory predicts the range of Poincar\'{e} numbers below which the instability is inhibited by viscosity very well. The simulation data starts deviating from the theoretically predicted growth rate at higher values of $\mathrm{Po}$, as the most unstable mode changes to modes that cannot be accurately captured in the box. The discrepancy here can thus, at least partially, be attributed to de-tuning. Higher-order effects at large $\mathrm{Po}$ may also play a role in causing this discrepancy; see, for example, the higher-order correction to the precessional instability growth rate in \citet[][]{Naing2011}{}{}, albeit in a triply periodic box. Therefore we conclude that we have accurately captured the growth of the precessional instability in isolation in both codes. We have also reproduced the previously-obtained results from simulations with stress-free impermeable walls at the top and bottom of the local box \citep{masonkerswell}.

We have also run multiple simulations to ensure the convective instability is captured accurately. The theoretical predictions, as well as the simulation data at the same $\mathrm{Ek}=5\cdot 10^{-5}$ in a 1-by-1-by-1 box, are plotted in the right-hand panel of Fig.~\ref{fig:lin_growth_conv_prec}. To calculate the theoretical maximum growth rate, we solved the dispersion relation for rotating convection with stress-free impenetrable walls. The simulation data from both codes agree well with each other and with the theoretically predicted growth rates. Thus, we conclude that both the precessional instability and the convective instability in isolation are accurately captured by the two codes.

\subsection{Parameter variations}
To examine the quantities of interest, namely the kinetic energy, energy injection rate and heat transport efficiency, the latter of which is represented by the Nusselt number, we will execute multiple parameter sweeps. We will vary the Rayleigh number, Ekman number and Poincar\'{e} number, to be able to independently vary the convective driving, rotational constraint, and precessional driving, respectively. The Rayleigh number is typically reported instead using the ``supercriticality" $R=\mathrm{Ra}/\mathrm{Ra}_c$ for clarity, where $\mathrm{Ra}_c$ is the critical value for onset of instability (determined numerically by solving the dispersion relation for rotating convection with stress-free walls). The range of this ratio studied at $\mathrm{Ek}=2.5\cdot10^{-5}$ is $R\in[0,15]$. We examine $\mathrm{Po}\in[0.01,1.10]$, which also includes the (probably) astrophysically-irrelevant, but fluid-dynamically interesting, case that the precession is faster than the rotation. Finally, we performed simulations with three different values of the Ekman number: $\mathrm{Ek}=[5\cdot10^{-5}, 2.5\cdot 10^{-5}, 10^{-5}]$. When varying Ek we set $\mathrm{Ra}=0$, i.e. there is no convective instability in this set of simulations. \textsc{dedalus} provides more accurate values of $K_{2D}$ and allows accurate creation of horizontal power spectra without needing interpolation, but it was found to run more slowly than \textsc{nek5000}. As a result, we  opt to use the faster \textsc{nek5000} code to perform most parameter sweeps, while executing selected simulations for interesting cases using \textsc{dedalus}. We have checked that the two codes also produce the same results in select non-linear simulations with the same parameters. A table of all simulations performed can be found in Appendix~\ref{app:sim_tables}.

\section{Analysis of illustrative simulations}
\label{sec:illustrative_simulations}

\subsection{Snapshots of the vertical vorticity}

\begin{figure*}
\subfloat[$t=0.344$, $\mathrm{Po}=0.1$, $\mathrm{Ra}=0$. \label{fig:Po0.1_1}]{
         \includegraphics[width=0.45\textwidth,trim={0 0 2cm 1cm},clip=true]{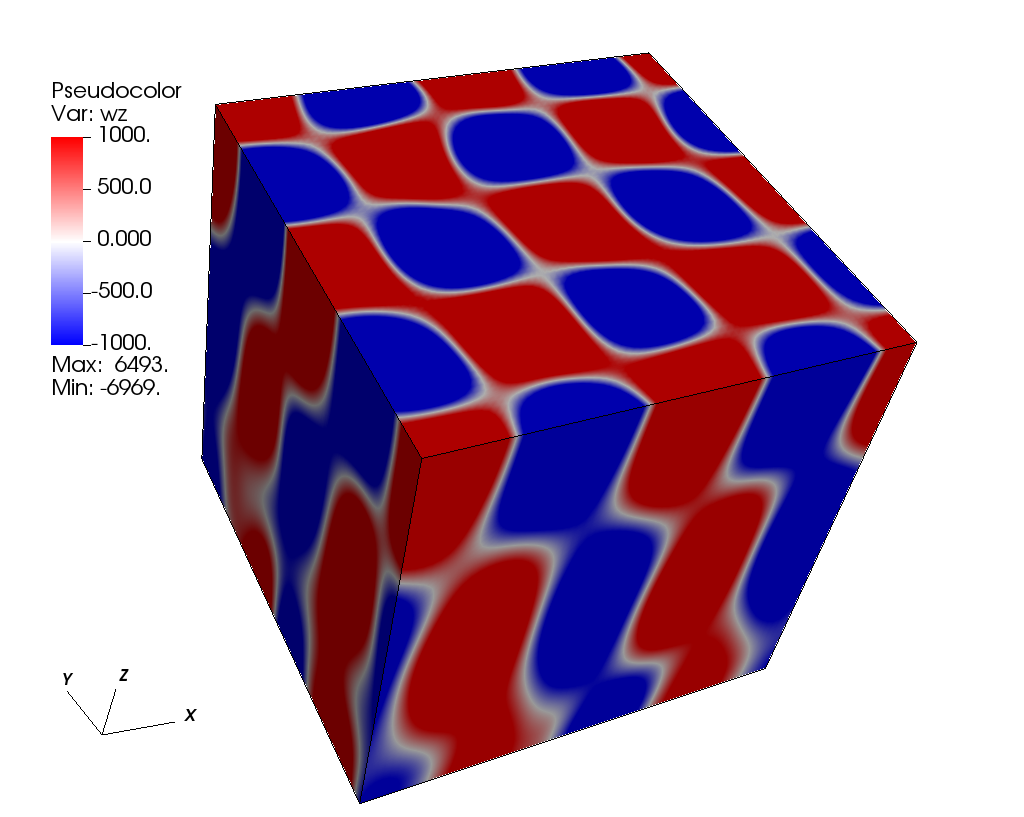}}
\hfill
\subfloat[$t=0.37$, $\mathrm{Po}=0.1$, $\mathrm{Ra}=0$. \label{fig:Po0.1_2}]{\includegraphics[width=0.45\textwidth,trim={0 0 2cm 1cm},clip]{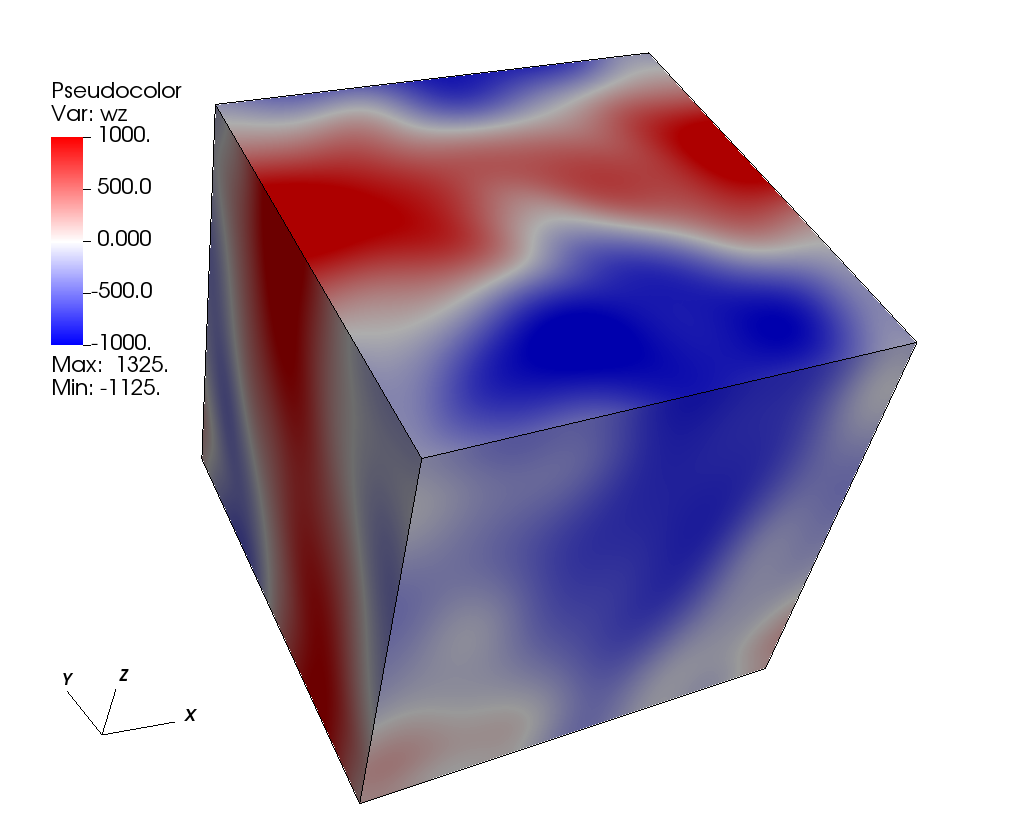}}\\

\subfloat[$t=0.062$, $\mathrm{Po}=0.2$, $\mathrm{Ra}=0$. \label{fig:Po0.2_1}]{\includegraphics[width=0.45\textwidth,trim={0 0 2cm 1cm},clip]{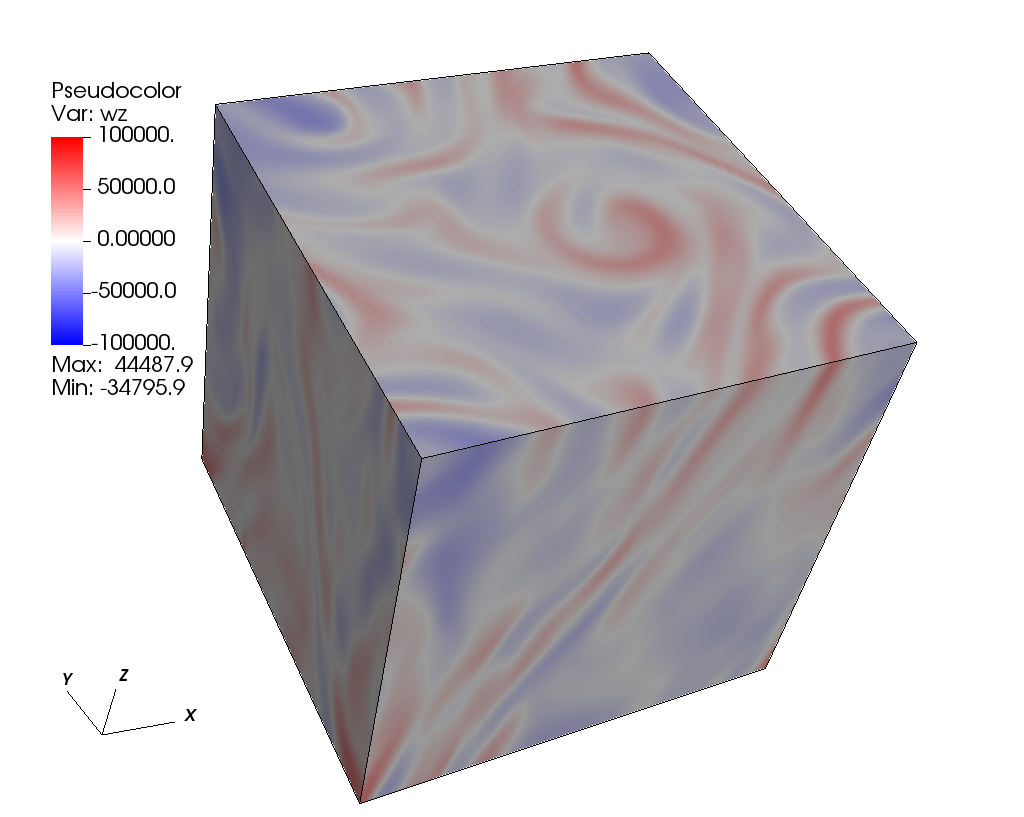}}
\hfill
\subfloat[$t=0.07$, $\mathrm{Po}=0.2$, $\mathrm{Ra}=0$. \label{fig:Po0.2_2}]{\includegraphics[width=0.45\textwidth,trim={0 0 2cm 1cm},clip]{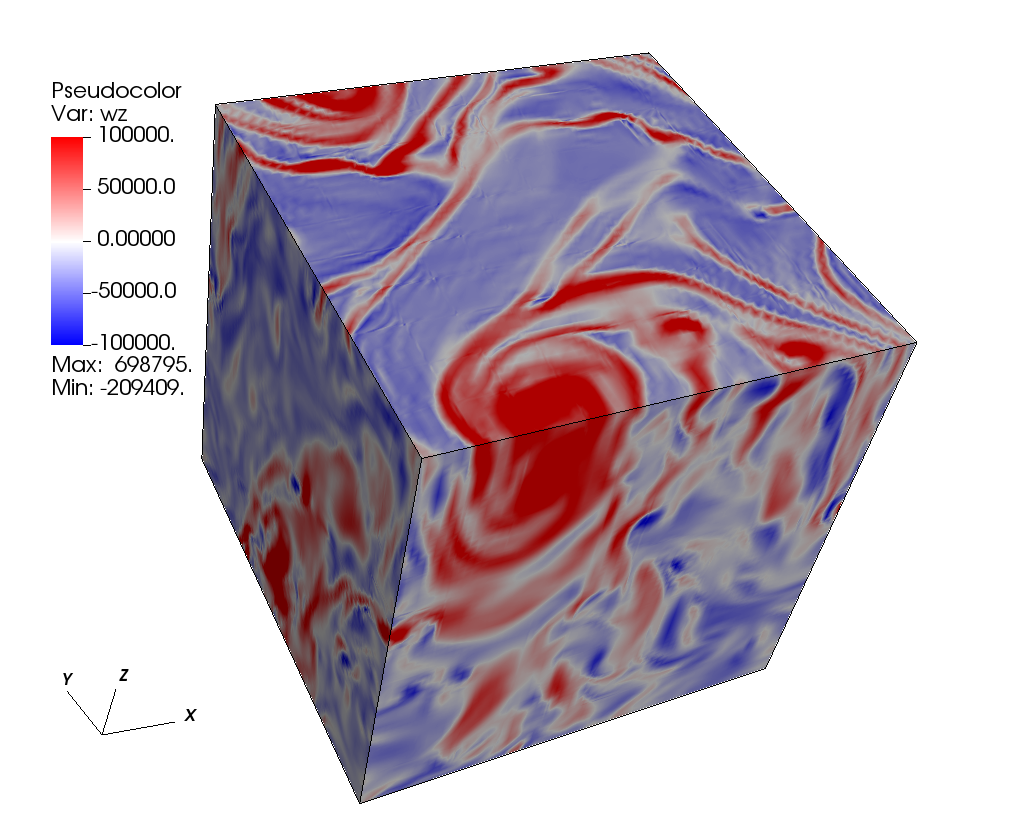}}\\

\subfloat[$t=0.05$, $\mathrm{Po}=0.1$, $\mathrm{Ra}=2\mathrm{Ra}_c$. \label{fig:Po0.1_Ra2}]{\includegraphics[width=0.45\textwidth,trim={0 0 2cm 1cm},clip]{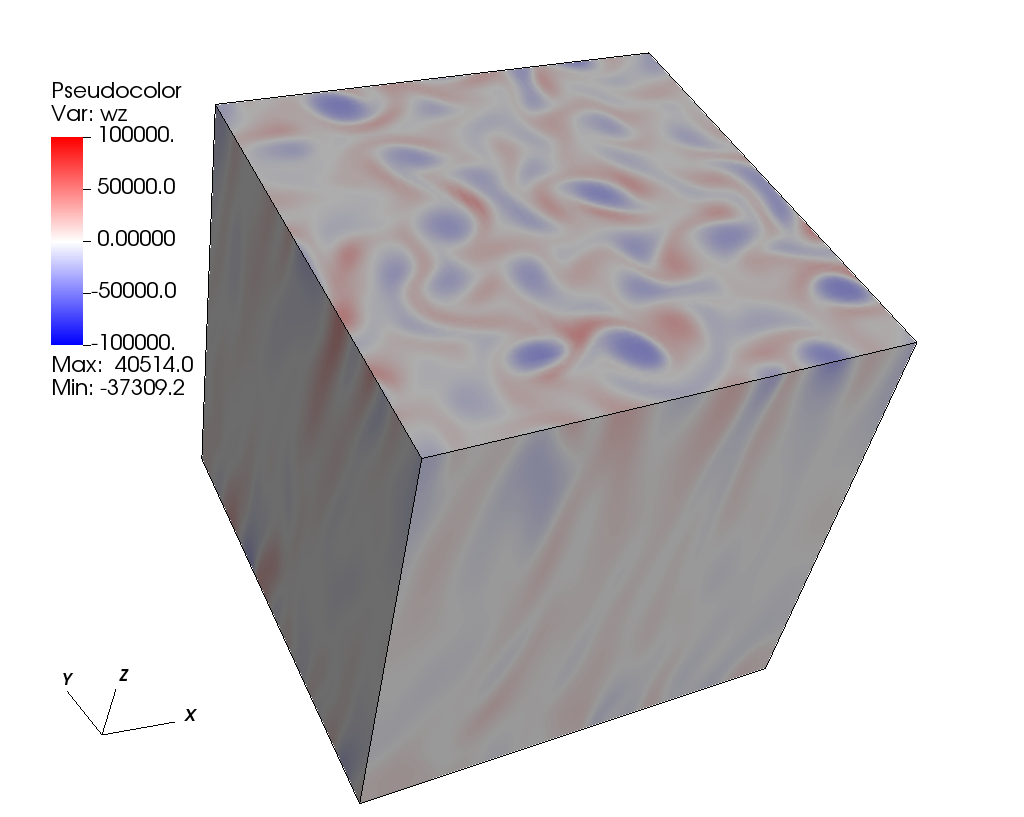}}
\hfill
\subfloat[$t=0.05$, $\mathrm{Po}=0.1$, $\mathrm{Ra}=6\mathrm{Ra}_c$. \label{fig:Po0.1_Ra6}]{\includegraphics[width=0.45\textwidth,trim={0 0 2cm 1cm},clip]{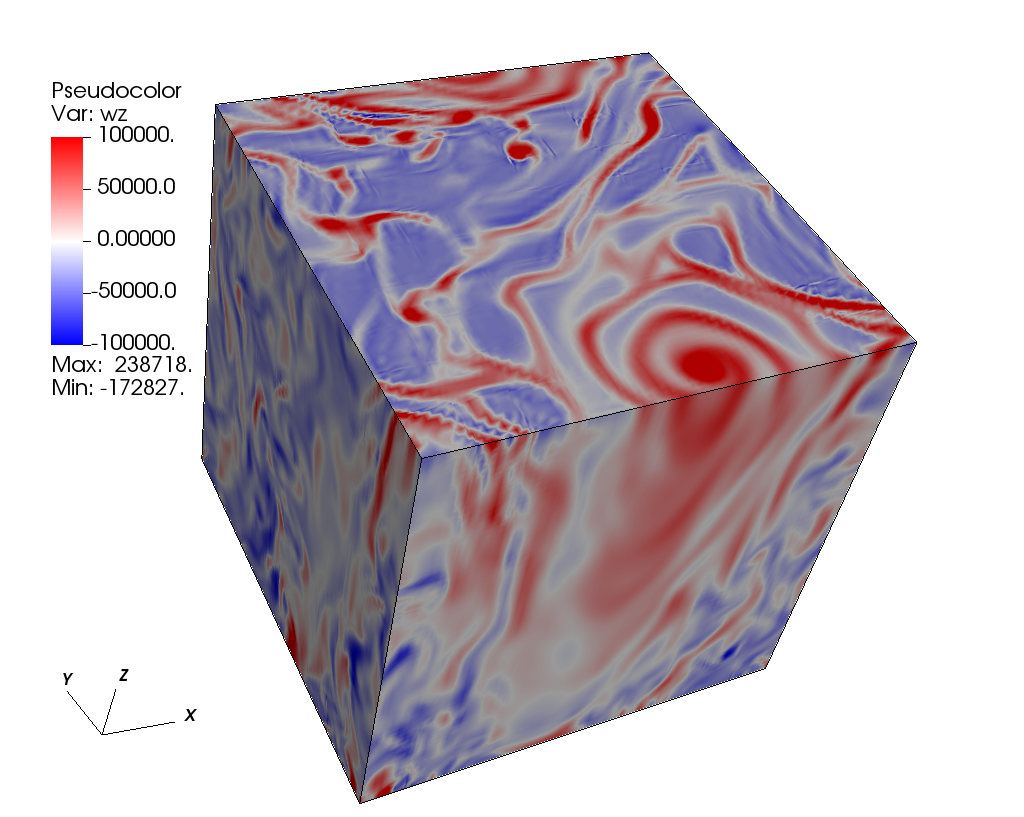}}\\

     \caption{Snapshots of the vertical vorticity $\omega_z$ of the flow in simulations executed using \textsc{nek5000} with $\mathrm{Ek}=2.5\cdot10^{-5}$. The top left and top right panels, both obtained from the same simulation, show snapshots taken during a burst of the energy injection and during the period of large-scale flow after the burst, respectively. The middle panels, both taken from a simulation with stronger precessional driving than the top panels, show vortices in the flow instead. A rapid ``secondary transition" appears to occur in this simulation, indicated by a large increase in vorticity and kinetic energy from the panel on the left to the one on the right. The bottom left panel shows a simulation with weak convective driving; the bursty behaviour of the precessional instability appears absent and instead the convection appears to dominate this flow. In the bottom right panel a snapshot of the simulation with stronger convective driving is shown. This flow strongly resembles that of the middle right panel, but it is unclear whether the vortex is driven by convection or precession.}
    \label{fig:Nek_snapshots}
\end{figure*}

We present snapshots of the vertical vorticity $\omega_z$ obtained from simulations using \textsc{nek5000} with $\mathrm{Ek}=2.5\cdot10^{-5}$ in Fig.~\ref{fig:Nek_snapshots}. This is the same Ekman number as utilised in \citet[][]{masonkerswell}{}{} in the strongly precessing case, after correcting for the factor of two which is present in our definition of the Ekman number and absent in theirs. The snapshots in Figs.~\ref{fig:Po0.1_1} and \ref{fig:Po0.1_2} are taken from a simulation of the precessional instability, in the absence of convection, with $\mathrm{Po}=0.1,\ \mathrm{Ra}=0$. Based on the previous results in \citet[][]{masonkerswell,Barker2016precession,Pizzi2022}{}{} we expect the flow in these simulations to be bursty in nature, similar to the elliptical instability. This bursty behaviour is indeed observed in these two snapshots. In Fig.~\ref{fig:Po0.1_1} we plot the vorticity at time $t=0.344$, given in thermal time units. This coincides with a burst in energy injection and kinetic energy of the precessional instability. These bursts are associated with the most unstable linear modes, with a horizontal form consisting of two full sinusoids in both horizontal directions, and a combination of the modes with $n_z=1$ and $n_z=2$ in the vertical direction. After the burst in energy injection has occurred, the most unstable linear modes break down and instead a large-scale flow emerges. This large-scale flow is portrayed in a snapshot taken at $t=0.37$ in Fig.~\ref{fig:Po0.1_2}. It does not look like the vortices we normally expect from inertial wave breakdown \citep[][]{Barker2016}; instead, it appears to be a large-scale sheared flow, the direction of which oscillates in time, and is likely related to the rotating shear introduced by the background flow. This large-scale flow inhibits growth of the most unstable linear modes of the precessional instability, and continues to do so until it is slowly viscously dissipated. Afterwards, the most unstable linear modes re-emerge in the flow and the cycle starts anew; the flow is maintained in this fashion throughout the entirety of this simulation.

The flow with $\mathrm{Po}=0.2,\ \mathrm{Ra}=0$ in Figs.~\ref{fig:Po0.2_1} and \ref{fig:Po0.2_2} features much stronger vorticities than in the simulation with $\mathrm{Po}=0.1,\ \mathrm{Ra}=0$. Note the change in colour scale in the middle and bottom panels compared to the one in the top panels. In Fig.~\ref{fig:Po0.2_1} we plot a snapshot taken at $t=0.06$, after the initial instability has saturated. We observe the formation of vortices in this simulation, as opposed to the more layer-like large-scale shear flow in Fig.~\ref{fig:Po0.1_2}. In these simulations, a ``secondary transition" occurs as the values of the vorticity, and associated with it the kinetic energy and energy injection, suddenly increase. This is visible by the larger values of the vorticity in the snapshot in Fig.~\ref{fig:Po0.2_2}, taken at $t=0.072$. The vortex in this snapshot appears to dominate the flow more than in Fig.~\ref{fig:Po0.2_1}. Upon examining the vertical vorticity as a function of depth, it was found that this vortex extends to the bottom of the box, but it is sheared, as we might expect from the precessional shear flow, and so the bottom of the vortex is located towards the middle of the horizontal plane in this particular snapshot. The direction of this shear, which we might consider to be the vortex axis, is precessing in time, on a timescale slightly longer than the precession period in this simulation. As a result, the strong vortex is not completely $z$-invariant, such that the 2D energy according to our definition does not fully capture the energies in these vortices. The development of this large scale vortex can be thought of as a modification to the background flow $\bm{U}_0$ that arises from the non-linear interaction with the inertial waves generated by the precessional instability. This modification occurs on large scales in our simulations, and can impact the precessional instability itself. The exact shape and impact on the precessional instability depends on the chosen boundary conditions and geometry, as the modification is by definition forced to conform to these.\\

We will now use these snapshots to get an idea of how convection modifies the behaviour of the flow. We start by examining a snapshot of the simulation with $\mathrm{Po}=0.1,\ \mathrm{Ra}=2\mathrm{Ra}_c$ in Fig.~\ref{fig:Po0.1_Ra2}. This snapshot was taken at $t=0.05$, after saturation of the initial instability and transients have died down. The flow in this snapshot is representative of the flow at all times in this simulation. The flow shows large-scale up and down flows, which appear somewhat sheared in the horizontal directions, and very small-scale but not vortex-like behaviour in the horizontal plane. These small-scale features appear, by eye, to be consistent with the size of the convectively unstable linear modes. We attribute this behaviour to the convective motions, but note that this close to onset the convection is unable to form a convective large-scale vortex, both with and without precession, as previously observed in \citet[][]{CelineLSV}{}{} without precession. The precessional instability appears to be weak or absent compared to these convective motions, as there is no clear evidence of its operation in the flow. Finally, we turn to even stronger convective driving in Fig.~\ref{fig:Po0.1_Ra6}, with $\mathrm{Po}=0.1,\  \mathrm{Ra}=6\mathrm{Ra}_c$ taken at $t=0.05$. Again, the initial transients have died down at this point in the simulation. The flow has saturated in a way that is very similar to the one in Fig.~\ref{fig:Po0.2_2}. We again observe a vortex in the flow, but the areas around the vortex appear to be more noisy. This noise is presumably associated with the convective eddies present in the simulation. It is unclear whether the vortex is of a convective nature, generated by the precessional instability, or both. We will examine the time series of the energy injection to examine whether the precessional instability is operating in these simulations with strong convection present. 

There appear to be signatures of the grid scale present in both Figs.~\ref{fig:Po0.2_2} and \ref{fig:Po0.1_Ra6}. These are prominent here, but are much less pronounced and usually absent in the velocity snapshots that we have used to compute these vorticity snapshots. The prominence of the features here is likely an artefact of how we have computed the vorticity on this grid, and the colour scale we have used to plot these particular results. A further discussion of the well-resolvedness of our simulations can be found in Appendix~\ref{app:snapshot_resolutions}.

\subsection{Time series of quantities of interest}

\begin{figure*}
\subfloat[$\mathrm{Po}=0.1$, $\mathrm{Ra}=0$. \label{fig:NEK_Po0.1}]{
         \hspace{-0.8mm}\includegraphics[width=0.43\textwidth,trim={2cm 1cm 2cm 2cm},clip=true]{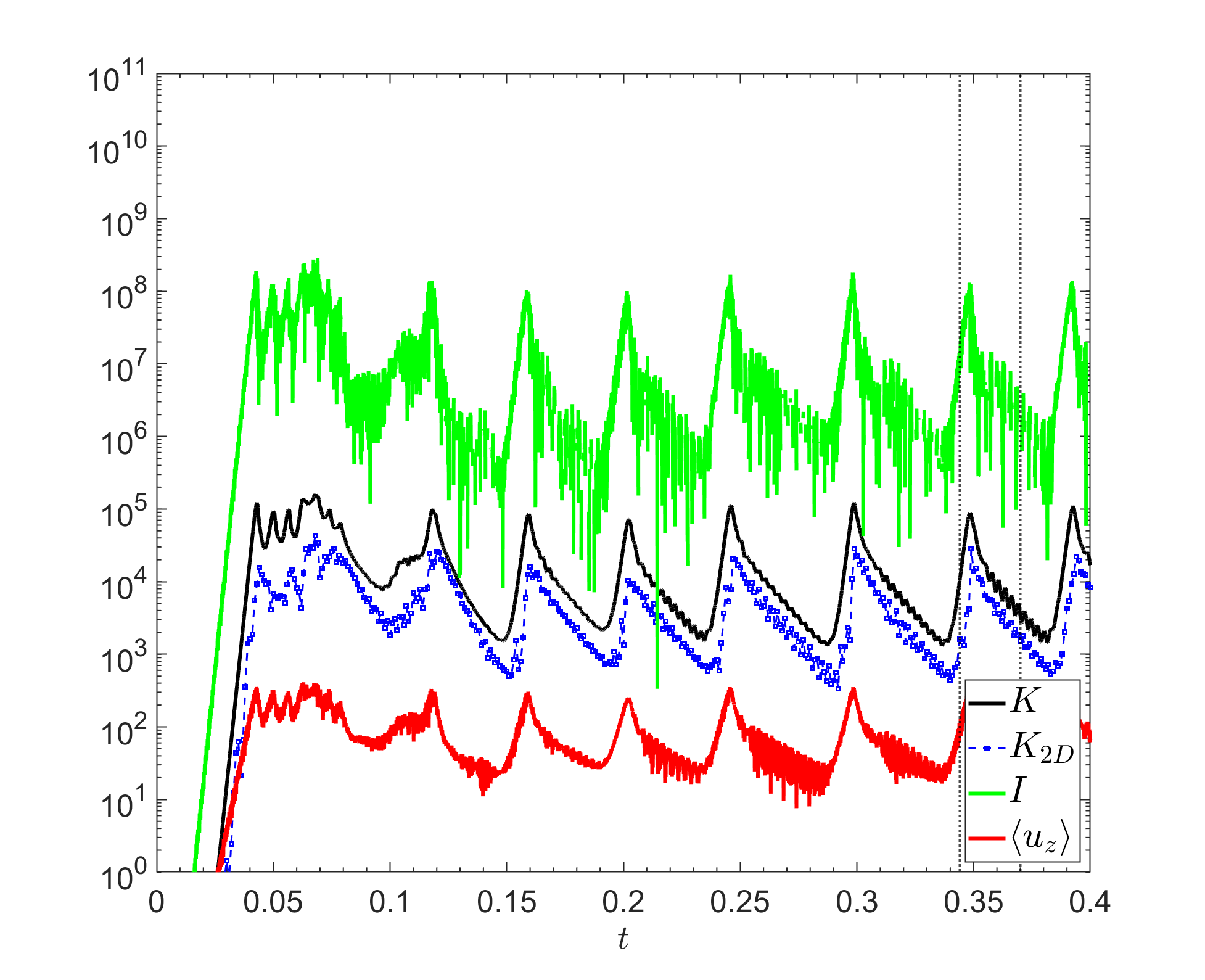}}
\hfill\subfloat[$\mathrm{Po}=0.2$, $\mathrm{Ra}=0$. \label{fig:NEK_Po0.2}]{ \includegraphics[width=0.43\textwidth,trim={2cm 1cm 2cm 2cm},clip=true]{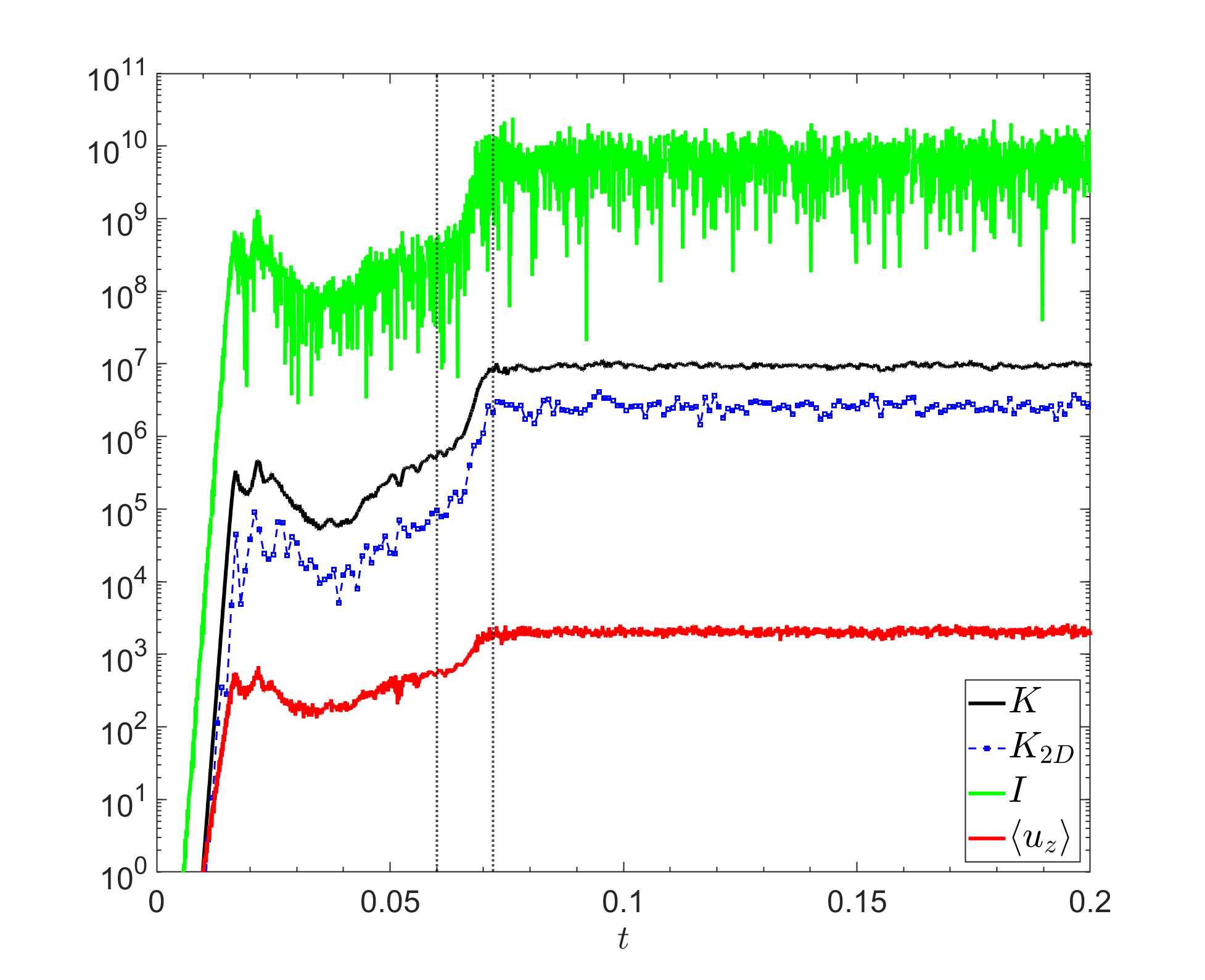}}\\

\subfloat[$\mathrm{Po}=0.1$, $\mathrm{Ra}=2\mathrm{Ra}_c$. \label{fig:NEK_Po0.1_Ra2}]{\includegraphics[width=0.43\textwidth,trim={2cm 1cm 2cm 2cm},clip=true]{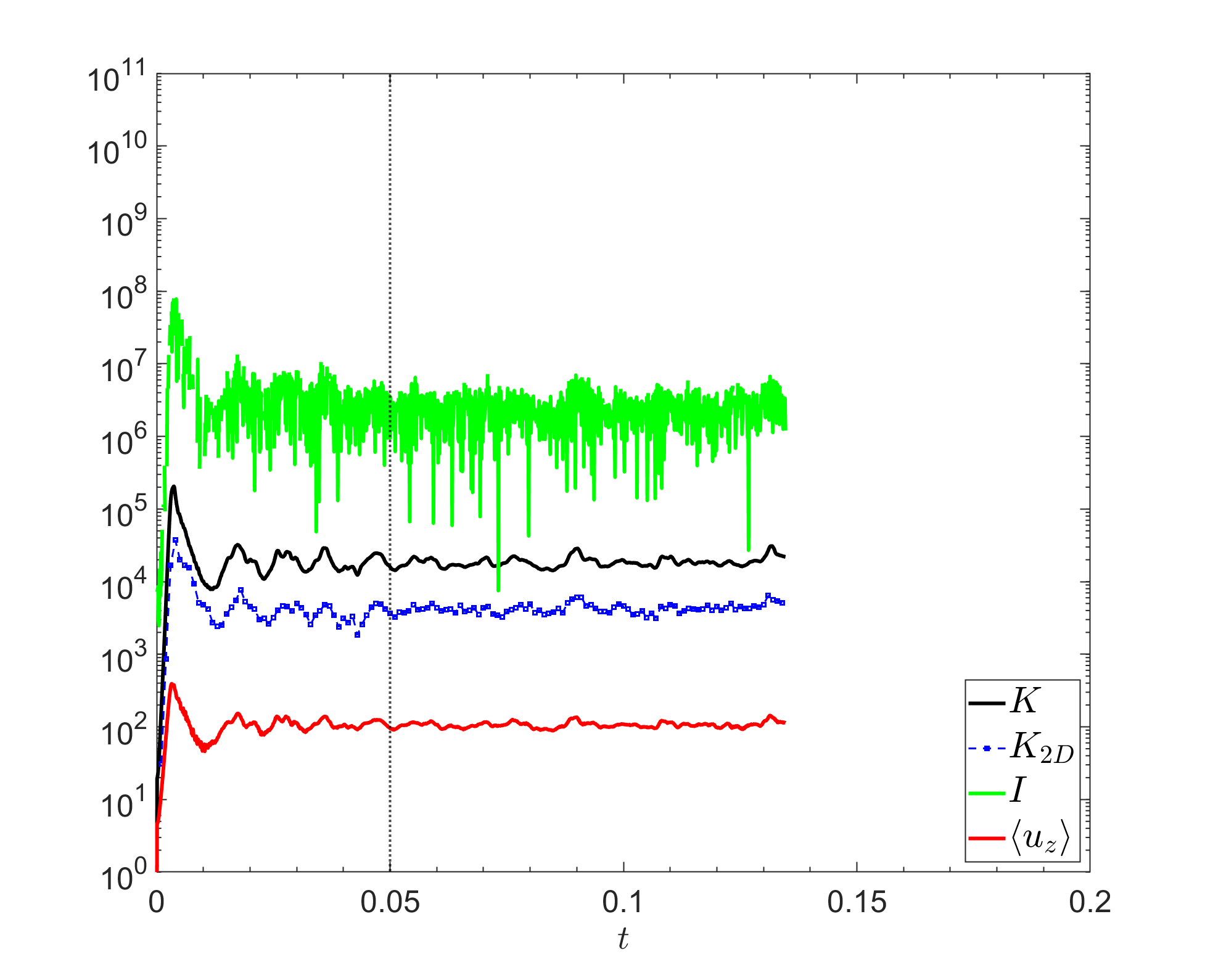}}
\hfill
\subfloat[$\mathrm{Po}=0.2$, $\mathrm{Ra}=2\mathrm{Ra}_c$. \label{fig:NEK_Po0.2_Ra2}]{\includegraphics[width=0.43\textwidth,trim={2cm 1cm 2cm 2cm},clip=true]{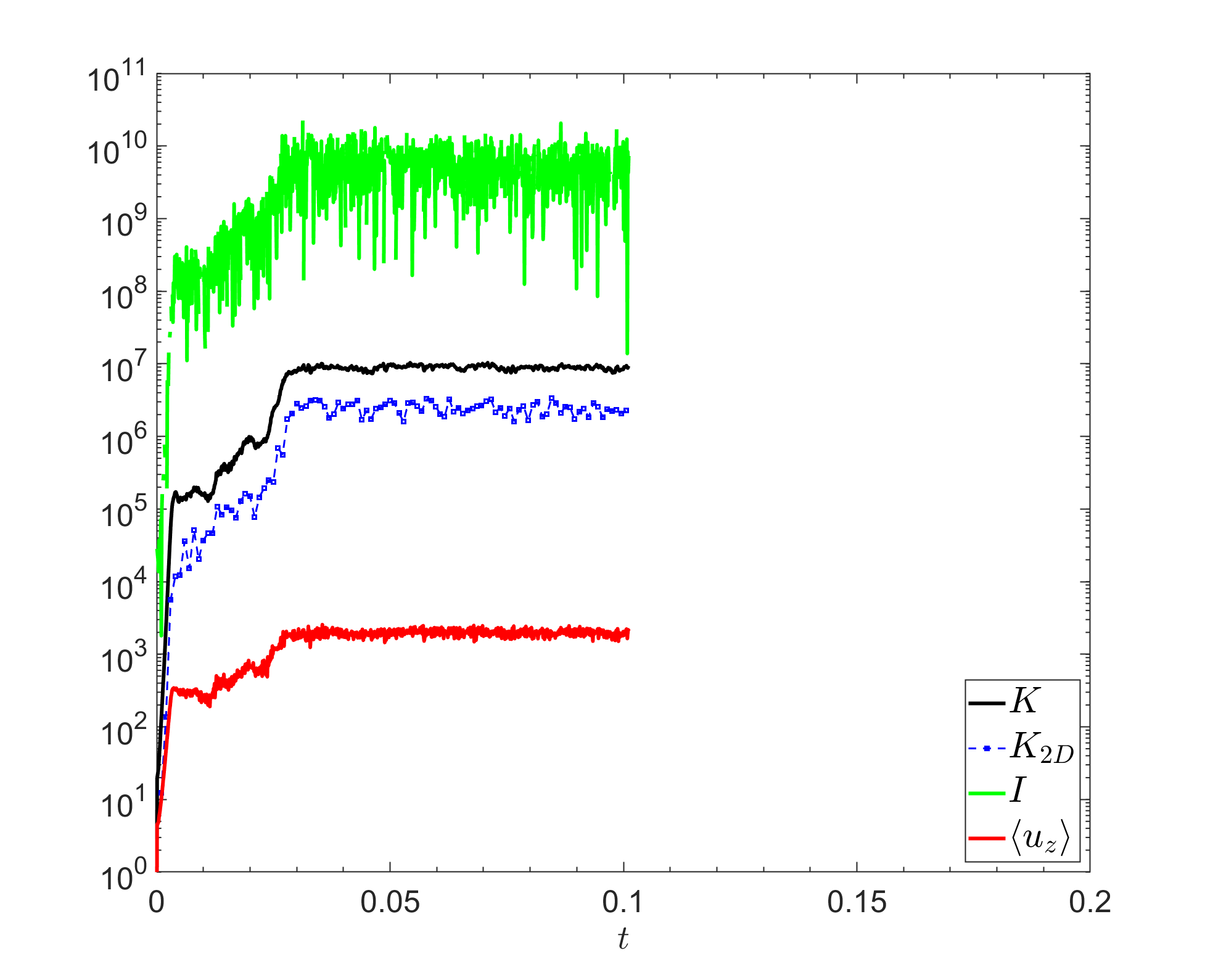}}\\

\subfloat[$\mathrm{Po}=0.1$, $\mathrm{Ra}=6\mathrm{Ra}_c$. \label{fig:NEK_Po0.1_Ra6}]{\includegraphics[width=0.43\textwidth,trim={2cm 1cm 2cm 2cm},clip=true]{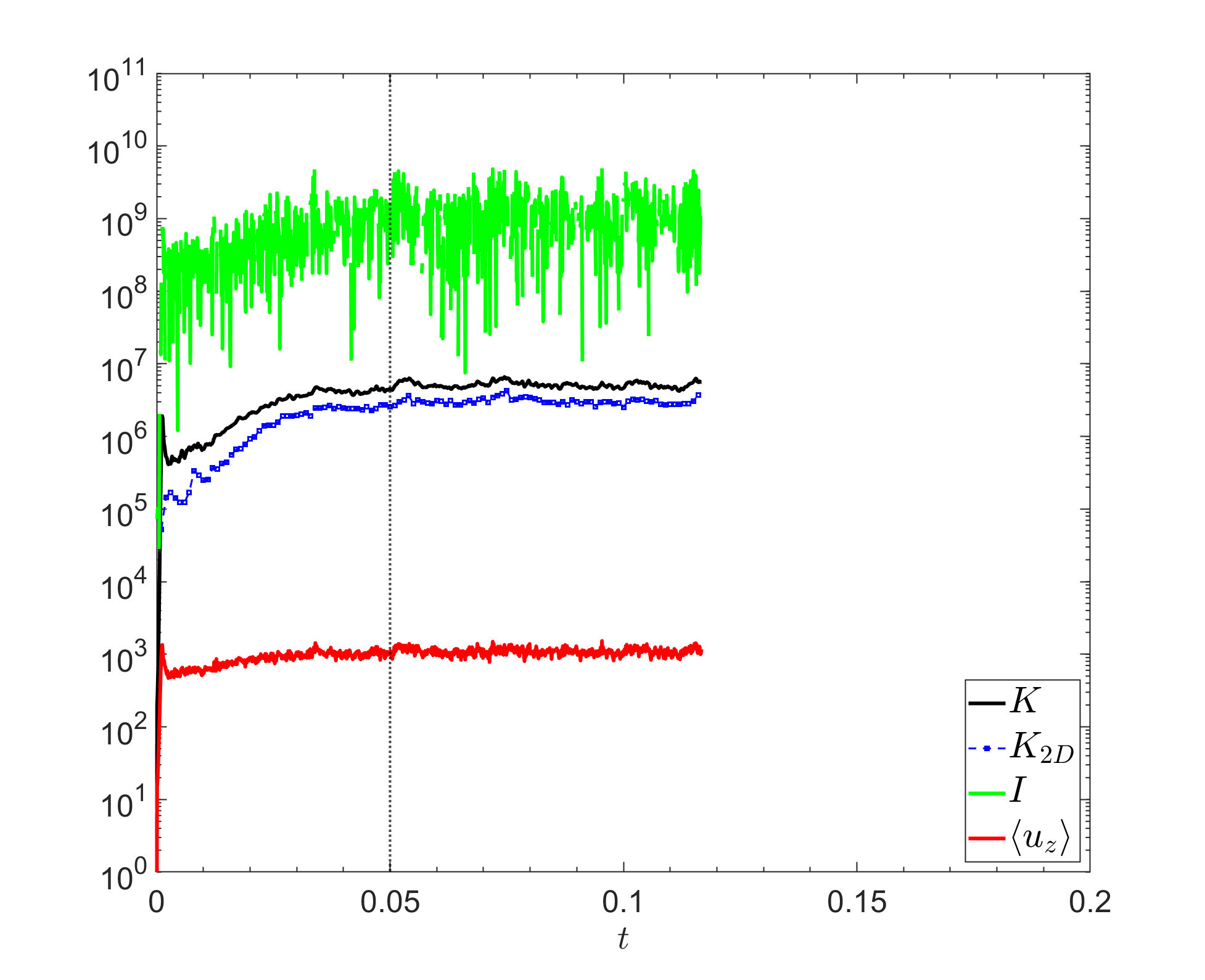}}
\hfill
\subfloat[$\mathrm{Po}=0.2$, $\mathrm{Ra}=6\mathrm{Ra}_c$. \label{fig:NEK_Po0.2_Ra6}]{\includegraphics[width=0.43\textwidth,trim={2cm 1cm 2cm 2cm},clip=true]{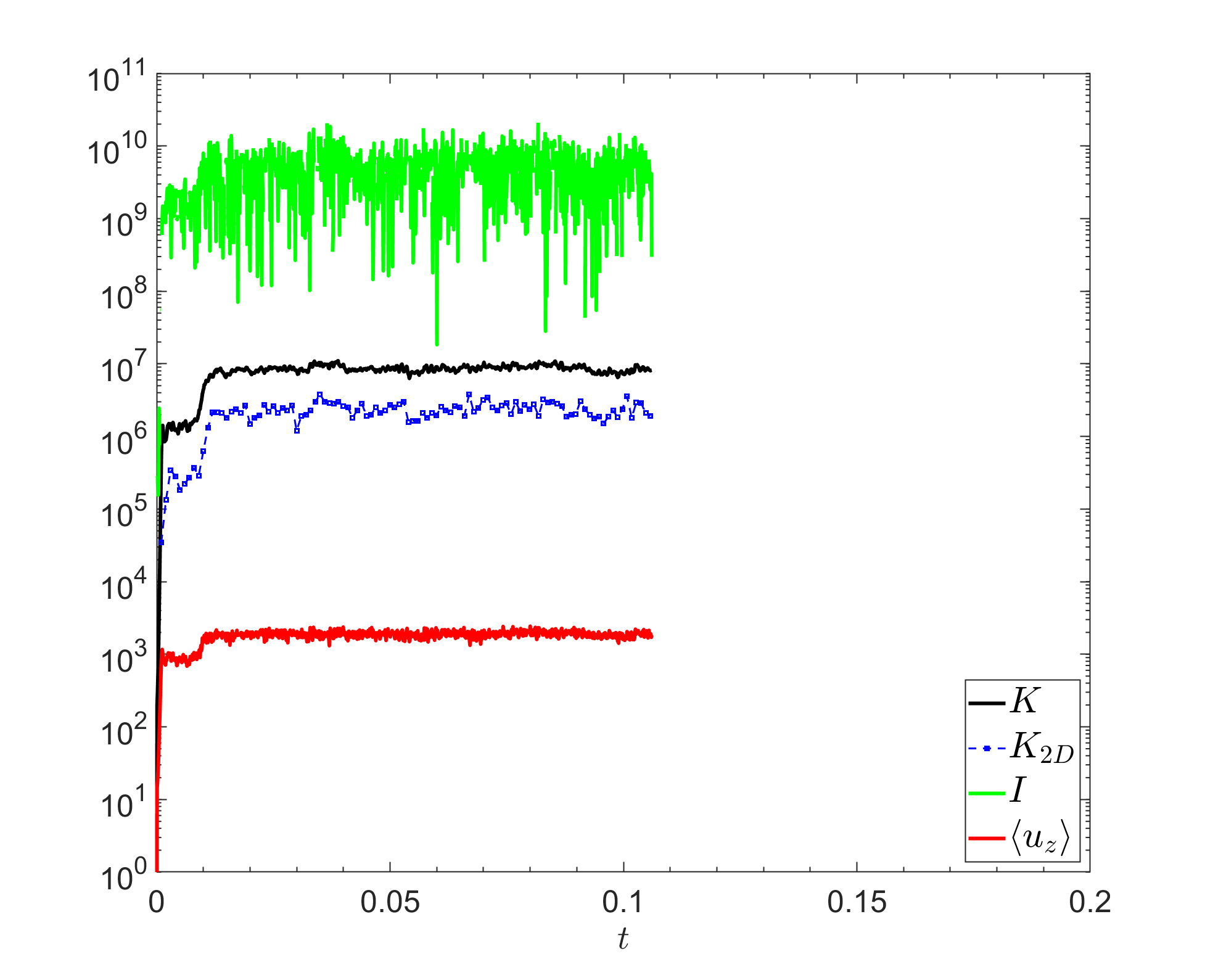}}\\

     \caption[Time series of the precessional instability and convection with fixed Ekman number executed using \textsc{nek5000}.]{Time series of the precessional instability and convection with $\mathrm{Ek}=2.5\cdot10^{-5}$, executed using \textsc{nek5000}. Vertical dotted-black lines correspond to the times at which the snapshots in Fig.~\ref{fig:Nek_snapshots} are taken. The top left and right panels show the precessional instability in isolation. Note the different $x$-axis ranges between the two panels. The simulation in the top left panel produces the expected bursty behaviour, while the top right panel with stronger precessional driving features a secondary transition to a continuously turbulent state. The introduction of convection inhibits the bursty behaviour in the figure on the middle left, displaying a continuous energy injection instead. The simulation on the middle right with stronger precessional driving than the one on the middle left shows similar behaviour to the panel on the top right. The simulation on the bottom left with strong convective driving shows evidence for a secondary transition that is more gradual than in the top right panel, so convection might allow the precessional instability to become continuously turbulent at lower values of the Poincar\'{e} number. Finally, the simulation on the bottom right also shows the secondary transition, but stronger convective driving appears to allow the transition to occur sooner.}
    \label{fig:Neksims}
\end{figure*}

We have so far observed quite diverse flows emerging in the flow snapshots, even when considering just the non-linear evolution of the precessional instability in isolation. To better understand the differences between these flows, we now examine time series of our quantities of interest, $K,\ K_{2D}$, $I$ and $u_z$. The latter of these is obtained by calculating the rms vertical velocity. These time series, obtained using \textsc{nek5000}, are portrayed in Fig.~\ref{fig:Neksims}. The times at which the snapshots in Fig.~\ref{fig:Nek_snapshots} are taken from these simulations have been marked with vertical dotted-black lines. All simulations have been executed with $\mathrm{Ek}=2.5\cdot10^{-5}$. In Fig.~\ref{fig:NEK_Po0.1} the simulation with $\mathrm{Po}=0.1,\ \mathrm{Ra}=0$ is presented. This simulation corresponds to the snapshots in the top two panels of Fig.~\ref{fig:Nek_snapshots}, showing the bursty behaviour. The peaks of the kinetic energy and energy injection correspond to flows that look similar to Fig.~\ref{fig:Po0.1_1}, while those flows found in Fig.~\ref{fig:Po0.1_2} occur during the troughs. The linearly unstable modes are associated with bursts in both total energy followed by bursts in the 2D energy with a slight delay. Furthermore, we observe that the energy injection is also bursty, and primarily associated with the 3D energy as one would expect. The vertical velocity appears to follow the kinetic energy quite closely in all panels of this figure.

In Fig.~\ref{fig:NEK_Po0.2} we plot the time series of the simulation executed with $\mathrm{Po}=0.2,\ \mathrm{Ra}=0$. Note the different $x$-axis range compared to Fig.~\ref{fig:NEK_Po0.1}. As we expect from the middle panels in Fig.~\ref{fig:Nek_snapshots}, the bursty behaviour is absent. Instead, the simulation goes through a regime of lower kinetic energy and energy injection from $t=0.03-0.06$ and then the flow goes through a rapid secondary transition into  a regime with the kinetic energy, energy injection and vertical velocity maintained at strongly elevated levels. We will denote this regime as the ``continuously turbulent" regime, in contrast to the bursty regime at lower values of Po.\\

We now turn to examine the modification of these flows and corresponding time series due to the introduction of convection. In Fig.~\ref{fig:NEK_Po0.1_Ra2} the time series of the simulation with $\mathrm{Po}=0.1,\ \mathrm{Ra}=2\mathrm{Ra}_c$ is shown. This simulation indeed shows the absence of the bursty behaviour, but has a kinetic energy roughly on the same order of magnitude as the one above. The precessional instability appears to not be operating in this figure, as we concluded from the snapshot. The continuous energy injection is similar to the energy injection in \citet[][]{deVries2023b} in simulations that featured only convection acting on the (elliptical) background flow like an effective viscosity. We therefore attribute this behaviour to the convection acting like an effective viscosity in damping the precessional flow as well. In Fig.~\ref{fig:NEK_Po0.2_Ra2} we show the simulation with parameters $\mathrm{Po}=0.2,\ \mathrm{Ra}=2\mathrm{Ra}_c$. Even though the convection is present we again observe the secondary transition, likely indicating the operation of the precessional instability in this simulation. The kinetic energy and energy injection appear to saturate at slightly lower values compared to Fig.~\ref{fig:NEK_Po0.2}. 

The time series corresponding to the simulation with $\mathrm{Po}=0.1, \ \mathrm{Ra}=6\mathrm{Ra}_c$ is shown in Fig.~\ref{fig:NEK_Po0.1_Ra6}. We observe a mixture of the continual ``convective energy injection" behaviour and the precessional instability's secondary transition. A transition akin to the precessional instability's secondary transition does arise like in Fig.~\ref{fig:NEK_Po0.2}, but it is much more gradual. So, neither from the snapshot of the flow, nor from the time series can we properly conclude whether this is due to the precessional instability, or a property of convection modified by precession. If it is indeed the precessional instability then the convection allows the secondary transition to appear at smaller values of Po. This can potentially enhance the energy injection and thus dissipation significantly, and is therefore important to examine further. Finally, we examine the time series of the simulation with $\mathrm{Po}=0.2,\ \mathrm{Ra}=6\mathrm{Ra}_c$ in Fig.~\ref{fig:NEK_Po0.2_Ra6}. We again notice a secondary transition, which occurs much faster than in the case of the top right and middle right panels, even though Po for all three simulations is the same. The energy and energy injections prior to this transition are larger due to the convective driving. We thus conclude that convection does not disrupt operation of the precessional instability; instead, it appears that convection allows the precessional instability to attain its continuously turbulent regime for lower values of Po. This is possibly caused by the increased value of the energy or 2D energy -- and thus presence of a vortex in the flow -- due to convection. An analysis of these 2D energies in the simulations executed using \textsc{Dedalus} is presented in Appendix~\ref{app:Dedalus_2D}. To understand the key processes operating in the flow, we now turn to analyse power spectra obtained using \textsc{Dedalus}.

\subsection{Horizontal energy spectra of the flow}

\begin{figure}
    \centering
    \includegraphics[width=0.8\linewidth]{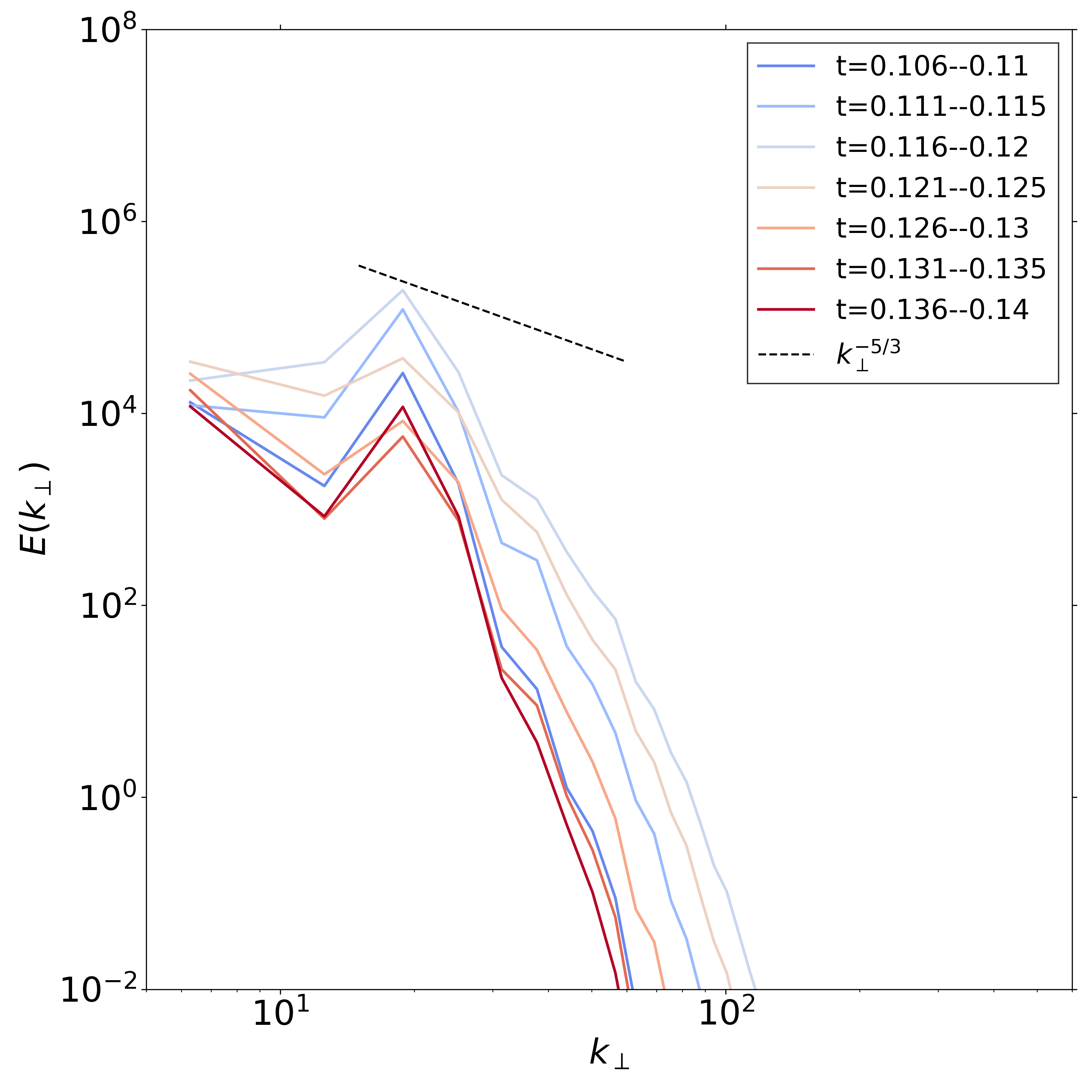}
    \caption[Horizontal energy spectra of a simulation with a ``bursty" precessional instability.]{Horizontal energy spectra taken at $z=0.5$ of the simulation with $\mathrm{Po}=0.1$, $\mathrm{Ra}=0$ from $t=0.106$ to $t=0.140$, corresponding to the interval between the vertical dotted lines in Fig.~\ref{fig:Ded_Po0.1}. The simulation goes through a burst and decay period in this interval. In the spectra this is indicated by the increase and subsequent decrease in the main energy injection spike in the wavenumber bin with $k_\perp=6\pi$, while the 2D energy, here located in the smallest wavenumber bin, lags slightly behind this spike. The spectra clearly do not match the Kolmogorov scaling as $k_\perp^{-5/3}$ in dashed-black and therefore this case is not turbulent.}
    \label{fig:spectrum_Ra0_Po01}
\end{figure}

We start by examining the horizontal energy spectra of the bursty simulation with $\mathrm{Po}=0.1,\ \mathrm{Ra}=0$ in Fig.~\ref{fig:spectrum_Ra0_Po01}, for which the corresponding time series can be found in Fig.~\ref{fig:Ded_Po0.1}. The horizontal energy spectrum is obtained at the mid-plane of the simulation, i.e. $z=0.5$, and is captured over one burst and decay period, the boundaries of which are denoted by dotted vertical lines in Fig.~\ref{fig:Ded_Po0.1} at $t=0.106$ and $t=0.140$. We have averaged these spectra every five output steps, with a time between outputs of $t=0.001$. The times corresponding to each spectrum are shown in the legend. Energy is preferentially injected by the precessional instability into the modes near $k_\perp=18.059$, binned into the wavenumber bin with $k_\perp=6\pi$. In all of the averaged energy spectra in this figure this injection is represented by the large spike in energy. The period under study starts with a significant fraction of the kinetic energy in the $K_{2D}$ modes compared to the total energies, which is represented in the energy spectrum in dark blue by the bin with the lowest horizontal wavenumber containing comparable energy to the wavenumbers where energy is injected. As the simulation evolves, the energy increases due to a burst in energy injection. This is represented by a large increase in the energy located in the primary energy injection spike plotted in progressively lighter blues, also adding energy in the smallest wavenumber bin corresponding to the large-scale flow. Towards the end of the burst energy is being transferred to smaller and smaller scales, as indicated by the widening of these spectra. After the burst finishes the energy starts decreasing again in the spectra plotted in progressively darker reds. Once these energies have decayed sufficiently, another burst can occur again and the cycle starts anew. Finally, we have plotted the turbulent Kolmogorov-like scaling of $k_\perp^{-5/3}$ in the dashed-black line in the figure for reference, but we note that no parts of the spectra agree with this line, as we might expect, since the flow does not appear turbulent at any stage from the snapshots.\\

\begin{figure}
    \centering
    \includegraphics[width=0.8\linewidth]{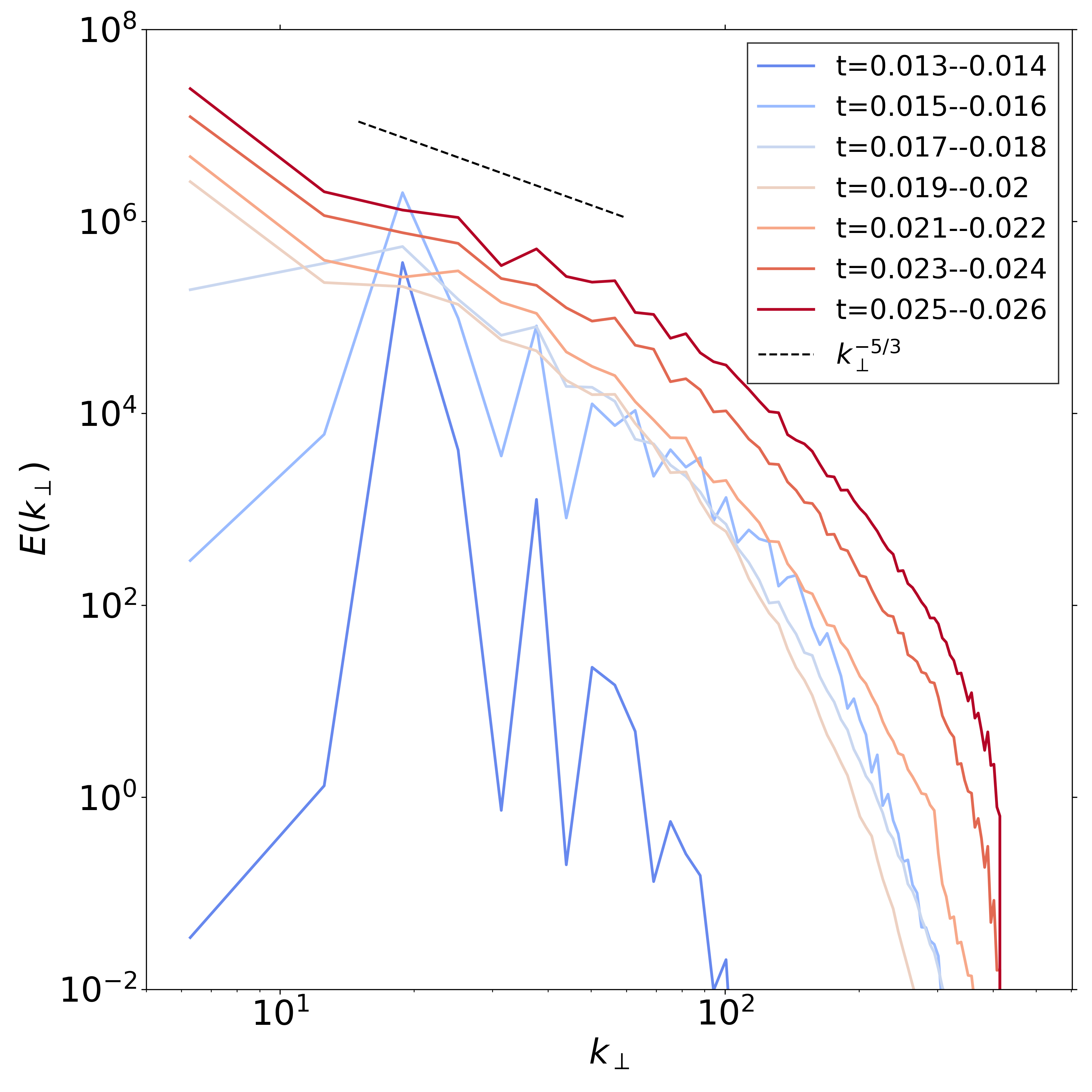}
    \caption[Horizontal energy spectra of the simulation with a ``continuously turbulent" precessional instability.]{Horizontal energy spectra taken at $z=0.5$ of the simulation with $\mathrm{Po}=0.2$, $\mathrm{Ra}=0$ from $t=0.013$ to $t=0.026$, corresponding to the interval between the vertical dotted lines in Fig.~\ref{fig:Ded_Po0.2}. The simulation goes through the end of the linear growth phase and the short low-energy turbulent phase followed by the secondary transition. The energy at the end of the linear growth phase, in the two spectra plotted in darkest blues, is predominantly located in the bin with the linearly most unstable wavenumbers, as well as integer multiples of this bin, possibly pointing to triadic resonances. In the spectrum taken from $t=0.019$ to $t = 0.02$ plotted in lightest red, the flow is dominated by $K_{2D}$. The flow shown in this and subsequent spectra is continuously turbulent due to the secondary transition, agreeing well with the Kolmogorov scaling as $k_\perp^{-5/3}$ in dashed-black at intermediate wavenumbers for the times indicated by the spectra plotted with progressively darker red colours.}
    \label{fig:spectrum_Ra0_Po02}
\end{figure}

Next, we turn our attention to the spectra of the simulation with $\mathrm{Po}=0.2,\ \mathrm{Ra}=0$ in Fig.~\ref{fig:spectrum_Ra0_Po02}, corresponding to the interval between the vertical dotted lines in Fig.~\ref{fig:Ded_Po0.2} from $t=0.013$ to $t=0.026$, in which we capture the final stages of the initial instability followed by the secondary transition. We have averaged the spectra in this figure over two output steps, again with a time between outputs of $t=0.001$. During the linear instability we again retrieve the primary peak of the most unstable linear modes in the wavenumber bin with $k_\perp=6\pi$, as well as a secondary peak in the bin with $k_\perp=12\pi$. As the instability saturates, plotted in progressively lighter blues, energy is moved to smaller and smaller scales, as well as into the largest-scale modes. The energy starts aligning to a more traditional turbulent shape while the energy in the lowest wavenumber bin keeps growing; the clear peak in the linear modes has vanished in the spectrum taken from $t=0.019$ to $t = 0.02$ plotted in the lightest red, and the $K_{2D}$ dominates the flow in this and all subsequent spectra. This seems to allow the secondary transition to occur, with energy subsequently being distributed across scales. We have again plotted the Kolgomorov scaling as $k_\perp^{-5/3}$ in dashed-black. It appears that the intermediate scales in this simulation agree well with this prediction. Thus, we can be reasonably justified in calling this regime the ``continuously turbulent" regime. We can use these energy spectra to examine whether the flow is spatially well-resolved, using the rule of thumb that the largest energy in one bin must be at least a factor of $10^3$ larger than the energy at the de-aliasing scale, which is clearly satisfied in these spectra. As the simulation continues evolving, it eventually achieves a statistically steady state, with spectra resembling the dark red spectrum in Fig.~\ref{fig:spectrum_Ra0_Po02}. We can also conclude that this simulation is well-resolved as the energy in the smallest wavenumber bin in these spectra is $\mathcal{O}(10^7)$ and the one in the largest wavenumber bin is $\mathcal{O}(10^2)$.\\

\begin{figure}
    \centering
    \includegraphics[width=0.8\linewidth]{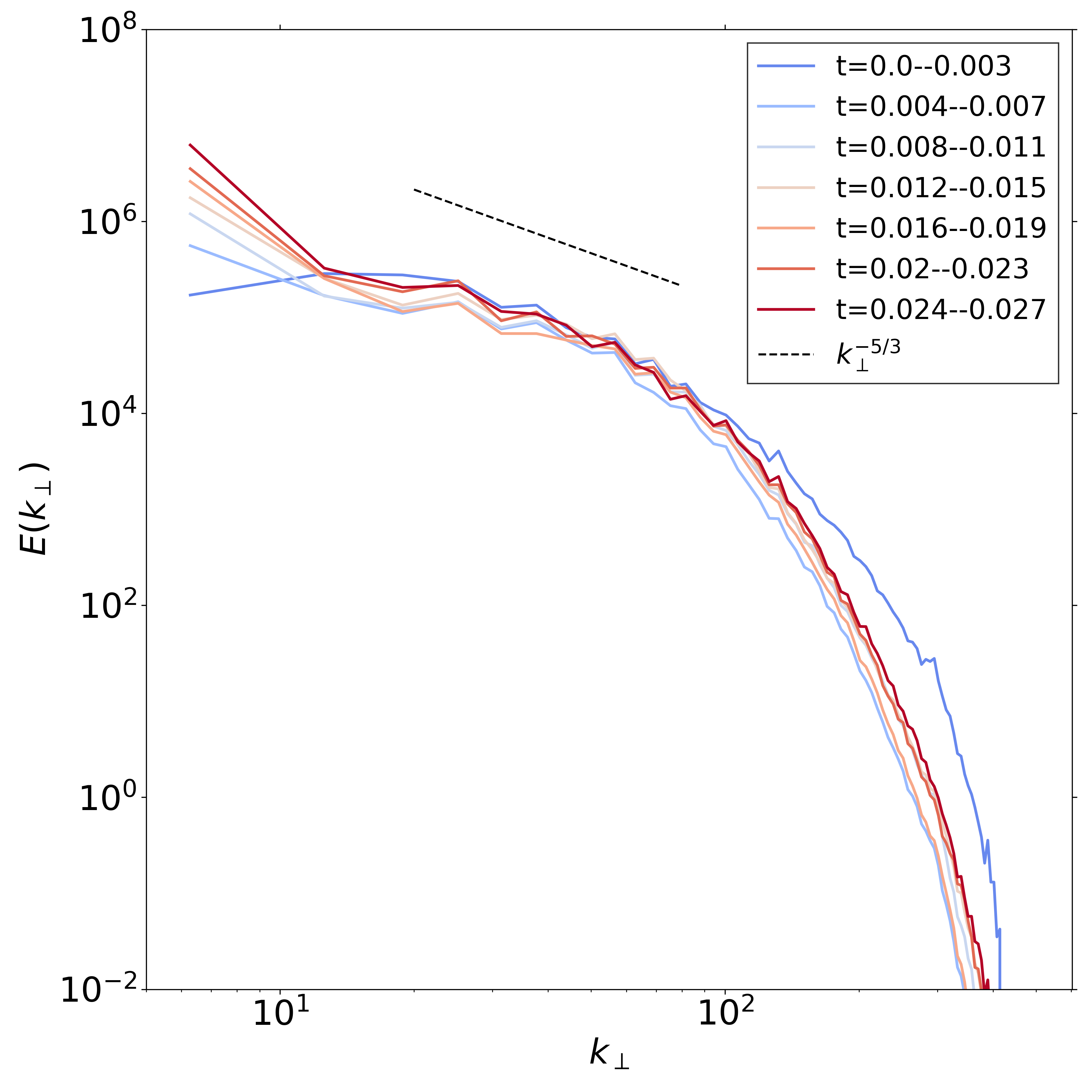}
    \caption[Horizontal energy spectra of the simulation with convection and the precessional instability interacting.]{Horizontal energy spectra taken at $z=0.5$ of the simulation with convection and the precessional instability interacting, with $\mathrm{Po}=0.1$, $\mathrm{Ra}=4\mathrm{Ra}_c$ from $t=0.0$ to $t=0.027$, corresponding to the interval between the start of the simulation and the vertical dotted line in the Fig.~\ref{fig:Ded_Po0.1_Ra4}. The simulation goes through the linear growth phase and the gradual transition to a higher energy state. No clear evidence of the precessional instability is present. Again, we find good agreement with the Kolmogorov scaling as $k_\perp^{-5/3}$ in dashed-black for the intermediate wavenumbers.}
    \label{fig:spectrum_Ra4_Po01}
\end{figure}

Finally, we turn to study the effect convection has on these flows by examining the energy spectra of the simulation with $\mathrm{Po}=0.1$, $\mathrm{Ra}=4\mathrm{Ra}_c$ in Fig.~\ref{fig:spectrum_Ra4_Po01}. The time we examine is denoted by the interval between the start of the simulation and the vertical dotted-black line in Fig.~\ref{fig:Ded_Po0.1_Ra4}, from $t=0$ to $t=0.027$. We average the spectra over four output steps, with an output step every $t=0.001$. The spectra appear turbulent from the start of the simulation, as we would expect from sufficiently strongly driven convection. There is a hint of energy injection by the precessional instability into the bin containing the precessionally unstable wavenumbers in the dark blue spectrum. As the simulation evolves the energy goes up predominantly in the smallest wavenumber bin and maintains roughly the same shape. The spectrum shows that this simulation is also well-resolved. As the simulation evolves and achieves a statistically steady state, the spectra resemble the dark red spectrum in Fig.~\ref{fig:spectrum_Ra4_Po01}, with the energy in the smallest wavenumber bin saturating at $\mathcal{O}(10^7)$, but with lower energies than the simulation in Fig.~\ref{fig:spectrum_Ra0_Po02}, while the energy in largest wavenumber bin saturates at $\mathcal{O}(10^0)$. We have also compared the energy spectra of this simulation with the ones from the purely convective simulation with $\mathrm{Po}=0$, $\mathrm{Ra}=4\mathrm{Ra}_c$ (not shown). These spectra look very similar to the ones in Fig.~\ref{fig:spectrum_Ra4_Po01}, so we have omitted them. There is, however, one crucial difference, the energy in the purely convective spectra stops growing. Most notably, the energy in the smallest wavenumber bin does not continue to grow like in Fig.~\ref{fig:spectrum_Ra4_Po01}, instead stagnating at an energy of $\approx10^6$, i.e. this bin saturates at lower energies.

We conclude that the energy spectra support the hypothesis that the 2D energy is -- at the very least -- related to the secondary transition to the continuously turbulent regime. We also find that convection assists in achieving this energy in a rapid manner in certain simulations that would otherwise be bursty in the absence of convection, such that even simulations with small values of the Po can achieve the continuously turbulent regime. However, the precessional instability must also be sufficiently strong to achieve this growth, as otherwise the precessional instability appears to be erased in favour of the turbulent ``effective viscosity" regime.

\section{Scaling laws for the quantities of interest}
\label{sec:scaling_laws}

We now turn to obtain scaling laws of the time-averaged values of the energy injection, $I$, rms vertical velocity, $u_z$, and Nusselt number, $\mathrm{Nu}$, as functions of the Rayleigh, Ekman and Poincar\'{e} numbers. We time-average these quantities starting from a suitable time in the simulations after a steady state has been reached. Because the continuously turbulent regime appears to be the final steady state in all simulations which feature it, we have taken time-averages only over this part of such simulations, neglecting any part of these simulations prior to the secondary transition.

\subsection{Scaling laws as a function of the Poincar\'{e} number}

We have identified three different qualitative regimes as a function of $\mathrm{Po}$ and $\mathrm{Ra}$. In the absence of convection and at low Po the flow is laminar, alternating between the most unstable mode and the $z$-invariant large-scale flow. The results in \citet[][]{Barker2016precession}{}{} indicate that the energy injection associated with this behaviour might be consistent with a scaling like $\mathrm{Po}^2$, as they appear to scale similarly to the laminar dissipation of the background flow $\bm{U}_0$; see the bottom left panel of Fig.~7 in \citet[][]{Barker2016precession}{}{}. Note that their measured dissipation, and the energy injection in our setup, does not include the laminar dissipation, such that this only gives a tentative indication how the energy injection in our simulations might scale with Po in the laminar regime, without explaining the mechanism behind it. If, however, convection is present at low Po, the flow appears much more turbulent due to convective action. The bursty behaviour has disappeared and a continuous energy injection is present because of the action of turbulent convection on the precessional flow. We would expect to see this behaviour scale as $\mathrm{Po}^2$ also and, more importantly, also depend on the Rayleigh number based on the results in \citet[][]{Craig2019effvisc,Craig2020effvisc,deVries2023b}. Finally, regardless of the presence of convection, these simulations indicate a continuously turbulent regime at high Poincar\'{e} numbers. We would expect the energy injection in this regime to scale as $\mathrm{Po}^3$. Even stronger convection might also suppress this regime or result in an effective viscosity whose rate of energy injection overshadows that of the precessional instability.

\begin{figure*}
         \centering
         \includegraphics[width=0.8\textwidth]{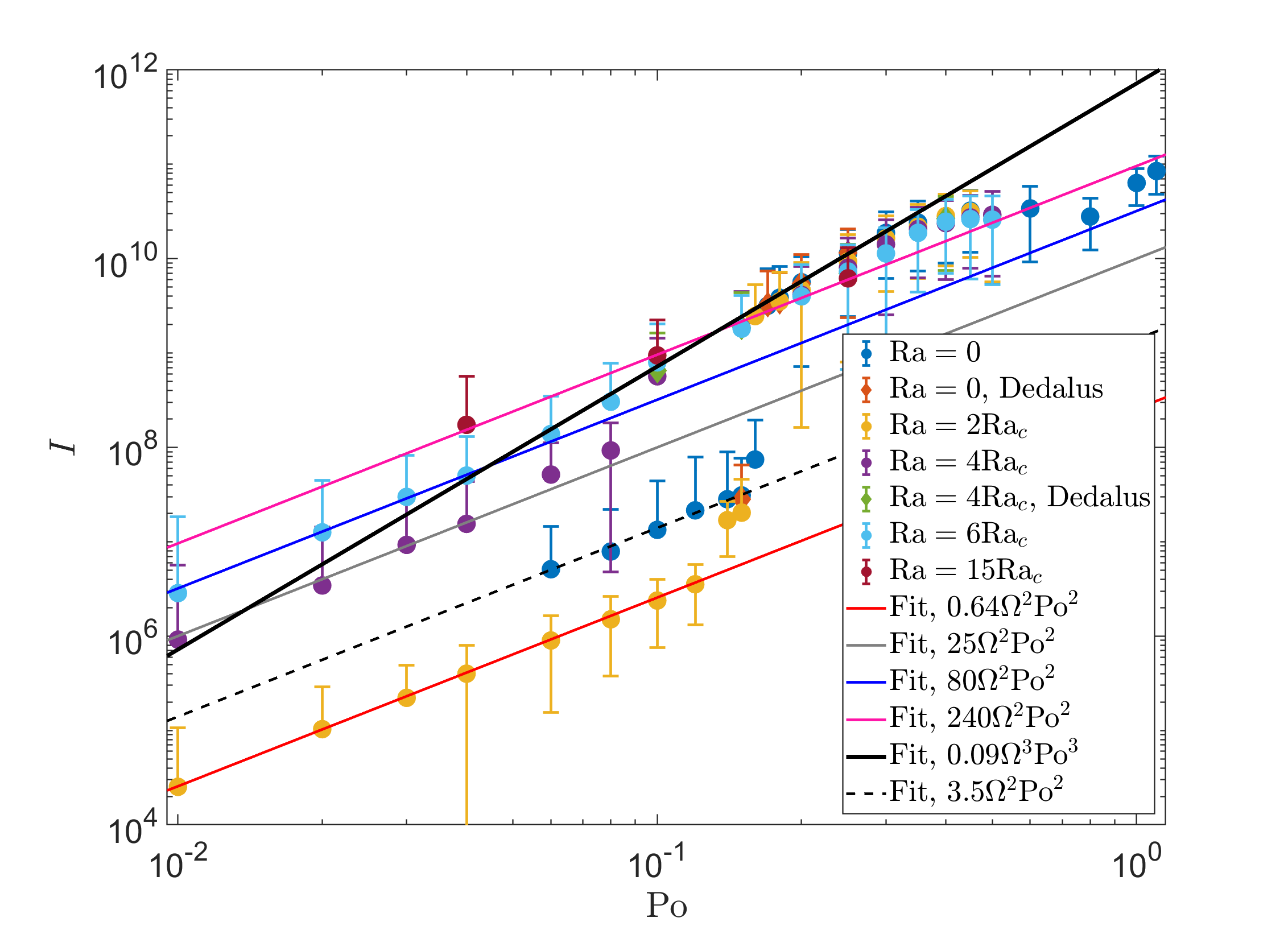}
        \caption[Time-averaged energy injection $ I $ with fixed Ekman number.]{Time-averaged energy injection $ I $ with fixed $\mathrm{Ek}=2.5\cdot 10^{-5}$. The simulations executed using \textsc{nek5000} are presented with circles, the ones executed using \textsc{dedalus} with diamonds. The laminar dissipation regime scaling as $\mathrm{Po}^2$ in the absence of convection and the effective viscosity regime also scaling as $\mathrm{Po}^2$ for simulations with convection present are found $\mathrm{Po}\lesssim0.1$. The transition regime of the purely precessional case is found around $\mathrm{Po}=0.16$. The continuously turbulent regime that all simulations agree with at moderately large values of $\mathrm{Po}$ has been fitted with a solid-black line scaling as $\mathrm{Po}^3$. Note that as the Poincar\'{e} number is increased beyond $\mathrm{Po}\approx0.35$ the energy injection plateaus, then appears to decrease, followed by a second increase around $\mathrm{Po}\approx1$.}
    \label{fig:I_tavg_funcPo}
\end{figure*}

The time-averaged values of the energy injection rate $ I$ as a function of $\mathrm{Po}$ with fixed $\mathrm{Ek}=2.5\cdot10^{-5}$ are plotted in Fig.~\ref{fig:I_tavg_funcPo}. In this figure numerous parameter sweeps at different values of the Rayleigh number are plotted; the runs executed using \textsc{nek5000} are plotted using circles, while the runs executed using \textsc{dedalus} are plotted using diamonds. The simulations we have executed using \textsc{dedalus} always show very good agreement with their \textsc{nek5000} counterparts, thus validating the results obtained using both codes. The simulations executed at $\mathrm{Ra}=0$ in blue circles start at $\mathrm{Po}=0.06$ because the precessional instability is not unstable below this value of Po with this value of the Ekman number. The data points in Fig.~\ref{fig:I_tavg_funcPo} are fitted using either of the obtained $\mathrm{Po}^2$ or $\mathrm{Po}^3$ scalings. We will define $D\equiv\zeta\Omega^3\mathrm{Po}^3$ and $D_{\mathrm{lam}}\equiv\Upsilon\Omega^2\mathrm{Po}^2$, with $\zeta$ and $\Upsilon$ the proportionality or efficiency ``factors" of both scaling laws respectively. We have defined $D_{\mathrm{lam}}$ for the dissipation scaling as $\mathrm{Po}^2$ reported in \citet[][]{Barker2016precession,Pizzi2022}{}{}. Note that we reserve the proportionality factor $\Upsilon$ for the precessional laminar scaling laws exclusively, and will not use this factor when fitting the effective viscosity. Furthermore, we have chosen an arbitrary scaling of this laminar energy injection as $\Omega^2$ for ease of comparison with the energy injection due to the effective viscosity, but, since there is no theoretical basis for this scaling, we will allow $\Upsilon$ to depend on $\Omega$. Lastly, $\zeta$ can also depend on the Rayleigh number.

At low values of $\mathrm{Po}$ the data points obtained with different values of the Rayleigh number agree well with the $\mathrm{Po}^2$ scaling. Indeed, even the dataset obtained at $\mathrm{Ra}=0$ agrees with this crude scaling with $\Upsilon=3.5$, which we have plotted in dashed-black. Although it should be stressed that this parameter sweep primarily agrees with this scaling law for a different reason than those parameter sweeps where convection is present. The energy injection rate at $\mathrm{Ra}=2\mathrm{Ra_c}$, represented with yellow circles and fitted in solid-red, is lower than at $\mathrm{Ra}=0$; the convection has prevented the precessional instability from operating, even though the energy injection from the convection acting on the background flow is actually lower than that of the precessional instability in isolation with $\mathrm{Ra}=0$. The energy injection in this regime increases with convective driving. A final point of note is that both the purple circles with $\mathrm{Ra}=4\mathrm{Ra_c}$, with fit in solid-grey, and the light blue circles with $\mathrm{Ra}=6\mathrm{Ra_c}$, with solid-blue fit, deviate from their respective fits as $\mathrm{Po}^2$ for $\mathrm{Po}\geq0.10$ and $\mathrm{Po}\geq0.06$ respectively. Instead, they then seem to agree with a fit as $\mathrm{Po}^3$ in solid-black, providing a tentative hint for the continuously turbulent scaling we have predicted previously. Finally, the two burgundy circles with $\mathrm{Ra}=15\mathrm{Ra_c}$ with associated solid-pink fit, do not agree with the solid-black fit. This is because the energy injection due to the convection interacting with the background flow is larger than the solid-black fit on this interval of $\mathrm{Po}$, because the convective driving for these parameters is quite strong.

The majority of the data points at $\mathrm{Po}\gtrsim0.15$ do not agree with the fits as $\mathrm{Po}^2$. Instead, a strong jump in energy injection in both the purely precessional simulations with $\mathrm{Ra}=0$ and the simulation with the weakest convective driving considered with $\mathrm{Ra}=2\mathrm{Ra}_c$ is present. Both of these scalings first deviate weakly from their $\mathrm{Po}^2$ fits, associated with the turbulent state without a secondary transition. Then, when the secondary transition is present in the simulations, the energy injection jumps up abruptly as a function of $\mathrm{Po}$ and aligns with the solid-black fit scaling as $\mathrm{Po}^3$. This abrupt jump occurs at different values of the Poincar\'{e} number, as previously observed in the $\mathrm{Ra}=4\mathrm{Ra}_c$ and $\mathrm{Ra}=6\mathrm{Ra}_c$ parameter sweeps. The abrupt jump occurs at $\mathrm{Po}=0.16$ for the $\mathrm{Ra}=2\mathrm{Ra}_c$ parameter sweep and at $\mathrm{Po}=0.17$ for the $\mathrm{Ra}=0$ parameter sweep. Thus we also observe in this figure that the secondary transition to the turbulent state is facilitated by the convection such that it occurs at lower $\mathrm{Po}$ for larger $\mathrm{Ra}$.

Once the turbulent scaling has been reached all simulations are, within the error bars, consistent with the solid black fit scaling as $\mathrm{Po}^3$ with $\zeta=0.09$ up to and including $\mathrm{Po}\approx0.35$. At larger values of the Poincar\'{e} number higher-order effects seem to arise, and the energy injection seems to plateau or possibly decrease slightly as a function of $\mathrm{Po}$, deviating from the solid-black fit. At even larger values of $\mathrm{Po}>0.5$ the energy injection does again increase a little bit, but the specific reason for this is unclear. It could be due to, for example, further higher-order effects, boundary layer effects, or possibly even the horizontal wavenumber decreasing sufficiently, as $\mathrm{Po}$ is increased, such that it is now no longer de-tuned as it selects a different mode available in the simulation. Finally, although all data points are consistent with the solid-black fit within error bars a trend does appear to arise as a function of the Rayleigh number. The energy injection rate decreases as the Rayleigh number is increased (which we will show in more detail in Sec.~\ref{sec:scaling_laws_Ra}). This decrease is expected to continue with increasing values of the Rayleigh number until the turbulent energy injection of the precessional instability is overshadowed by the energy injection due to the effective viscosity. This has already happened to the burgundy circle with $\mathrm{Ra}=15\mathrm{Ra}_c$ at $\mathrm{Po}=0.25$ where the energy injection agrees with the $\mathrm{Po}^2$ fit in solid-pink, even though it is in the continuously turbulent regime and visual inspection of the time series of this simulation (not shown) reveals the secondary transition.

\begin{figure}
    \centering
    \includegraphics[width=\linewidth]{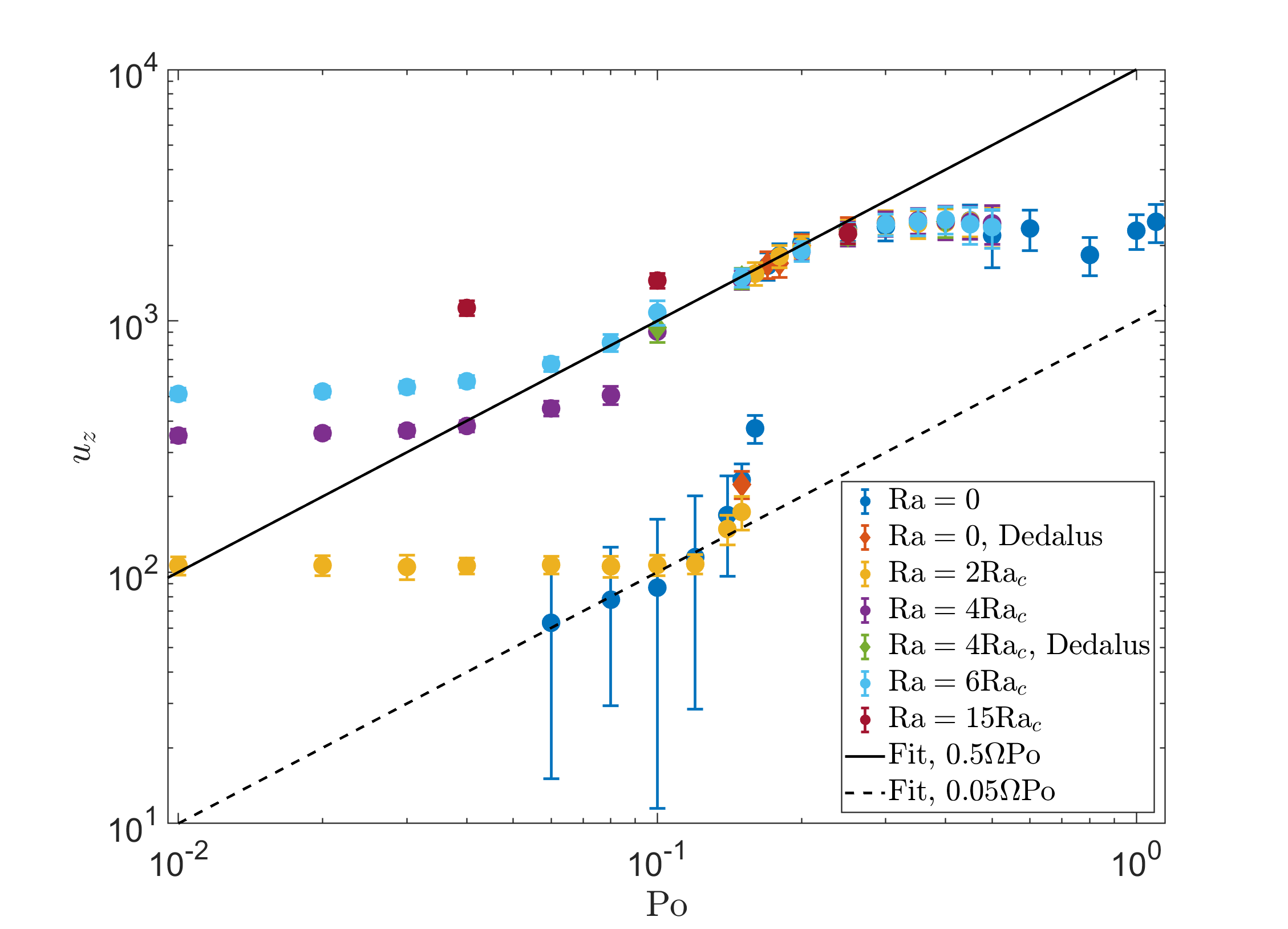}
    \caption[Same as Fig.~\ref{fig:I_tavg_funcPo}, instead showing the time-averaged rms vertical velocity $ u_z$.]{Same as Fig.~\ref{fig:I_tavg_funcPo}, instead showing the time-averaged rms vertical velocity $ u_z$. The vertical velocity in the absence of convection agrees with two different fits in the laminar (dashed-black line), and continuously turbulent regimes (solid-black line), respectively. Both fits are consistent with scaling as $\mathrm{Po}$. The velocities in the presence of convection match the convective velocity (i.e., with Po $=0$), until the continuously turbulent precessional regime arises in the simulation (Po $\gtrsim 0.16$), at which point they align with the continuously turbulent scaling in the absence of convection. The vertical velocity plateaus around $\mathrm{Po}\approx0.30$.}
    \label{fig:uz_tavg_funcPo}
\end{figure}

Having examined the time-averaged energy injection we now examine the vertical velocity. We examine the time-averaged rms vertical velocity $ u_z $ in Fig.~\ref{fig:uz_tavg_funcPo}. The velocities largely follow the same patterns as the energy injection: at low Poincar\'{e} numbers the convection dominates and the vertical velocity is therefore independent of $\mathrm{Po}$. The vertical velocities instead scale with the Rayleigh number. As $\mathrm{Po}$ is increased these values start deviating from the purely convective result, with clearly noticeable departure from $\mathrm{Po}\gtrsim0.06$ for $\mathrm{Ra}=4\mathrm{Ra}_c$ and $\mathrm{Ra}=6\mathrm{Ra}_c$. At even larger values of $\mathrm{Po}$ these data points instead start aligning with what appears to be a linear scaling with $\mathrm{Po}$ in solid-black. This scaling was predicted in Sec.~\ref{sec:diagnostics} to derive the energy injection rate. The solid-black fit again tends to agree with the results in the continuously turbulent regime after the abrupt jump at all values, up until $\mathrm{Po}\approx0.30$, after which it appears to plateau again. Note that the burgundy data points corresponding to the parameter sweep with $\mathrm{Ra}=15\mathrm{Ra}_c$ only agree in the continuously turbulent regime; in the convectively dominated regime the velocities of this sweep are much larger than predicted by the $\mathrm{Po}$ scaling in solid-black. The purely precessional results with $\mathrm{Ra}=0$ in blue data points at low $\mathrm{Po}$ are consistent with a different fit than the one shown in solid-black. This fit, plotted in dashed-black, also scales as $\mathrm{Po}$. The vertical velocities associated with this fit are about one order of magnitude smaller than those in the continuously turbulent regime.

\begin{figure}
    \centering
    \includegraphics[width=\linewidth]{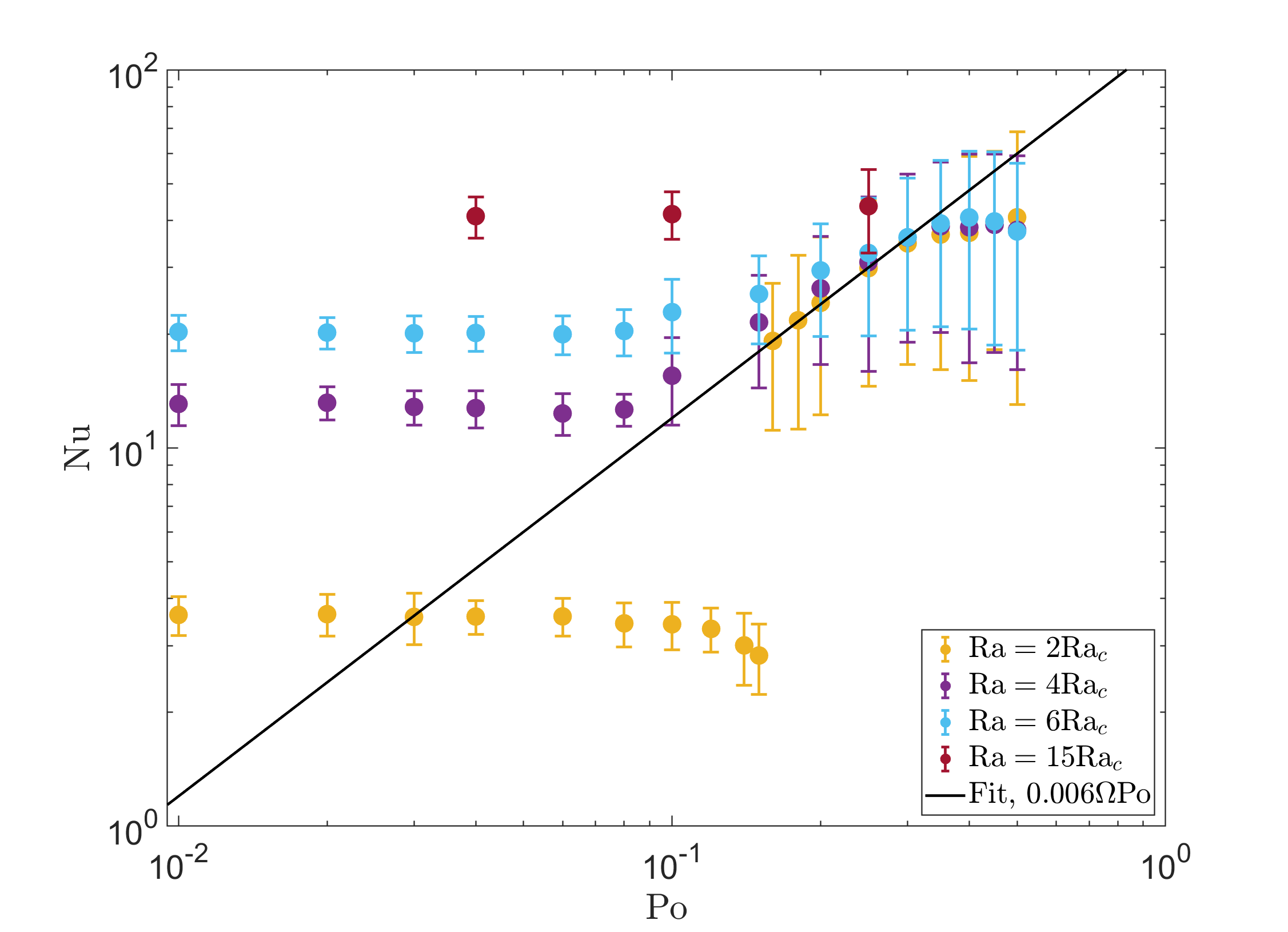}
    \caption[Same as Fig.~\ref{fig:I_tavg_funcPo}, instead showing the time-averaged Nusselt number $ \mathrm{Nu}$.]{Same as Fig.~\ref{fig:I_tavg_funcPo}, instead showing the time-averaged Nusselt number, $ \mathrm{Nu}$. The parameter sweep with $\mathrm{Ra}=0$ has not been plotted as the Nusselt number is then one by definition. Nu appears to decrease slightly in the effective viscosity regime as the 2D energy gets stronger with stronger precessional driving for the $\mathrm{Ra}=2\mathrm{Ra}_c$ case. Then it jumps or increases to match the continuously turbulent regime, where the Nusselt number appears to collapse onto a linear scaling consistent with $\mathrm{Po}$ like the vertical velocity. Nu plateaus around $\mathrm{Po}\approx0.35$.}
    \label{fig:Nu_tavg_funcPo}
\end{figure}

The Nusselt number in the convective, $\mathrm{Ra}\neq0$, simulations is plotted in Fig.~\ref{fig:Nu_tavg_funcPo}. For these convective parameter sweeps the Nusselt number follows the same general behaviour as the vertical velocity. Nu starts at the convective value for small Po, and then remains almost constant until it starts to transition into the continuously turbulent precessional regime. In the continuously turbulent precessional regime Nu is higher than in the absence of precession for most of the studied Rayleigh numbers, as also found in \citet[][]{WeiTilgner2013}{}{}. In the case of $\mathrm{Ra}=4\mathrm{Ra}_c$ and particularly in the case of $\mathrm{Ra}=6\mathrm{Ra}_c$ this transition is very gradual. The Nusselt number in simulations with $\mathrm{Ra}=2\mathrm{Ra}_c$ prior to the transition to the fully turbulent regime exhibits a downward trend. This is likely a result of the large-scale flow forming a vortex and becoming more energetic in these simulations as Po is increased. After the abrupt jump the results seem to agree roughly with a linear scaling with $\mathrm{Po}$ in the solid-black fit, just like the convective velocity. The Nusselt number once again deviates from this scaling from $\mathrm{Po}\approx0.35$ onwards. Note that the results obtained at $\mathrm{Ra}=15\mathrm{Ra}_c$ do not agree with the solid-black fit, as the inherent Nu due to convection is larger than any effects of the precessional instability in this parameter sweep. However, we do notice a slight increase of the convective heat transport in this parameter sweep with increasing values of $\mathrm{Po}$. We suspect this is because the convection is inhibited less by rotation due to the introduction of precession, which results in an effective rotation axis inclined with respect to the direction of gravity. This weakens the Taylor-Proudman constraint (along $z$), and thus allows for more efficient vertical heat transport. 

\subsection{Scaling laws as a function of the rotation rate}

\begin{figure*}
\subfloat[$I$ as a function of $\mathrm{Po}$ at different $\mathrm{Ek}$. \label{fig:I_tavg_func_Po_Ek}]{
         \includegraphics[width=0.48\textwidth]{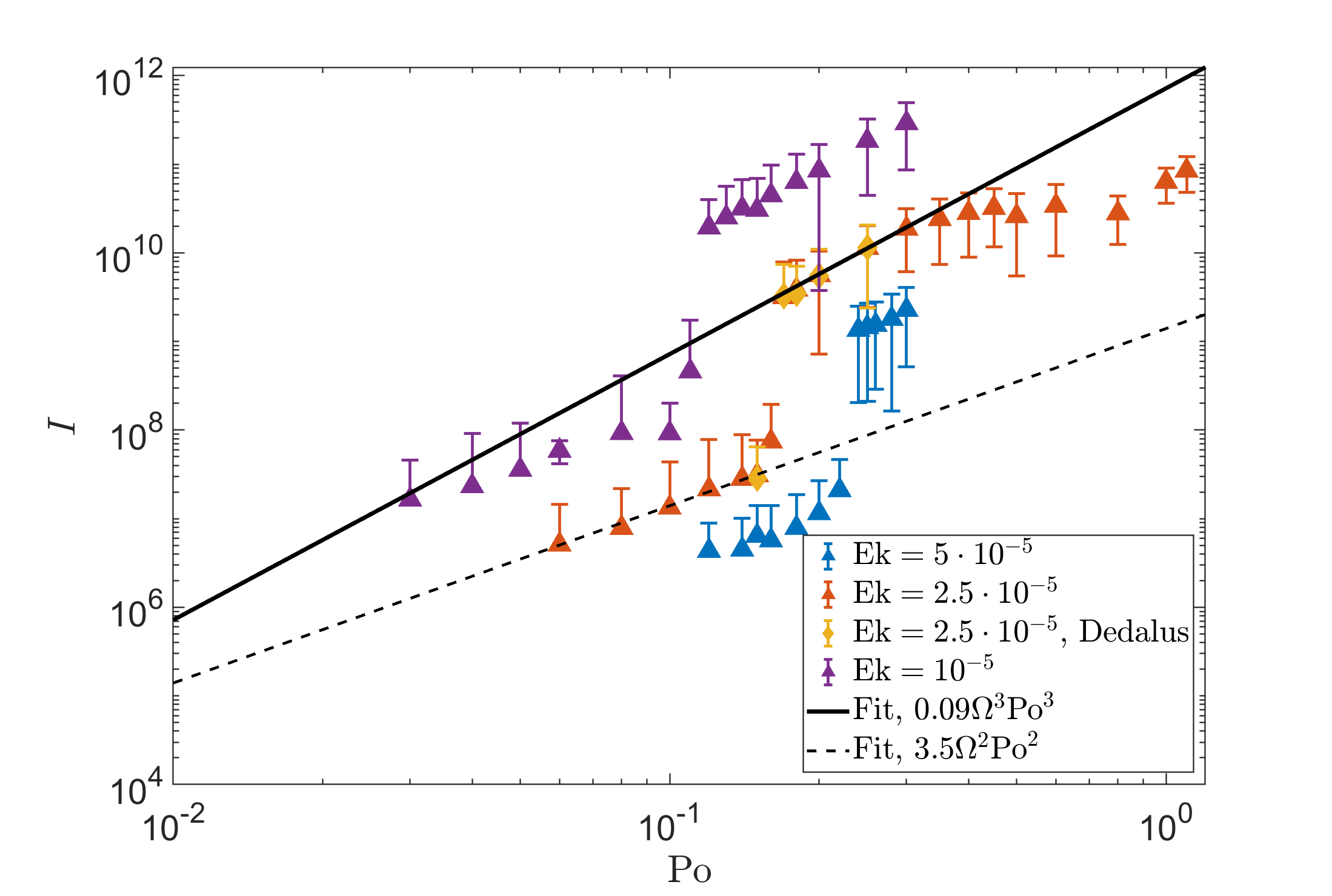}}
\hfill
\subfloat[Same as Fig.~\ref{fig:I_tavg_func_Po_Ek}, but rescaling the $x$-axis with $\mathrm{Ek}^{4/10}$. \label{fig:I_tavg_func_Po_Ek_xaxis}]{\includegraphics[width=0.48\textwidth]{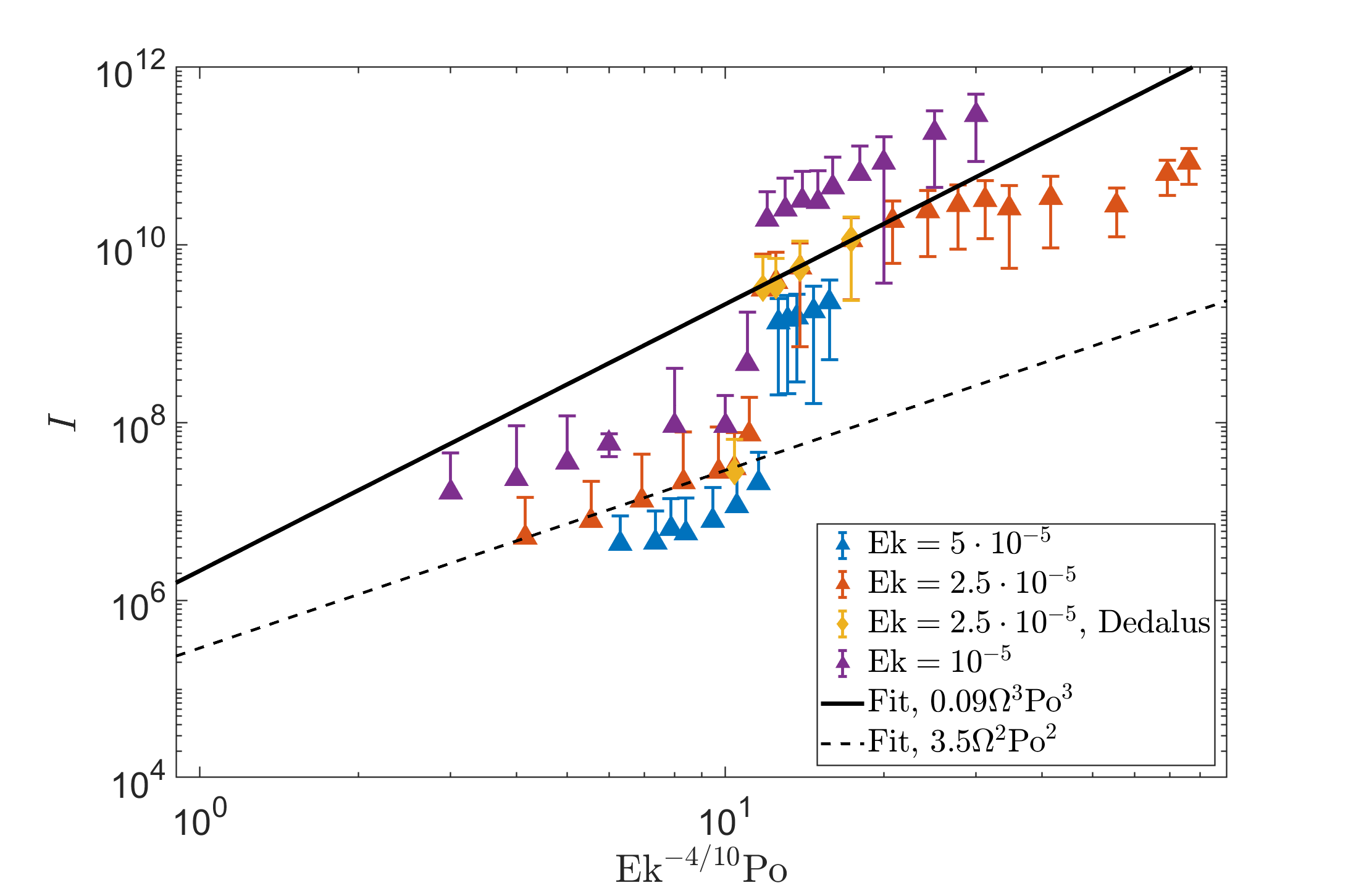}}\\

\subfloat[Same as Fig.~\ref{fig:I_tavg_func_Po_Ek}, but rescaling the $y$-axis with $\Omega^{3}$. \label{fig:I_tavg_func_Po_Ek_yaxis_cube}]{\includegraphics[width=0.48\textwidth]{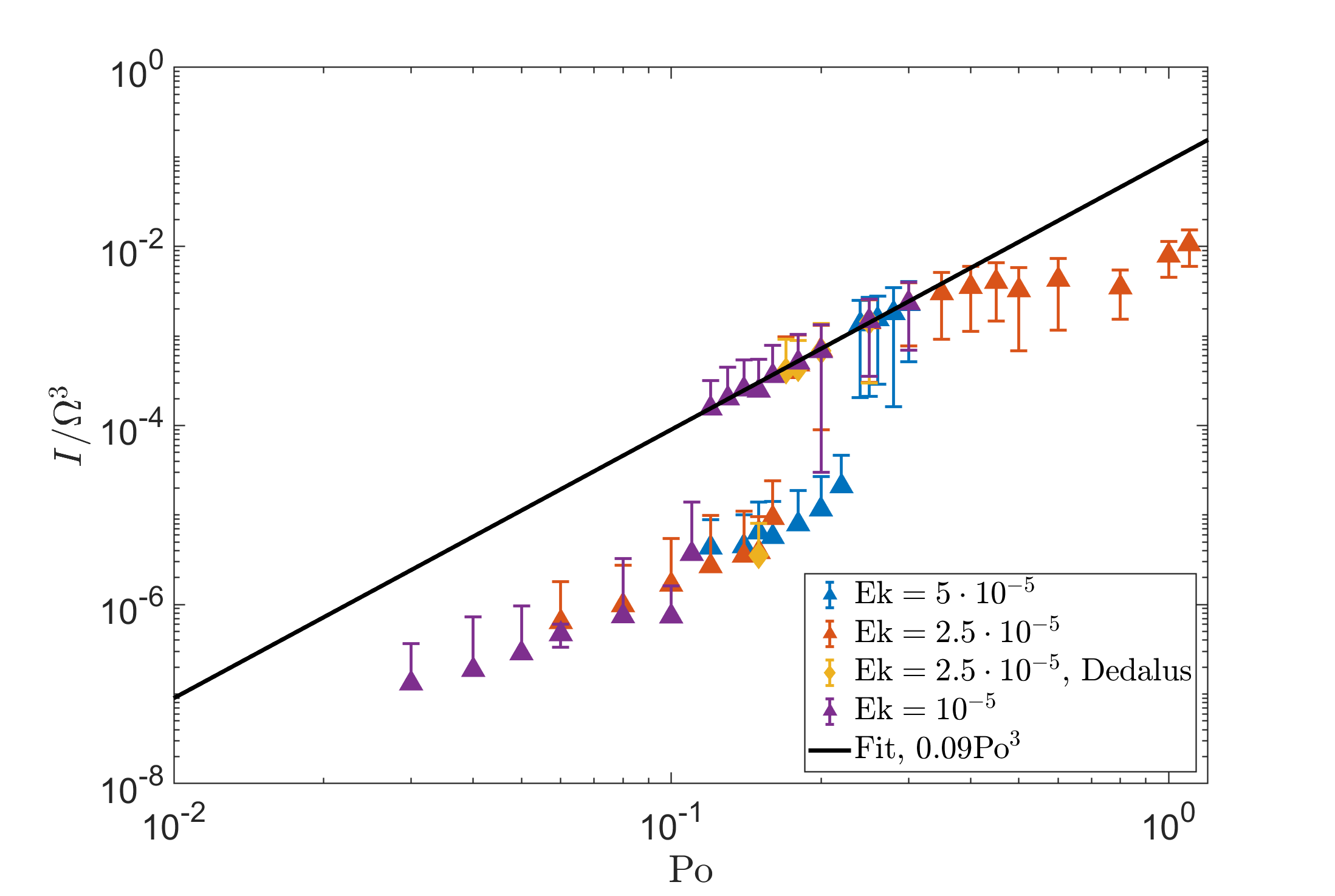}}
\hfill
\subfloat[Same as Fig.~\ref{fig:I_tavg_func_Po_Ek}, but rescaling the $y$-axis with $\Omega^{2.5}$. \label{fig:I_tavg_func_Po_Ek_yaxis_sqr}]{\includegraphics[width=0.48\textwidth]{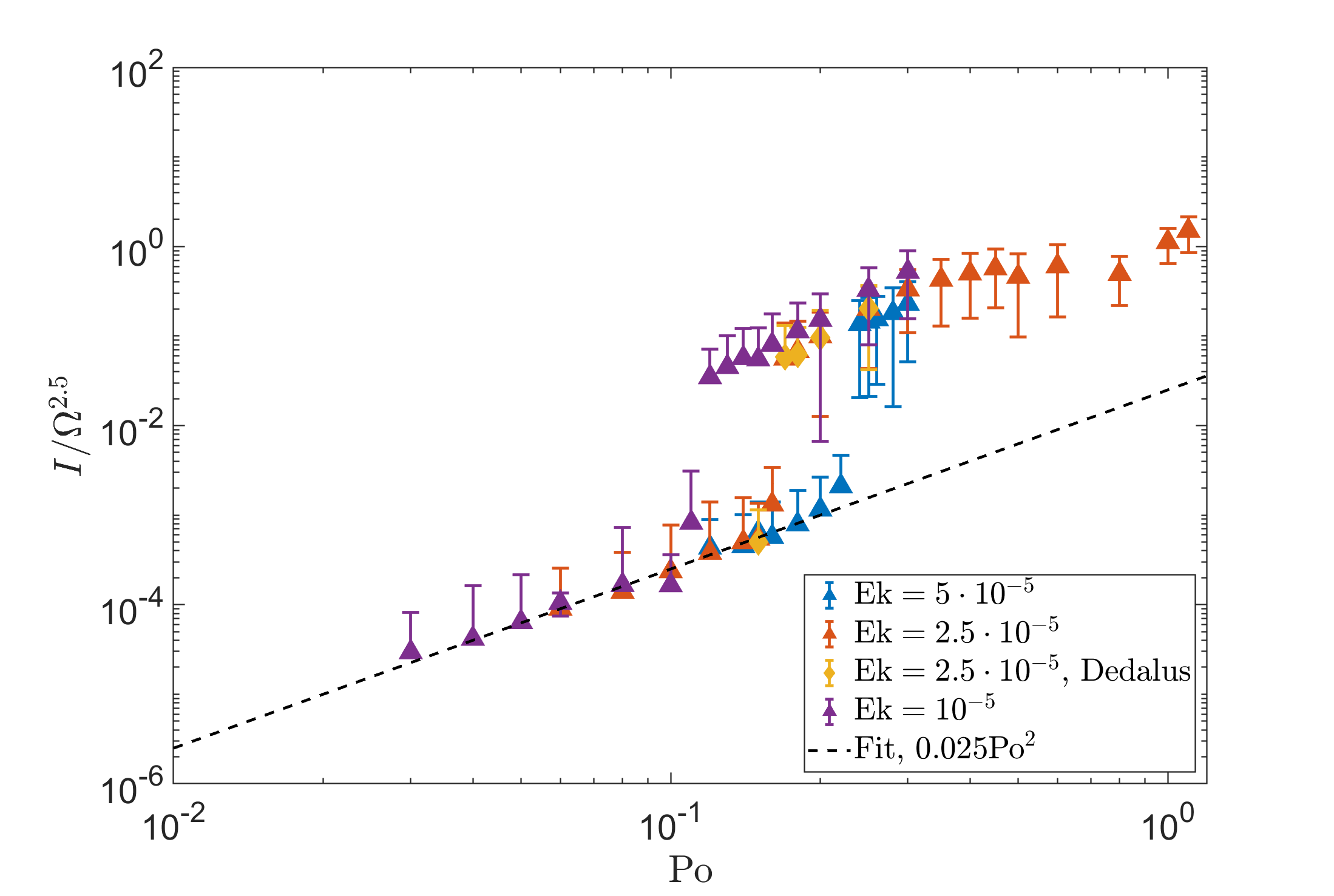}}

     \caption{The time-averaged energy injection $ I$ as a function of the Poincar\'{e} number at different values of the Ekman number, in the absence of convection. The chosen values of the Ekman number are $\mathrm{Ek}=[5\cdot 10^{-5},2.5\cdot10^{-5},10^{-5}]$ in blue, orange and purple triangles respectively. The \textsc{dedalus} results are plotted in yellow diamonds. The continuously turbulent scaling in solid-black and the laminar scaling in dashed-black from Fig.~\ref{fig:I_tavg_funcPo} are reproduced in the top-left panel. It shows that the transition to the continuously turbulent regime decreases with increasing Po, while the energy injection itself increases with increasing $\Omega$. Good agreement of the location of the transition in all three cases is found when rescaling the $x$-axis with $\mathrm{Ek}^{-4/10}$ in the top-right panel. In the bottom-left panel we have rescaled the $y$-axis with $\Omega^3$, and find that all three parameter sweeps collapse to the solid-black fit of the continuously turbulent regime. In the bottom-right panel we rescale with $\Omega^{5/2}$ instead and find that now the parameter sweeps collapse to the laminar scaling in dashed-black.}
    \label{fig:I_tavg}
\end{figure*}

 To start our examination of the effects of varying the rotation rate on the precessional instability -- in the absence of convection -- we have performed parameter sweeps as a function of Po with three different values of the Ekman number, $\mathrm{Ek}=[5\cdot10^{-5},2.5\cdot10^{-5},10^{-5}]$, obtained by varying the rotation rate. The results are plotted in   Fig.~\ref{fig:I_tavg_func_Po_Ek}. These three parameter sweeps are indicated in blue, orange and purple triangles, respectively. The \textsc{dedalus} simulations are plotted in yellow diamonds. The solid-black and dashed-black fits from Fig.~\ref{fig:I_tavg_funcPo} are reproduced in this figure. The energy injection increases as the Ekman number decreases, but the same general behaviour is maintained across all three parameter sweeps. All three parameter sweeps feature a laminar energy injection that appears to be consistent with a fit scaling as $\mathrm{Po}^2$. The energy injection of all three parameter sweeps starts deviating slightly from the laminar scaling and then abruptly jumps to one consistent with $\mathrm{Po}^3$ scaling. The value of Po at which this abrupt jump occurs decreases with decreasing Ekman number.

In Fig.~\ref{fig:I_tavg_func_Po_Ek_xaxis} the $x$-axis of Fig.~\ref{fig:I_tavg_func_Po_Ek} has been rescaled in such a way that the jump occurs at roughly the same value for all three parameter sweeps at different values of the Ekman number. The best agreement (by eye) was found by rescaling the $x$-axis to $\mathrm{Ek}^{-4/10}\mathrm{Po}$. This rescaling seems to indicate that the continuously turbulent regime of the precessional instability appears to emerge when $\mathrm{Ek}^{-4/10}\mathrm{Po}\gtrsim11$. Both the effects of changing the viscosity and the rotation rate to alter the Ekman number in these simulations has been tested, and it was found that it is the Ekman number that governs this transition, with the same location of the transition being found independently of varying the viscosity or rotation rate. The energy injection rate, however, depends on the tidal frequency, and thus it depends on the rotation rate and not the Ekman number. In astrophysical systems, with values of Po that are not too small and very small values of the Ekman number, we might thus expect to be in the continuously turbulent regime for the precessional instability in isolation. 

In Fig.~\ref{fig:I_tavg_func_Po_Ek_yaxis_cube} the $y$-axis of Fig.~\ref{fig:I_tavg_func_Po_Ek} has been rescaled instead, to test the scaling of the energy injection with the rotation rate (tidal frequency). The energy injection rate of all three parameter sweeps was rescaled by dividing each by their respective rotation rate cubed; the predicted scaling of the energy injection in the continuously turbulent regime. Upon rescaling, the data points which are located in the continuously turbulent region collapse and agree remarkably well with the fit of the energy injection rate in solid-black as given in Fig.~\ref{fig:I_tavg_funcPo}, with $\zeta=0.09$. The data points located in the laminar regime in this panel do not collapse under this rescaling, and the dashed-black fit corresponding to the laminar regime is omitted in this figure. Finally, in Fig.~\ref{fig:I_tavg_func_Po_Ek_yaxis_sqr} the energy injection rate is rescaled by dividing by $\Omega^{5/2}$ for each parameter sweep. This differs from the choice made previously in which the energy injection in the laminar regime scales as $\Omega^2$. However, the data points collapse more uniformly and better agreement with the fit in dashed-black is achieved when the $y$-axis is rescaled by $\Omega^{5/2}$ instead. For consistency, the original definition in which the energy injection scales as $\Omega^2$ in the laminar regime is maintained and thus the proportionality factor is $\Upsilon=0.025\Omega^{1/2}$. All three parameter sweeps in the laminar regime agree very well with the $\mathrm{Po}^2$ fit in dashed-black. The energy injection rate associated with the precessional instability in isolation, in this setup, therefore scales as:
\begin{equation}
   I =
    \begin{cases}
      0.025\Omega^{5/2}\mathrm{Po}^2 &\ \mathrm{for} \quad \mathrm{Ek}^{-4/10}\mathrm{Po}\lesssim11,\\
      0.09\Omega^{3}\mathrm{Po}^3 &\ \mathrm{for} \quad \mathrm{Ek}^{-4/10}\mathrm{Po}\gtrsim11.\\
    \end{cases} 
    \label{eq:Iscalings_OmegaPo}
\end{equation}

The scaling for the turbulent regime is diffusion-free and can therefore be readily extrapolated to planetary conditions. The transition between the two regimes, however, depends on the diffusivities, as does the scaling in the laminar regime. Care should therefore be taken when applying these results to planetary conditions, as the behaviour may be different at planetary diffusivities.

\subsection{Scaling laws as a function of the Rayleigh number}
\label{sec:scaling_laws_Ra}

The final aim of this work is to constrain the impact that convection has in decreasing the turbulent energy injection, as well as constraining the energy injection due to the turbulent effective viscosity of convection acting on the precessional flow. To this end, we plot the results we have obtained in three large parameter sweeps as a function of the Rayleigh number with $\mathrm{Po}=[0.04,0.10,0.25]$. The fitting of the precessional instability is complicated by the convection acting on the tidal flow like an effective viscosity and extracting energy from the precessional flow in our simulations, such that at high values of the Rayleigh number we expect the energy injection of the effective viscosity to overshadow that of the precessional instability. 

\begin{figure*}
    \centering
    \includegraphics[width=0.9\linewidth]{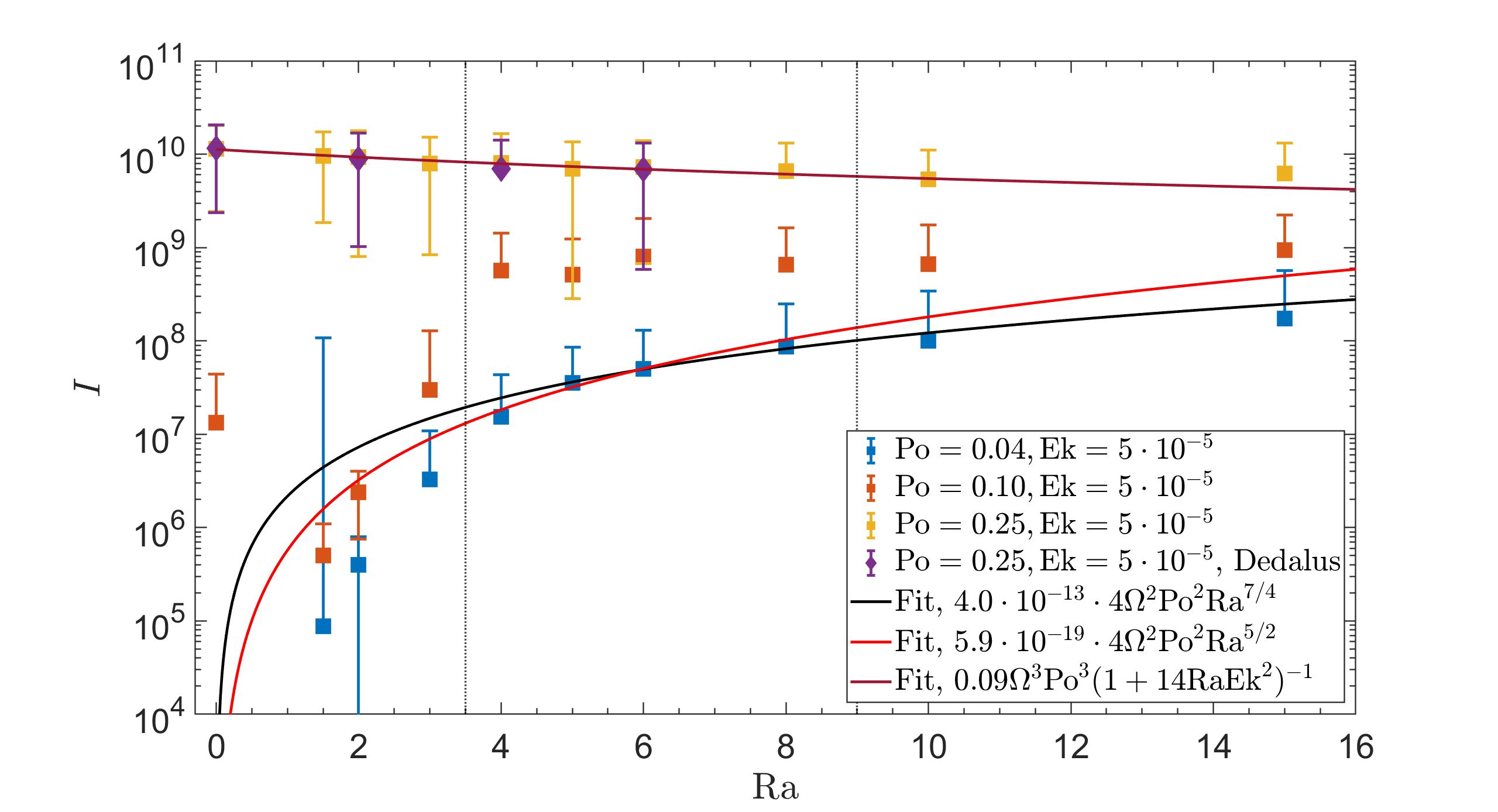}
    \caption[The time-averaged energy injection $ I $ as a function of the Rayleigh number at different values of the Poincar\'{e} number.]{Time-averaged energy injection $ I $ as a function of the Rayleigh number at different values of the Poincar\'{e} number. The parameter sweeps executed using \textsc{nek5000} with $\mathrm{Po}=[0.04,0.10,0.025]$ are plotted in blue, orange and yellow squares respectively. The \textsc{dedalus} results with $\mathrm{Po}=0.25$ are plotted in purple diamonds. The agreement between results obtained using both codes is very good. Results with $\mathrm{Po}=0.04$ portray the effective viscosity for all simulation values plotted, and we have therefore fitted them using the prescriptions of the effective viscosity given in Eq.~\eqref{eq:effviscscalings} in the high and intermediate frequency regimes in solid-red and solid-black respectively. We again observe good agreement for $\mathrm{Ra}>3\mathrm{Ra}_c$. The parameter sweep with $\mathrm{Po}=0.10$ is split into three parts by the vertical dotted-black lines. In the left third, when $\mathrm{Ra}\neq0$ the results follow the same pattern as the blue squares, and thus portray the effective viscosity. In the middle third the results instead follow the same pattern as the yellow squares and thus portray the convectively enabled continuously turbulent regime. Finally, in the right third the energy injection due to the effective viscosity is larger than that of the continuously turbulent precessional instability and thus the results portray the effective viscosity regime again. The results with $\mathrm{Po}=0.25$ show that the convection does indeed decrease the energy injection due to the continuously turbulent precessional instability, which is fitted in the solid-burgundy line.}
    \label{fig:I_tavg_funcRa}
\end{figure*}

The behaviour of the three parameter sweeps presented as a function of the Rayleigh number is plotted in Fig.~\ref{fig:I_tavg_funcRa}. There is a clear difference between the parameter sweeps with $\mathrm{Po}=0.04$ and $0.10$ with $\mathrm{Po}=0.25$. The energy injection in the former two increases as the Rayleigh number is increased, while the energy injection in the latter appears to decrease as $\mathrm{Ra}$ is increased. This behaviour is expected from Fig.~\ref{fig:I_tavg_funcPo}; in the effective viscosity regime larger Rayleigh numbers increase the energy injection, whereas in the continuously turbulent regime larger Rayleigh numbers decrease the energy injection. The parameter sweep with $\mathrm{Po}=0.04$ in blue squares portrays only the effective viscosity, as the precessional instability is expected to be unable to operate at this value of the Poincar\'{e} number, unless the convective driving is very strong, in which case the precessional instability is likely to be overshadowed by the effective viscosity regardless. Therefore, we have fitted this parameter sweep using the scalings of the effective viscosity in Eq.~\eqref{eq:effviscscalings}. In particular we have used the scalings of the high and intermediate frequency regimes in solid-red and solid-black, respectively. The energy injection agrees well with the high frequency prediction when $\mathrm{Ra}\gtrsim3\mathrm{Ra}_c$, until the convective driving becomes too strong, such that the simulations enter the intermediate frequency regime, agreeing well with the solid-black fit instead of the solid-red fit. From the fits we have obtained, we find the following scalings for the intermediate and high frequency regimes:
\begin{equation}
  \nu_{\mathrm{eff}} =
    \begin{cases}     0.091\mathrm{Ra}^{7/4}\mathrm{Ek}^{2}\mathrm{Pr}^{-1/4}\kappa^{3/2}d^{-1}\omega^{-1/2} & \mathrm{intermediate}\ \mathrm{frequency},\\
      0.380\mathrm{Ra}^{5/2}\mathrm{Ek}^{2}\mathrm{Pr}^{1/2}\kappa^{3}d^{-4}\omega^{-2}& \mathrm{high}\ \mathrm{frequency},
    \end{cases} 
    \label{eq:effvisc_fits}
\end{equation}
\noindent with $\omega=\Omega$. We assume that the fits for $u_c$ and $l_c$ as obtained previously in \citet[][]{deVries2023b} still apply. This is a valid assumption in the absence of precession, but as seen in Fig.~\ref{fig:uz_tavg_funcPo}, the convective velocities do increase as $\mathrm{Po}$ is increased. Upon reproducing the low frequency regime scaling from Eq.~\eqref{eq:effviscCraig_Ch4} for completeness (since we do not observe it here), and reproducing the frequencies at which these regimes apply, we find that the effective viscosity is described well by:
\begin{equation}
  \nu_{\mathrm{eff}} =
    \begin{cases}
      5u_{c}l_{c} & \frac{| \omega|}{\omega_c}\lesssim 10^{-2},\\
      0.78u_{c}l_{c}\Big(\frac{\omega_c}{\omega}\Big)^{\frac{1}{2}} & \frac{| \omega|}{\omega_c}\in[10^{-2},5],\\
      11.5u_{c}l_{c}\Big(\frac{\omega_c}{\omega}\Big)^2 & \frac{| \omega|}{\omega_c}\gtrsim 5.
    \end{cases} 
    \label{eq:effvisc_altered}
\end{equation}
It appears that the scalings are fairly consistent between the elliptical \citep[in][]{deVries2023b} and precessional background flows, but the specific flow appears to introduce different prefactors, even when its form is already taken into account in the definition of the effective viscosity in Eq.~\eqref{eq: I_and_nueff}.

Furthermore, there appear to be two different behaviours of the parameter sweep with $\mathrm{Po}=0.10$ in Fig.~\ref{fig:I_tavg_funcRa} and as such we have not performed any fits to this set. Instead, we have plotted two vertical dotted lines to indicate the boundaries between the three regimes that appear to be present in this sweep. In the leftmost third of the figure the sweep follows the effective viscosity, when the convection is unstable, following the exact same pattern as the sweep in blue squares. In the middle third the energy injection no longer follows the same behaviour as the blue squares and instead appears to be weakly decreasing with increasing Rayleigh number. This agrees with the observations in Fig.~\ref{fig:I_tavg_funcPo}; the convective driving has allowed the precessional instability to achieve the continuously turbulent regime. The orange squares in this middle third between the two dashed lines follow the same behaviour as the parameter sweep with $\mathrm{Po}=0.25$ in yellow squares. Finally, the energy injection due to the convection acting as an effective viscosity on the background flow overtakes the energy injection of the continuously turbulent precessional instability in the right-most part of the figure, and the two orange squares again follow the same trend as the blue squares.\\

Finally, we examine the energy injection in the parameter sweep with $\mathrm{Po}=0.25$ plotted in the yellow squares, as well as purple diamonds from the same simulations executed using \textsc{dedalus}. Good agreement is again found between the \textsc{dedalus} and \textsc{nek5000} results. In this continuously turbulent regime, a continuous downward trend is present until $\mathrm{Ra}=15\mathrm{Ra_c}$, where the energy injection of the turbulence acting like an effective viscosity again overtakes the precessional turbulence. We have fitted all yellow squares bar this last data point, by modifying the fit obtained for the continuously turbulent regime in Eq.~\eqref{eq:Iscalings_OmegaPo} according to:
\begin{equation}
     I  =0.09\Omega^3\mathrm{Po}^3(1+\Xi\mathrm{Ra}\mathrm{Pr}^{-1}\mathrm{Ek}^2)^{-1},
\end{equation}
 \noindent with $\Xi$ the fitting parameter to be determined. We have opted for $\mathrm{Ra}\mathrm{Pr}^{-1}\mathrm{Ek}^2$ instead of $\mathrm{Ra}$ to maintain the essence of $N^2/\Omega^2$ in this fit, which is a dimensionless and diffusion-free ratio that most often arises in the linear stability analysis of \citet[][]{Benkacem2022}, for example. The proportionality factor $\zeta=I/\Omega^3\mathrm{Po}^3$ thus depends on the convective driving, here represented by the Rayleigh number. Upon fitting this to the data, we find: 
\begin{equation}
     I  =0.09\Omega^3\mathrm{Po}^3(1+14\mathrm{Ra}\mathrm{Pr}^{-1}\mathrm{Ek}^2)^{-1}.
     \label{eq:I_prec_final}
\end{equation}
 \noindent This fit is plotted in solid-burgundy in Fig.~\ref{fig:I_tavg_funcRa}. No clear theoretical basis for this prescription has been identified, but it agrees very well with the data. Moreover, the parameter sweep is narrow, and is hindered by the presence of the effective viscosity in the simulations. We therefore caution the reader that this prescription may not be the correct one outside of the explored parameter regime. However, it clearly demonstrates the qualitative result that the energy injection due to the continuously turbulent precessional instability decreases as the convective driving is increased.

\section{Astrophysical applications}
\label{sec: astro app}

\subsection{Simple estimates of the tidal dissipation}
\label{sec: simple estimates}

Here we will attempt to obtain simple estimates for the tidal quality factor $Q'_{2,1,0}$ due to the precessional instability using the scaling laws obtained in Sec.~\ref{sec:scaling_laws}. We choose to work with the scaling of the energy injection rate $I\propto\zeta\mathrm{Po}^3$ for the precessional instability because the Ekman number in Jupiter and Jupiter-like planets is expected to be so small \citep[][]{Jupiterparams2004}, such that the condition in Eq.~\eqref{eq:Iscalings_OmegaPo} is satisfied if the Poincar\'{e} number is indeed $\mathcal{O}(10^{-4})$ in Hot Jupiters \citep{Barker2016precession}. In addition, using the parameters in \citet[][]{Jupiterparams2004} we find that the convective modification of the precessional dissipation according to Eq. ~\eqref{eq:I_prec_final} is negligible. The tidal dissipation due to the precessional instability, after re-introducing dimensional units and multiplying by the volume of the planet, is given by: 
\begin{equation}
    D=\zeta M_1 R_1^2 \Omega^3\mathrm{Po}^3,
    \label{eq:dissp_dimd_ellip}
\end{equation}
where $M_1$ is the planetary mass, and $R_1$ is the planetary radius. We have chosen to equate the size of our Cartesian box with the planetary radius in this expression. Although crude, it provides a first estimate of tidal dissipation rates, in the absence of global (spheroidal shell) simulations with convection, which is broadly consistent with our simulations. Next, the energy stored in the tidal response ($E_0$) is (to within an $O(1)$ factor):
\begin{equation}
    E_0\sim\frac{G M_1^2}{R_1}\epsilon^2,
\end{equation}
with $G$ the gravitational constant and $\epsilon$ the tidal amplitude due to tides raised by a body $M_2$ inside the planet, given by:
\begin{equation}
    \epsilon=\frac{M_2}{M_1}\left(\frac{R_1}{a}\right)^3=\left(\frac{M_2}{M_1+M_2}\right)\left(\frac{P_{\mathrm{dyn}}}{P_{\mathrm{orb}}}\right)^2,
\end{equation}
with $P_{\mathrm{orb}}$ the orbital period and $P_{\mathrm{dyn}}=2\pi\sqrt{\frac{R_1^3}{G M_1}}$ the dynamical or free-fall period of the body, which for Jupiter is $\approx2.8$ hours. The tidal quality factor $Q'_{2,1,0}$ is then given by \citep{Barker2016precession}:
\begin{equation}
     Q'_{2,1,0}\sim\frac{3}{2k_2}\frac{\Omega E_0}{D}\approx\frac{1}{\zeta\mathrm{Po}^3}\frac{3}{2k_2}\left(\frac{M_2}{M_1+M_2}\right)^2\frac{P_{\mathrm{dyn}}^2 \, P_{\mathrm{rot}}^2}{P_{\mathrm{orb}}^4}.
     \label{eq:Qprecessional_incomp}
\end{equation}
The rotation period $P_{\mathrm{rot}}=2\pi/\Omega$ is equal to the tidal period for this component of the tide. We can find the Poincar\'{e} number using $\mathrm{Po}=P_{\mathrm{rot}}/P_{\mathrm{p}}$, where $P_{\mathrm{p}}$ is the precession period given by \citep[e.g.][]{Kopal1959,Eggleton2001}{}{}:
  \begin{equation}
     P_{\mathrm{p}}\approx0.2\ \mathrm{yrs}\ \bigg(\frac{P_{\mathrm{rot}}}{1\ \mathrm{ d}}\bigg)    \bigg(\frac{P_{\mathrm{orb}}}{1\ \mathrm{ d}}\bigg)^2 
      \left(\frac{M_1+M_2}{M_2}\right)\frac{(1-e^2)^{3/2}}{\cos i},
     \label{eq:Pprec_estimate}
 \end{equation}
with $i$ the inclination or obliquity of the body's spin axis to its orbital rotation axis, and $e$ the orbital eccentricity. We have substituted characteristic values for Jupiter: $r_g^2\approx0.26$, $k_2\approx0.56$, the latter being obtained through Juno observations \citep[e.g.][]{Durante2020}{}{} and theoretical calculations \citep[][]{Lai2021,Dewberry2022}{}{}. We set $P_{\mathrm{orb}}= P_{\mathrm{rot}} = 1$ day, to mimic spin-orbit synchronisation already having occurred. We have chosen not to insert values of the eccentricity and inclination but instead included them in the expression for completeness. For the Poincar\'{e} number we find:
 \begin{equation}
     \mathrm{Po}\approx0.01 \bigg(\frac{1\ \mathrm{ d}}{P_{\mathrm{orb}}}\bigg)^2 
      \left(\frac{M_2}{M_1+M_2}\right)\frac{\cos i}{(1-e^2)^{3/2}}.
     \label{eq:Po_estimate}
 \end{equation}
Next, we rewrite Eq.~\eqref{eq:Qprecessional_incomp} by substituting $\mathrm{Po}$ using Eq.~\eqref{eq:Po_estimate}, to find:
\begin{equation}
     Q'_{2,1,0}\approx2\cdot10^5\,\frac{0.09}{\zeta}\left(\frac{M_1+M_2}{M_2}\right)    \bigg(\frac{P_{\mathrm{rot}}}{1\ \mathrm{ d}}\bigg)^2    \bigg(\frac{P_{\mathrm{orb}}}{1\ \mathrm{ d}}\bigg)^2 
     \left(\frac{(1-e^2)^{3/2}}{\cos i}\right)^3.
     \label{eq:Q_comp_prec}
\end{equation}
Our simple estimate therefore indicates that a Jupiter-like planet on a short-period orbit results in a low tidal quality factor of $Q'\sim10^5$, of the same magnitude as was found for the elliptical instability \citep[][]{deVries2023b}.

\subsection{Detailed computations of the tidal dissipation}
\label{sec:Detailed}

To provide a more detailed estimate of the tidal dissipation and associated tidal quality factor due to the precessional instability, as well as the effective viscosity, in a (Hot) Jupiter-like planet, we require models for its internal structure, i.e. profiles of pressure and density (and other quantities) as functions of radius. As in \citet[][]{deVries2023b,Lazovik2024}, we use a modified version of the test suite case \textit{make\_planets} of the \textsc{mesa} code \citep[][]{Paxton2011, Paxton2013, Paxton2015, Paxton2018, Paxton2019, Jermyn2022} with the \textsc{mesasdk} \citep[][]{richard_townsend_2022_7457723} to generate 1D interior profiles of giant planets, as also used previously \citep[e.g.][]{Muller2020,Muller2023}. 

We summarise here the parameters used to generate these models. The initial Jupiter model has a radius of $2R_J$ and a mass of $1M_J$, of which $10$ Earth-masses are located in an inert core with density $10\ \mathrm{g}\ \mathrm{cm}^{-3}$. The model is evolved for 4.5 Gyr to mimic the age of Jupiter and employs a constant surface irradiation of $5\cdot10^4$ erg $\mathrm{cm}^{-2}\ \mathrm{s}^{-1}$, similar to what Jupiter receives from the Sun, which is deposited at a column depth of 300 $\mathrm{g}\ \mathrm{cm}^{-2}$ (about $0.7$ bar). A Hot Jupiter model was also created with the same parameters except that we increased the surface heating to represent the irradiation of a one-day planet around a Sun-like star of $10^9$ erg $\mathrm{cm}^{-2}\ \mathrm{s}^{-1}$. Furthermore, we incorporate additional interior heating with uniform rate $0.05$ erg $\mathrm{cm}^{-3}\ \mathrm{s}^{-1}$ throughout the fluid envelope, which can be thought to represent the impact of tidal heating or Ohmic dissipation (or other mechanisms) that could possibly inflate a number of Hot Jupiters. This allows us to determine the effects of the increased radius (and stronger convection) of a puffy Hot Jupiter on the tidal dissipation rates. The \textsc{mesa} code by default treats convection using non-rotating mixing-length theory (MLT), for which we use the Cox prescription \citep[][]{Cox1968}.

 Details of the calculations of the relevant quantities from these models can be found in \citet[][]{deVries2023b}. We summarise the methods here. The convective velocities and lengthscales (mixing-lengths) obtained using \textsc{mesa} are calculated using flux-based non-rotating MLT. Thus, we must convert these convective velocities and lengthscales to RMLT by introducing a correction scaling with the convective Rossby number. We will denote quantities calculated using non-rotating MLT with tildes. The converted RMLT versions of the effective viscosity are thus:
\begin{equation}
  \nu_{\mathrm{eff}} \propto
    \begin{cases}
      5\Tilde{u}_c\Tilde{l}_c\Tilde{\textrm{Ro}}_c^{4/5}  & \frac{| \omega|}{\omega_c}\lesssim 10^{-2},\\
     0.5\Tilde{u}_c\Tilde{l}_c\Tilde{\textrm{Ro}}_c^{3/5} \bigg(\frac{\Tilde{u}_c/\Tilde{l}_c }{\omega}\bigg)^{\frac{1}{2}} & \frac{| \omega|}{\omega_c}\in[10^{-2},5],\\
      \frac{25}{\sqrt{20}}\Tilde{u}_c\Tilde{l}_c \bigg(\frac{\Tilde{u}_c/\Tilde{l}_c }{\omega}\bigg)^2 & \frac{| \omega|}{\omega_c}\gtrsim 5.
    \end{cases} 
    \label{eq:effviscCraig_RMLT}
\end{equation}

The Rossby numbers in both models were found to be much smaller than one, and hence convection is highly rotationally constrained, which justifies the use of RMLT (over MLT) in giant planets \citep[see][]{deVries2023b}. Likewise, it was found that the fast tides regime is the relevant one in both models, except for the final percent or so of the radius, which approaches the surface stable layer.

\subsection{Tidal dissipation rates in Jupiter and Hot Jupiters}
\label{sec: tidaldiss}

Using the radial profiles of $\nu_{\mathrm{eff}}$ obtained in \citet[][]{deVries2023b} we compute the resulting damping of the precessional flow and the associated tidal quality factors $Q'$ in our planetary models. We follow the procedure outlined in Sec. 2.1 of \citet[][]{Barker2020}{}{}, which follows e.g. \citet[][]{OL2004,OgilvieIW,Ogilvie2014}{}{}. Only the degree $l$ matters for this calculation, such that the tidal quality factor associated with the effective viscosity is the same regardless of the value of the order $m$, and thus should be the same for the convection acting on the precessional flow with $l=2,m=1,n=0$ as the $l=m=n=2$ tidal flow in \citet[][]{Craig2019effvisc,Craig2020effvisc,deVries2023b}. We therefore only compute this quantity using an imposed tidal potential given by $\Psi_{2,2,2}$ and apply this to find $Q'_{2,1,0}$, which should be the same as $Q'_{2,2,2}$. The only modification compared to \citet[][]{Barker2020}{}{} is that we account for the rotational dependence of $\nu_{\mathrm{eff}}$ and $\omega_c$ as described above, otherwise we employ their Eq.~(27) to obtain $\nu_{\mathrm{eff}}(r)$ in the various different frequency regimes. The resulting tidal quality factor is \citep[][]{Ogilvie2014}{}{}:
 \begin{equation}
    Q'_{l,m,n}=\frac{3(2l+1)R_1^{2l+1}}{16\pi G}\frac{|\omega||A|^2}{D_\nu},
\end{equation}
\noindent where $A\propto \epsilon$ is the amplitude of the tidal perturbation (so that the ratio $D_\nu/|A|^2$ and hence $Q'_{l,m,n}$ are independent of the tidal amplitude). We have already determined an expression for $Q_{2,1,0}'$ from the precessional instability in Eq.~\eqref{eq:Q_comp_prec}.

To further put our results in context, we will also compute the tidal quality factor resulting from the dissipation of linearly-excited inertial waves in convective regions by applying the frequency-averaged formalism of \citet{OgilvieIW} to our planetary models. We follow the approach outlined in Section 3.1 of \citet{Barker2020}, fully accounting for the planetary structure. This prediction for $\langle Q'_{l,m,\mathrm{IW}} \rangle$ provides a tidal frequency-independent ``typical level of dissipation" due to inertial waves according to linear theory. This is thought to be representative of the dissipation of inertial waves excited by linear tidal forcing, i.e.~not via the precessional instability. The frequency-averaged estimate of the tidal dissipation and associated tidal quality factor $Q'_{2,1,0}$, due to linearly-excited inertial waves for obliquity tides is, however, unreliable, because the tidal frequency of $-\Omega$ also resonates with what is known as the spin-over mode \citep[][]{OgilvieIW}{}{}. This mode should not be dissipated viscously in a spherical body, even though the frequency-averaged formalism predicts a large dissipation associated with the spin-over mode resonance. The frequency-average of the tidal dissipation consequently massively overestimates the resulting dissipation \citep[][]{Ogilvie2014}{}{}. When shifting into the precessing frame, where this resonance vanishes, as in \citet[][]{LinOgilvie2017}{}{}, the associated viscous dissipation purely due to this mode also disappears. The tidal dissipation associated with the $l=2$, $m=1$, $n=0$ component calculated this way depends on the size of the convective region, but generally turns out to have a similar order of magnitude as the frequency-averaged dissipation of the $l=2$, $m=2$, $n=2$ component \citep[][]{LinOgilvie2017}{}{}. Therefore, we opt not to plot $\langle Q'_{2,1,\mathrm{IW}}\rangle$, as it is unreliable \citep[see also][]{DamianiMathis2018}. Instead we point to $\langle Q'_{2,2,\mathrm{IW}} \rangle$, which we will denote as $\langle Q'_{\mathrm{IW}}\rangle$ going forwards, to give us an idea of the magnitude of the frequency-averaged dissipation due to linearly-excited inertial waves due to the obliquity tide.

\begin{figure*}
     \centering
     \begin{subfigure}[b]{0.48\textwidth}
         \centering
    \includegraphics[width=\linewidth, 
    trim=0cm 0cm 1cm 0.5cm,clip=true]{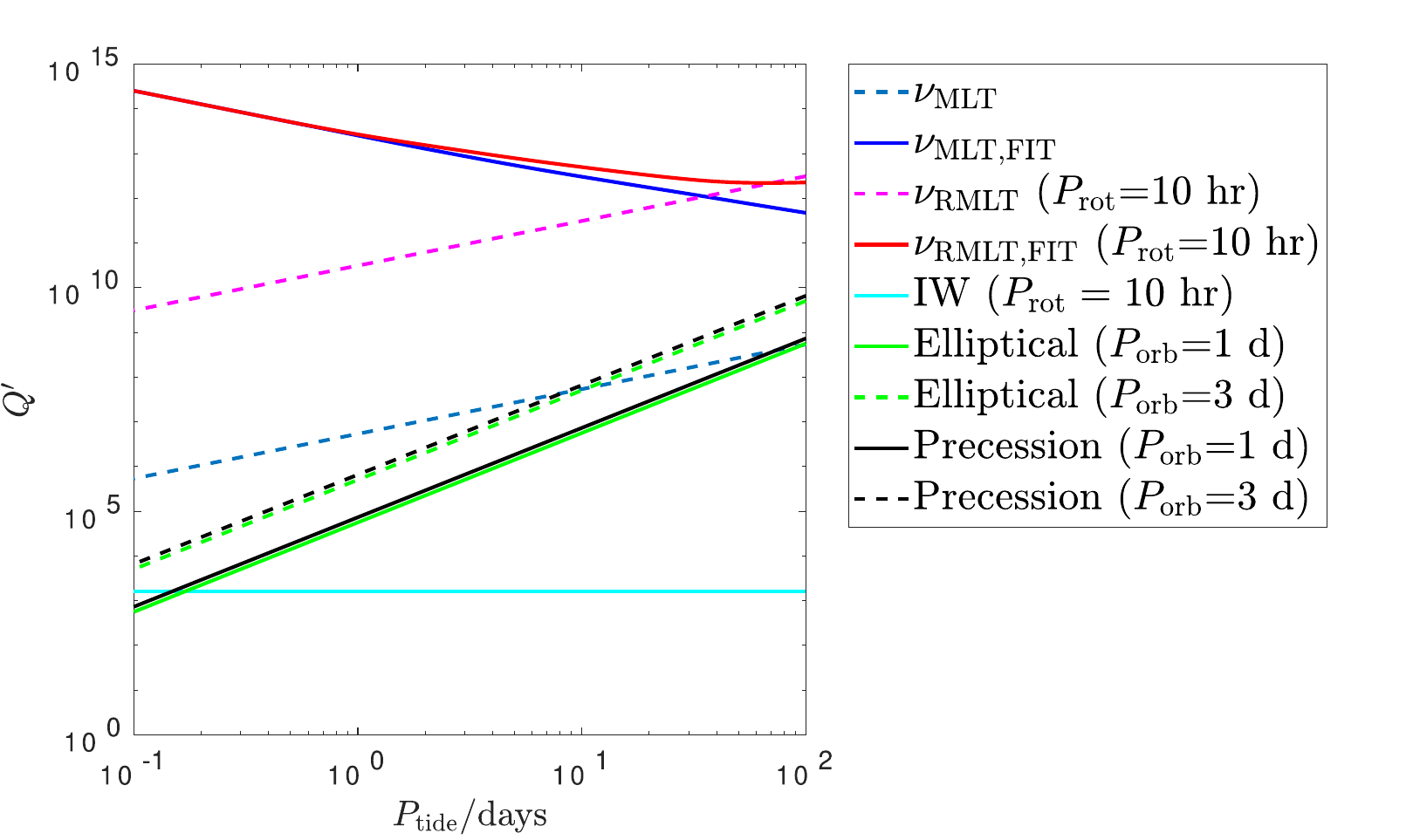}
    \caption{Tidal quality factor $Q'$ in the Jupiter model\label{fig:Q_jup}}
    
     \end{subfigure}
     \hfill
     \begin{subfigure}[b]{0.48\textwidth}
        \centering
    \includegraphics[width=\linewidth, 
    trim=0cm 0cm 1cm 0.5cm,clip=true]{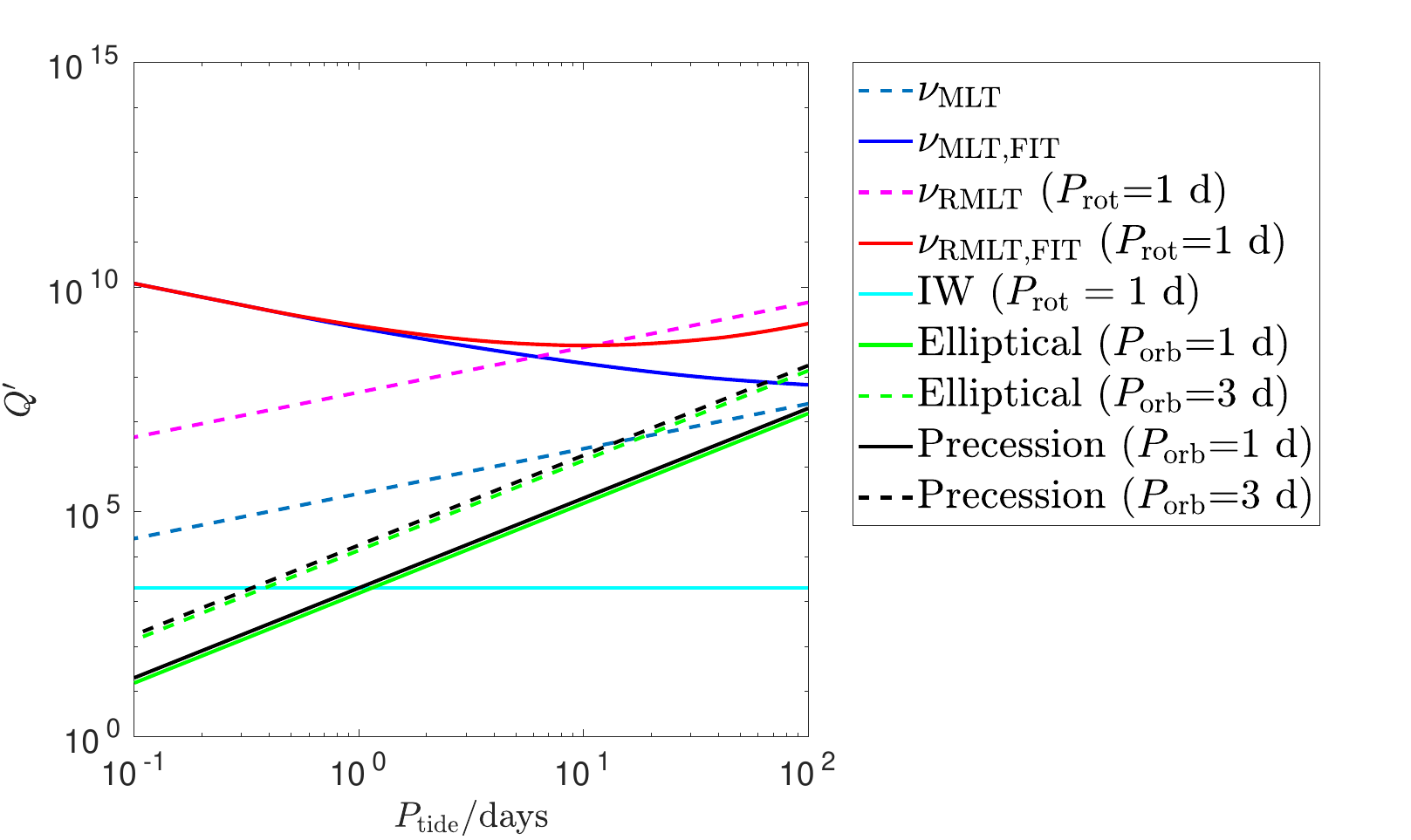}
    \caption{Tidal quality factor $Q'$ in the Hot Jupiter model}
    \label{fig:Q_HJ}
     \end{subfigure}
     \caption{Tidal quality factor $Q'$ as a function of tidal period for a myriad of mechanisms. In both panels, MLT and RMLT predictions for $Q'$ due to convective damping of equilibrium tides, using an effective viscosity with no tidal frequency reduction (low frequency regime), are shown in dashed-blue and dashed-magenta respectively. The frequency-reduced effective viscosities in solid-blue and solid-red for MLT and RMLT respectively indicate that the frequency reduction for fast tides significantly reduces the effectiveness of the dissipation. The elliptical instability in solid-green and dashed-green lines and the precessional instability in solid-black and dashed-black for two different orbital periods, and the (linear) frequency-averaged inertial wave dissipation in solid-cyan, are also plotted. Note that the inertial wave mechanism only applies when $|\omega|\leq|2\Omega|$, such that this mechanism is only valid when $P_{\mathrm{tide}}\geq{P_{\mathrm{rot}}}/{2}$. Inertial waves are considerably more dissipative than equilibrium tide damping by turbulent viscosity, whether they are linearly or non-linearly (i.e.\ via elliptical or precessional instabilities) excited. The elliptical and precessional instabilities are predicted to be dominant for the shortest tidal periods, and linear excitation of inertial waves is dominant for longer periods. The Hot Jupiter model has smaller $Q'$ (hence more efficient dissipation) for all dissipation mechanisms due to the larger radius and slower rotation.}
\label{fig:Q_Jup_HJ}
\end{figure*}

We show $Q'$ in Fig.~\ref{fig:Q_Jup_HJ} as a function of the tidal period for the precessional instability in black. We have incorporated $Q'$ due to the precessional instability together with our estimates for  $Q'$ obtained for the elliptical instability in \citet[][]{deVries2023b}. We set $e=0$, such that we consider a circular orbit, and $\cos{i}=1$, when considering the precessional instability. The latter implies that the spin and orbit are aligned, in which case the precessional instability does not operate. However, this is representative of the $Q'$ at small inclinations when $\cos{i}\approx1$, while providing a lower bound on $Q'_{2,1,0}$ due to the precessional instability. We show two predictions for the precessional instability, one with $P_{\mathrm{orb}}=1$ day and the other with $P_{\mathrm{orb}}=3$ days. When considering tidal dissipation due to the effective viscosity acting on the precessional background flow in the RMLT regime, i.e., when considering $Q'_{2,1,0}$ for $\nu_{\mathrm{RMLT}}$, the tidal frequency is set by the spin frequency, and thus only the tidal frequency corresponding to the rotation rate used to compute this curve should be considered for this background flow. The convective damping of the precessional flow by an effective viscosity in the Jupiter model in Fig.~\ref{fig:Q_jup} is an inefficient tidal dissipation mechanism in giant planets and leads to large $Q'$ and thus large tidal timescales. The low tidal frequency regime in dashed-blue and dashed-magenta for MLT and RMLT, respectively, indicate their strongest dissipation when the tidal frequency is large. If RMLT applies, as is expected, $Q'$ is still $\mathcal{O}(10^9)$ if we neglect the frequency reduction of $\nu_{\mathrm{eff}}$ for fast tides, thus the dissipation (and resulting tidal evolution) is weak. The combination of low, intermediate and high tidal frequency regimes for $\nu_{\mathrm{eff}}$ with the fitted prefactors in Eq.~\eqref{eq:effviscCraig_Ch4} dubbed $\nu_{\mathrm{FIT}}$ in solid-blue and solid-red, indicates that the high tidal frequency regime significantly impacts the effective viscosity, particularly when $P_{\mathrm{tide}}$ is small. Thus, we find weak tidal dissipation due to convection acting as an effective viscosity on the precessional flow throughout the entire range of tidal frequencies considered.

The precessional instability on a 1 day orbit in solid-black on the other hand, is an efficient dissipation mechanism, particularly when the tidal frequency, and thus for the precessional instability the rotation rate of the planet, is high. Note that according to our calculations the precessional instability results in values of $Q'$ that are similar to those due to the elliptical instability. The precessional instability on a 3 day orbit in dashed-black is weaker than the 1 day orbit prediction, but would still predict more effective dissipation than the (irrelevant) slow tides MLT effective viscosity for almost all of the parameter range considered. 

The most efficient mechanism in this model, except for the very highest tidal frequencies, is the frequency-averaged dissipation due to inertial waves shown in solid-cyan, which produces $Q'= \mathcal{O}(10^3)$ for our chosen rotation period. Since the rotation period is known, we would thus predict a typical value
\begin{align}
    Q' \approx 2 \cdot 10^3 \left(\frac{P_{\mathrm{rot}}}{10 \mathrm{hr}}\right)^2
    \label{eq:QIW_J}
\end{align}
for tidal dissipation due to inertial waves in the Jupiter model. Note that the inertial wave mechanism only applies when $|\omega|\leq|2\Omega|$, such that this mechanism is only valid when $P_{\mathrm{tide}}\geq\frac{P_{\mathrm{rot}}}{2}$.

The Hot Jupiter model in Fig.~\ref{fig:Q_HJ}, on the other hand, has a larger radius, stronger convection, and is rotating somewhat more slowly, so it has much higher effective viscosities and is impacted to a lesser extent by rotation. As a result, all mechanisms are more efficient. The elliptical and precessional instabilities are predicted to be particularly efficient for short orbital periods, e.g.~1 day orbit prediction for $Q'=\mathcal{O}(10^3)$ when the tidal period -- and thus rotation period for the precessional instability -- is 1 day. The increase in dissipation here due to the precessional instability (and elliptical instability) stems from the large radius of the Hot Jupiter, resulting in $\epsilon\approx0.095$, also increasing the value of $\mathrm{Po}$. Radius inflation and internal heating, as well as the marginally decreased rotation rate, allow the convective damping of equilibrium tides to operate more efficiently than in the Jupiter-like model in the top panel. However, once again the inertial wave mechanisms are predicted to be substantially more dissipative than the effective viscosity acting on equilibrium tides. For the linear dissipation of inertial waves in the Hot Jupiter model we find:
\begin{align}
    Q' \approx 2 \cdot 10^3 \left(\frac{P_{\mathrm{rot}}}{1 \mathrm{d}}\right)^2,
    \label{eq:QIW_HJ}
\end{align} which is predicted to be dominant for $P_{\mathrm{tide}}\gtrsim 2$ days.

 To estimate how quickly a precessing planet may align its spin and orbit, we define an alignment timescale $\tau_{i}$, like the one in \citet[][]{Barker2016precession}, as an e-folding timescale, i.e. the time in which the quantity under consideration decreases\footnote{Or increases, depending on the sign of $Q'_{2,1,0}$, which depends on the sign of the tidal frequency, and the sign in the relevant equation.} by a factor of $\mathrm{e}$:
\begin{equation}
    \tau_{i} \approx \frac{8}{9}\frac{Q'_{2,1,0}r_g^2}{2\pi}\left(\frac{M_1+M_2}{M_2}\right)^{2}    \frac{P^4_{\mathrm{orb}}}{P_{\mathrm{rot}}P^2_{\mathrm{dyn}}}(1-e^2)^{-1/2}.
\end{equation}
We substitute in characteristic values for Jupiter and the Sun. For the computations using the Jupiter model, we have set $P_{\mathrm{tide}}=P_{\mathrm{rot}}=10 \ \mathrm{hrs}$. We have computed values of $Q'$ from Eq.~\eqref{eq:Q_comp_prec} at the specified tidal periods for the precessional instability. This yields $Q'_{2,1,0} \approx 3\cdot10^4$ upon setting the inclination to $i=30^\circ$ and the eccentricity to zero. The timescale is then estimated to be:
\begin{equation}
    \tau_{i} \approx 6\cdot10^2\ \mathrm{yrs}\ \bigg(\frac{Q'_{2,1,0}}{3\cdot10^4}\bigg)\bigg(\frac{P_{\mathrm{orb}}}{1\ \mathrm{ d}}\bigg)^4\bigg(\frac{10\ \mathrm{h}}{P_{\mathrm{rot}}}\bigg)\bigg(\frac{2.8\ \mathrm{h}}{P_{\mathrm{dyn}}}\bigg)^2.
    \label{eq:tau_i_estimate_prec}
\end{equation}
 Using Eq.~\eqref{eq:tau_i_estimate_prec} we provide estimates of the orbital period for which the alignment timescale is 1 Gyr in Table~\ref{table:period_estimates}. A 1 Gyr timescale is chosen because it allows sufficient time for spin-orbit alignment of the Hot Jupiter to occur, and thus potentially be observable, while its (Sun-like) host star is on the main sequence. We produced these estimates for both the Jupiter-like and Hot Jupiter-like models. The computations are done using either a fixed rotation period or by setting the rotation period equal to the orbital period, i.e. $P_{\mathrm{rot}}=P_{\mathrm{tide}}=P_{\mathrm{orb}}$. These are referred to as non-synchronised and synchronised, respectively. In the former case, we have set for the Jupiter model $P_{\mathrm{rot}}=P_{\mathrm{tide}}=10 \ \mathrm{hrs}$, and used Jupiter's dynamical period. For the Hot Jupiter-like model we have used $P_{\mathrm{rot}}=P_{\mathrm{tide}} = 1 \ \mathrm{d}$, and computed the dynamical period to be $P_{\mathrm{dyn}} \approx 6.8 \ \mathrm{hrs}$.

These estimates of orbital periods are obtained in a rather crude way as both cases omit evolution of the rotation period in tandem with evolution of the planetary obliquity, and are therefore for illustrative purposes only. The efficiencies of all mechanisms based on inertial waves increase as the rotation rate is increased. As a result, the non-synchronised case can be thought of as an upper bound on the dissipation and critical orbital period (lower bound on $Q'$), while the synchronised case represents a plausible lower bound. They indicate the upper bounds that planets with orbital periods up to 11 days for the Jupiter model, or 23 days for the inflated Hot Jupiter model, can achieve spin-orbit alignment within 1 Gyr due to the precessional instability. The estimated values of $Q'$ for the effective viscosity mechanism are taken from the solid-red lines in Fig.~\ref{fig:Q_Jup_HJ}, estimating $Q'\sim9\cdot10^{13}\left(\frac{10  \ \mathrm{ hrs}}{P_{\mathrm{tide}}}\right)$ for the Jupiter model, and $Q'\sim4\cdot10^{9}\left(\frac{1  \ \mathrm{d}}{P_{\mathrm{tide}}}\right)$ for the Hot Jupiter model. The rotation and tidal periods are sufficiently low such that the effective viscosity is likely still in the RMLT fast tides regime. The resulting orbital periods for which the effective viscosity can achieve significant tidal evolution are extremely small in the non-synchronised case because the tidal frequencies are high. For the Hot Jupiter model in the synchronised case spin-orbit alignment can be achieved out to the same orbital distance as for the precessional instability due to the high values of the effective viscosity itself in this model, as well as the low tidal frequencies. This can only be achieved, however, if another mechanism is able to first synchronise the spin and orbit at these orbital periods, because the effective viscosity is unable to achieve spin-orbit synchronisation in a planet with rapid initial rotation, and thus high tidal frequency. The frequency-averaged inertial wave dissipation based on Eqs~\eqref{eq:QIW_J} and \eqref{eq:QIW_HJ} appears to be the most effective mechanism according to these estimates. Because this mechanism is most efficient when the rotation of the body is fast, the orbital period estimates for the non-synchronised case are particularly high. However, even the lower-bound synchronised estimate indicates alignment for Hot/Warm Jupiters with orbital periods up to 53 days. Although we caution that for such Warm Jupiters, the radius inflation is likely to lessened, and thus the Jupiter model might be more applicable. Furthermore, we caution that, as the orbital parameters evolve, the Poincar\'{e}  number changes, and thus the regime the flow is in could change. According to Eq.~\eqref{eq:Po_estimate} the Poincar\'{e}  number drops if the orbital separation gets larger,  the planetary obliquity increases or the eccentricity gets larger, which has not been taken into account in these estimates.

 \begin{table}
 \caption{Table of orbital periods in days at which the estimates of the alignment timescales in Eq.~\eqref{eq:tau_i_estimate_prec}, in Jupiter-like or Hot Jupiter-like planets, are 1 Gyr. The tidal dissipation mechanisms are the precessional instability (PI), effective viscosity of convection (eff. viscosity) and frequency-averaged inertial waves (IW). In the non-synchronised (non-sync) Jupiter-like estimates we have used $P_{\mathrm{tide}}=P_{\mathrm{rot}} = 10 \ \mathrm{hrs}$, while for the non-synchronised Hot Jupiter-like model we have used $P_{\mathrm{rot}}=P_{\mathrm{tide}} = 1 \ \mathrm{d}$. For the synchronised models (sync) we have set $P_{\mathrm{rot}}=P_{\mathrm{tide}}=P_{\mathrm{orb}}$. We have taken values of $Q'$ from Eqs.~\eqref{eq:Q_comp_prec},\eqref{eq:QIW_J} and \eqref{eq:QIW_HJ}, as well as Fig.~\ref{fig:Q_Jup_HJ}. We have set an initial inclination $i=30^\circ$ and the eccentricity to zero.}
\label{table:period_estimates}
\centering
\begin{tabular}{|l|l|l|l|}
\hline
                            & PI                                              & eff. viscosity  & IW    \\ \hline

Jupiter alignment non-sync            & 11 d                                            & 0.16 d               & 74 d  \\ \hline
Hot Jupiter alignment non-sync        & 23 d                                            & 3.8 d                & 142 d  \\ \hline

Jupiter alignment sync           & 6.9 d                                            & 0.06 d               & 26 d  \\ \hline
Hot Jupiter alignment sync        & 14 d                                            & 14 d                & 53 d  \\ \hline
\end{tabular}
\end{table}

\section{Discussion and conclusion}
\label{sec:discussion}
\subsection{Conclusions}

We have simulated the precessional instability in a local Cartesian model and its interaction with rotating Rayleigh-B\'{e}nard convection. The precessional instability excites inertial waves in convective regions of planets and stars, and can potentially drive turbulence. Our goal was to explore this instability as a possible mechanism of tidal dissipation in convective regions of giant planets, which could play an important role for spin-orbit evolution in planetary (and stellar) systems. Hence, we have attempted to quantify and obtain scaling laws for the (tidal) energy transfers from the precessional flow, and its resulting dissipation, both with and without convection, as well as to understand the dynamics of the flow.

The precessional instability in isolation in our model displays two main types of behaviour. In simulations with small values of the Poincar\'{e} number, bursty behaviour is observed, consisting of alternating periods of energy injection into the most unstable linear modes and periods where oscillating large-scale flows dominate. The formation of large-scale vortices is not observed in these cases. Meanwhile, at larger values of $\mathrm{Po}$ we observe a slightly more turbulent state with a large-scale vortex forming, which, when sufficiently powerful, allows a rapid secondary transition to a continuously turbulent state in which the flow itself is characterised by (sheared) vortices. We expect that this transition has been seen before by \citet[][]{masonkerswell}{}{} using the same setup, who describe a transition to a strongly energetic state in their 3D simulations at sufficiently large values of $\mathrm{Po}$. This phenomenon therefore appears to be important in this local model of the precessional instability.

We have furthermore identified that the introduction of convection replaces these two regimes. At small values of $\mathrm{Po}$, a predominantly convective regime is found, where the energy injection rate resembles that associated with convection acting on the laminar precessional flow as an effective viscosity. For sufficiently large Poincar\'{e} numbers, an abrupt secondary transition to the continuously turbulent regime of the precessional instability is again observed. However, the presence of convection lowers the critical Poincar\'{e} number that is required to achieve the continuously turbulent regime, in which case the secondary transition is much more gradual. On the other hand, convection also reduces the energy injection in the continuously turbulent regime of the precessional instability, resulting in a lower energy injection with increasing Ra. Convection therefore allows a slightly weakened version of the continuously turbulent regime to be reached at lower values of $\mathrm{Po}$.

We fit the energy injection scalings of the precessional instability in isolation as functions of $\mathrm{Po}$. The energy injection rate in the laminar regime scales as $\Upsilon\Omega^{2}\mathrm{Po}^2$, with a proportionality factor $\Upsilon=0.025\Omega^{1/2}$. In the continuously turbulent regime, the energy injection rate, and hence tidal dissipation rate, scales as $\zeta\Omega^3\mathrm{Po}^3$, with $\zeta=0.09$. The scaling law $\Omega^3\mathrm{Po}^3$ was also identified for the precessional instability in triply periodic box simulations \citep{Barker2016precession,Pizzi2022}. In our simulations of precessional instability in isolation, the condition to achieve the continuously turbulent regime can be described by $\mathrm{Ek}^{-4/10}\mathrm{Po}\gtrsim11$. Based on this condition, we might expect that the continuously turbulent regime is the correct one for Hot Jupiters, with very small values of the microscopic Ekman number and not too small values of the Poincar\'{e} number. When considering the presence of convection, the laminar regime is replaced by the energy injection rate of the convection acting like an effective viscosity scaling as $4\Omega^2\mathrm{Po}^2$, which also depends on the Rayleigh number. In both the continuously turbulent regime and the laminar regime, the vertical velocity scales linearly with $\mathrm{Po}$, although with different proportionality factors. The velocities are enhanced over the convective velocities in the absence of precession. The Nusselt number is enhanced by the precession \citep[this was previously observed in a different model by][]{WeiTilgner2013}{}{}, except at very small Rayleigh numbers. The Nusselt number also scales linearly with $\mathrm{Po}$ in the continuously turbulent regime, as long as the inherent convective heat transport is not too large to overshadow this scaling. Fits to the sustained energy injection in the laminar regime in convective simulations as a function of the Rayleigh number show good agreement with the intermediate and high frequency regimes of the effective viscosity damping the background precessional flow. The prefactors of these regimes are very close to those previously found for the background elliptical (or oscillatory simple shear) flow in \citet[][]{Craig2020effvisc,deVries2023b}. Furthermore, we have attempted to fit the modification of the energy injection due to the presence of convection in the continuously turbulent regime. A proportionality factor that depends on the convective driving as\footnote{Note that $\mathrm{Ra}\mathrm{Pr}^{-1}\mathrm{Ek}^2$ is related to $N^2/\Omega^2$, which is a dimensionless and diffusion-free ratio important for the precessional instability.} $\zeta=0.09(1+14\mathrm{Ra}\mathrm{Pr}^{-1}\mathrm{Ek}^{2})^{-1}$ is consistent with the data. This illustrates that convective driving reduces the energy injected by the precessional instability, even though no theoretical basis for this particular form of modification has yet been found.

By computing an expression for the (modified) tidal quality factor $Q'$ due to the precessional instability and applying it to the \textsc{mesa} models computed in \citet[][]{deVries2023b}, we predict that the precessional instability is efficient for very short orbital and tidal periods (with $Q'\sim 10^3$ in Hot Jupiters for periods of order one day), but that its efficiency falls off rapidly with increasing (tidal and orbital) periods. We have also computed $\langle Q'_{\mathrm{IW}} \rangle$ arising from the frequency-averaged dissipation due to linearly excited inertial waves (as opposed to their ``non-linear excitation" by the precessional instability) in ``realistic models" of giant planets \citep[following][]{OgilvieIW,Barker2020}. We find that inertial waves are by far the most efficient mechanism studied here, either those excited by the precessional instability for short orbital and tidal periods, or by the linear frequency-averaged dissipation. Based on our Hot Jupiter model, we predict an upper bound of 23 days for the orbital period of Hot/Warm Jupiters that should be aligned (i.e., to have zero \textit{planetary} obliquities) on timescales shorter than 1 Gyr due to the precessional instability. We also find an upper bound for these same planets to be aligned within 1 Gyr due to the linear frequency-averaged inertial wave mechanism out to even longer orbital periods of up to 142 days, but we caution that the evolution of the spin period has not been taken into account in this calculation. Lower bounds for these orbital periods assuming spin-orbit synchronisation are 14 days due to the precessional instability and 53 days due to the frequency-averaged inertial wave mechanism respectively. Hence, inertial waves (whether they are excited by precessional instability or are directly tidally forced) are likely to be the dominant contributors of tidal spin-orbit alignment in giant planets \citep[see also][]{deVries2023b,Lazovik2024}.

\subsection{Future work}

Future work should study in more detail the modification of convection by precession. Both a linear stability analysis of convection in the presence of precession, as well as an analysis of the precessional instability in the presence of stratification would be worthwhile. Next, a more detailed analysis of convective quantities such as the Nusselt number and convective velocity in non-linear simulations would be interesting, to more clearly constrain the effect that precession has on convective motions, although it appears that precession acts primarily to enhance convection in this setup. Additionally, simulations using local models with different aspect ratios should be performed. Boxes that are larger in the horizontal direction allow larger energies in the convectively generated vortices and possibly also in the precessional vortices. This higher energy ceiling in the simulation may result in the secondary transition taking longer, or possibly multiple transitions will occur during one simulation. The continuously high energies in the vortices appears to have been important in the suppression of the elliptical instability \citep[][]{deVries2023}, but their effects on the precessional instability are unclear. Furthermore, we have not probed the effects of stable stratification in non-linear simulations. Finally, a larger parameter sweep at larger values of the rotation rate, and thus smaller values of the Ekman number that are closer to astrophysical values, would be useful. This would reduce the effective viscosities, both because the tidal frequency would increase and the convective velocities and lengthscales would decrease. If the effective viscosity is less powerful, while the energy injection is larger due to larger rotation rates, the energy injection due to the precessional instability can be probed at larger Rayleigh numbers because it would no longer be overshadowed by the energy injection due to the effective viscosity. This would allow the effects of convection on the energy injection by the precessional instability to be constrained in a broader region of parameter space.

In our model, the local box is located at the poles of the planet. It is important to examine the effects of different latitudes on the interactions of the precessional and convective instabilities. However, the precessional instability makes this particularly difficult, as the precessional flow would clash with the impenetrable boundaries at lower latitudes in boxes that are angled such that gravity still points in the $z$-direction. As such, global simulations that are sufficiently turbulent and rapidly rotating to capture regimes similar to those we have explored would probably be the only method of achieving this. Thus, we propose simulations similar to \citet[][]{WeiTilgner2013}, but focussing on the tidal dissipation. Such a study with global simulations (in oblate spheroids or triaxial ellipsoids, for example) would further shed light on the modification of the background flow by interactions with the inertial waves excited by the precessional instability, and the effect of this modification on the resulting turbulence and tidal dissipation. Such global simulations also have the advantage that the inertial waves are no longer constrained by the (artificial) aspect ratio of the box. Furthermore, one might find that the inertial waves, excited by the precessional instability, can lead to wave attractors if one studies them in spherical shells \citep[e.g.][]{hollerbach_kerswell_1995,Noir2001}{}{}, which can be thought of as mimicking a giant planet with a dilute core that is sufficiently stably stratified. 

There are strong magnetic fields present in Jupiter, and there are observations tentatively indicating that a number of Hot Jupiters also possess strong magnetic fields \citep[][]{Cauley_HJ_strongB}. Therefore it is important to study the effects of magnetic fields in our simulations, as they could have significant effects on tidal dissipation. Magnetic fields may prevent LSV formation \citep[][]{Mak2017}{}{} by the precessional instability, and therefore allow continuous operation of the resulting energy transfers \citep[][]{Barker2014}{}{}. The precessional instability seems to achieve the continuously turbulent regime even at small values of $\mathrm{Po}$ when magnetic fields are introduced \citep[][]{Barker2016precession}{}{}, but it would be interesting to see if this is maintained if convection is also present. Furthermore, since our results point to a reduction in the energy injected by the precessional instability in the presence of convection, but the results in \citet[][]{Barker2016precession}{}{} point to enhanced energy injection in the presence of magnetic fields, magnetic fields could remedy this weakening of the precessional instability as a tidal dissipation mechanism. Furthermore, it would be important to study the precessional instability as a dynamo mechanism \citep{Malkus1968,Kerswell1996,Tilgner2005,Wu2008, LeBars_prec_2015} with convection also present, to study whether it is still efficient as a dynamo mechanism. On the other hand, our results, like those of \citet[][]{WeiTilgner2013}{}{}, also point to more vigorous convection in the presence of precessional flows, and it would be useful to study whether precession could act to enhance rotating convection as a dynamo mechanism. 

A final avenue of future work is related to the analysis of tidal dissipation rates using planetary models. It would be worthwhile to modify the equation of state in the interior models obtained with \textsc{mesa} in a manner akin to \citet[][]{Muller2020}{}{}, which would allow us to obtain an extended dilute core, and to measure the impact of such a core on tidal dissipation rates. Furthermore, a stably stratified dilute core might provide an important additional contribution to tidal dissipation by permitting the excitation of internal (inertia-)gravity waves \citep[e.g.][]{F2016,Andre2019,Christina_2020,Pontin_thesis,Pontin_2023,Lin2023,Dewberry2023,Pontin_2023b,Dhouib2024} as well as modifying the efficiency of inertial wave excitation in the overlying convective envelope \citep[][]{Pontin_thesis,Pontin_2023b}{}{}. Finally, studying how $Q'$ evolves with planetary and orbital evolution for each of the mechanisms explored in Sec.~\ref{sec: astro app} would be worthwhile, research which has been started in the investigation of \citet[][]{Lazovik2024}{}{}. For self-consistency, one might also consider evolving irradiation fluxes in tandem with the structural evolution.

\section*{Acknowledgements}

We would like to thank the referee for their careful reading of the manuscript and for their constructive comments that helped us to improve the paper. NBV was supported by EPSRC studentship 2528559 and STFC grant ST/Y002164/1. AJB and RH were supported by STFC grants ST/S000275/1,  ST/W000873/1 and UKRI1179. AJB and RH thank the Isaac Newton Institute for Mathematical Sciences, Cambridge, for support and hospitality, during the programmes ``Anti-diffusive dynamics: from sub-cellular to astrophysical scales'' (AJB) and ``Frontiers in dynamo theory: from the Earth to the stars'' (RH). RH's visit to the Newton Institute was also supported by a grant from the Heilbronn Institute. Simulations were undertaken on ARC4, part of the High Performance Computing facilities at the University of Leeds, and the DiRAC Data Intensive service at Leicester, operated by the University of Leicester IT Services, which forms part of the STFC DiRAC HPC Facility (\href{www.dirac.ac.uk}{www.dirac.ac.uk}). The equipment was funded by BEIS capital funding via STFC capital grants ST/K000373/1 and ST/R002363/1 and STFC DiRAC Operations grant ST/R001014/1. DiRAC is part of the National e-Infrastructure.

%%%%%%%%%%%%%%%%%%%%%%%%%%%%%%%%%%%%%%%%%%%%%%%%%%
\section*{Data Availability}

The simulation data used in this article will be shared on reasonable request to the corresponding author.

%%%%%%%%%%%%%%%%%%%% REFERENCES %%%%%%%%%%%%%%%%%%

\bibliographystyle{mnras}
\bibliography{references}

\begin{thebibliography}{}
\makeatletter
\relax
\def\mn@urlcharsother{\let\do\@makeother \do\$\do\&\do\#\do\^\do\_\do\%\do\~}
\def\mn@doi{\begingroup\mn@urlcharsother \@ifnextchar [ {\mn@doi@}
  {\mn@doi@[]}}
\def\mn@doi@[#1]#2{\def\@tempa{#1}\ifx\@tempa\@empty \href
  {http://dx.doi.org/#2} {doi:#2}\else \href {http://dx.doi.org/#2} {#1}\fi
  \endgroup}
\def\mn@eprint#1#2{\mn@eprint@#1:#2::\@nil}
\def\mn@eprint@arXiv#1{\href {http://arxiv.org/abs/#1} {{\tt arXiv:#1}}}
\def\mn@eprint@dblp#1{\href {http://dblp.uni-trier.de/rec/bibtex/#1.xml}
  {dblp:#1}}
\def\mn@eprint@#1:#2:#3:#4\@nil{\def\@tempa {#1}\def\@tempb {#2}\def\@tempc
  {#3}\ifx \@tempc \@empty \let \@tempc \@tempb \let \@tempb \@tempa \fi \ifx
  \@tempb \@empty \def\@tempb {arXiv}\fi \@ifundefined
  {mn@eprint@\@tempb}{\@tempb:\@tempc}{\expandafter \expandafter \csname
  mn@eprint@\@tempb\endcsname \expandafter{\@tempc}}}

\bibitem[\protect\citeauthoryear{Akinsanmi, Barros, Santos, Oshagh  \&
  Serrano}{Akinsanmi et~al.}{2020}]{Akinsanmi2020}
Akinsanmi B.,  Barros S. C.~C.,  Santos N.~C.,  Oshagh M.,   Serrano L.~M.,
  2020, \mn@doi [Monthly Notices of the Royal Astronomical Society]
  {10.1093/mnras/staa2164}, 497, 3484

\bibitem[\protect\citeauthoryear{{Andr{\'e}}, {Mathis}  \&
  {Barker}}{{Andr{\'e}} et~al.}{2019}]{Andre2019}
{Andr{\'e}} Q.,  {Mathis} S.,   {Barker} A.~J.,  2019, \mn@doi [Astronomy \&
  Astrophysics] {10.1051/0004-6361/201833674}, \href
  {https://ui.adsabs.harvard.edu/abs/2019A&A...626A..82A} {626, A82}

\bibitem[\protect\citeauthoryear{Ascher, Ruuth  \& Spiteri}{Ascher
  et~al.}{1997}]{Ascher1997}
Ascher U.~M.,  Ruuth S.~J.,   Spiteri R.~J.,  1997, \mn@doi [Applied Numerical
  Mathematics] {10.1016/S0168-9274(97)00056-1}, 25, 151

\bibitem[\protect\citeauthoryear{Barker}{Barker}{2016a}]{Barker2016}
Barker A.~J.,  2016a, \mn@doi [Monthly Notices of the Royal Astronomical
  Society] {10.1093/mnras/stw702}, 459, 939

\bibitem[\protect\citeauthoryear{Barker}{Barker}{2016b}]{Barker2016precession}
Barker A.~J.,  2016b, \mn@doi [Monthly Notices of the Royal Astronomical
  Society] {10.1093/mnras/stw1172}, 460, 2339

\bibitem[\protect\citeauthoryear{{Barker}}{{Barker}}{2020}]{Barker2020}
{Barker} A.~J.,  2020, \mn@doi [Monthly Notices of the Royal Astronomical
  Society] {10.1093/mnras/staa2405}, \href
  {https://ui.adsabs.harvard.edu/abs/2020MNRAS.498.2270B} {498, 2270}

\bibitem[\protect\citeauthoryear{Barker}{Barker}{2022}]{Barker2022}
Barker A.~J.,  2022, \mn@doi [The Astrophysical Journal Letters]
  {10.3847/2041-8213/ac5b63}, 927, L36

\bibitem[\protect\citeauthoryear{{Barker}}{{Barker}}{2025}]{B2025}
{Barker} A.~J.,  2025, \mn@doi [arXiv e-prints] {10.48550/arXiv.2504.10941},
  \href {https://ui.adsabs.harvard.edu/abs/2025arXiv250410941B} {p.
  arXiv:2504.10941}

\bibitem[\protect\citeauthoryear{Barker \& Lithwick}{Barker \&
  Lithwick}{2013}]{Barker2013}
Barker A.~J.,  Lithwick Y.,  2013, \mn@doi [Monthly Notices of the Royal
  Astronomical Society] {10.1093/mnras/stt1561}, 435, 3614

\bibitem[\protect\citeauthoryear{Barker \& Lithwick}{Barker \&
  Lithwick}{2014}]{Barker2014}
Barker A.~J.,  Lithwick Y.,  2014, \mn@doi [Monthly Notices of the Royal
  Astronomical Society] {10.1093/mnras/stt1884}, 437, 305

\bibitem[\protect\citeauthoryear{Barker, Dempsey  \& Lithwick}{Barker
  et~al.}{2014}]{Barker2014RMLT}
Barker A.~J.,  Dempsey A.~M.,   Lithwick Y.,  2014, \mn@doi [The Astrophysical
  Journal] {10.1088/0004-637x/791/1/13}, 791, 13

\bibitem[\protect\citeauthoryear{Barker, Braviner  \& Ogilvie}{Barker
  et~al.}{2016}]{BBO2016}
Barker A.~J.,  Braviner H.~J.,   Ogilvie G.~I.,  2016, \mn@doi [Monthly Notices
  of the Royal Astronomical Society] {10.1093/mnras/stw701}, 459, 924

\bibitem[\protect\citeauthoryear{Barnes \& Fortney}{Barnes \&
  Fortney}{2003}]{Barnes2003}
Barnes J.~W.,  Fortney J.~J.,  2003, \mn@doi [The Astrophysical Journal]
  {10.1086/373893}, 588, 545

\bibitem[\protect\citeauthoryear{Benkacem, Salhi, Khlifi, Nasraoui  \&
  Cambon}{Benkacem et~al.}{2022}]{Benkacem2022}
Benkacem N.,  Salhi A.,  Khlifi A.,  Nasraoui S.,   Cambon C.,  2022, \mn@doi
  [Physical Review E] {10.1103/PhysRevE.105.035107}, 105, 035107

\bibitem[\protect\citeauthoryear{Biersteker \& Schlichting}{Biersteker \&
  Schlichting}{2017}]{Biersteker2017}
Biersteker J.,  Schlichting H.,  2017, \mn@doi [The Astronomical Journal]
  {10.3847/1538-3881/aa88c2}, 154, 164

\bibitem[\protect\citeauthoryear{Bodenheimer, Lin  \& Mardling}{Bodenheimer
  et~al.}{2001}]{Bodenheimer_2001}
Bodenheimer P.,  Lin D.~N.~C.,   Mardling R.~A.,  2001, \mn@doi [The
  Astrophysical Journal] {10.1086/318667}, 548, 466

\bibitem[\protect\citeauthoryear{Boyd}{Boyd}{2001}]{boyd2001chebyshev}
Boyd J.~P.,  2001, {Chebyshev and Fourier spectral methods}.
Dover Publications

\bibitem[\protect\citeauthoryear{Braviner}{Braviner}{2015}]{Braviner_thesis}
Braviner H.~J.,  2015, Stellar and Planetary Tides at Small Orbital Radii,
  Doctoral dissertation, University of Cambridge

\bibitem[\protect\citeauthoryear{Bryan et~al.,}{Bryan et~al.}{2020}]{Bryan2020}
Bryan M.~L.,  et~al., 2020, \mn@doi [The Astronomical Journal]
  {10.3847/1538-3881/ab76c6}, 159, 181

\bibitem[\protect\citeauthoryear{Bryan, Chiang, Morley, Mace  \& Bowler}{Bryan
  et~al.}{2021}]{Bryan2021}
Bryan M.~L.,  Chiang E.,  Morley C.~V.,  Mace G.~N.,   Bowler B.~P.,  2021,
  \mn@doi [The Astronomical Journal] {10.3847/1538-3881/ac1bb1}, 162, 217

\bibitem[\protect\citeauthoryear{{Burns}, {Vasil}, {Oishi}, {Lecoanet}  \&
  {Brown}}{{Burns} et~al.}{2020}]{Dedalus2020}
{Burns} K.~J.,  {Vasil} G.~M.,  {Oishi} J.~S.,  {Lecoanet} D.,   {Brown} B.~P.,
   2020, \mn@doi [Physical Review Research] {10.1103/PhysRevResearch.2.023068},
  \href {https://ui.adsabs.harvard.edu/abs/2020PhRvR...2b3068B} {2, 023068}

\bibitem[\protect\citeauthoryear{Carter \& Winn}{Carter \&
  Winn}{2010}]{Carter2010}
Carter J.~A.,  Winn J.~N.,  2010, \mn@doi [The Astrophysical Journal]
  {10.1088/0004-637X/709/2/1219}, 709, 1219

\bibitem[\protect\citeauthoryear{{Cauley}, {Shkolnik}, {Llama}  \&
  {Lanza}}{{Cauley} et~al.}{2019}]{Cauley_HJ_strongB}
{Cauley} P.~W.,  {Shkolnik} E.~L.,  {Llama} J.,   {Lanza} A.~F.,  2019, \mn@doi
  [Nature Astronomy] {10.1038/s41550-019-0840-x}, \href
  {https://ui.adsabs.harvard.edu/abs/2019NatAs...3.1128C} {3, 1128}

\bibitem[\protect\citeauthoryear{{C{\'e}bron}, {Le Bars}, {Moutou}  \& {Le
  Gal}}{{C{\'e}bron} et~al.}{2012}]{Cebron_terrestrial}
{C{\'e}bron} D.,  {Le Bars} M.,  {Moutou} C.,   {Le Gal} P.,  2012, \mn@doi
  [Astronomy \& Astrophysics] {10.1051/0004-6361/201117741}, 539, A78

\bibitem[\protect\citeauthoryear{{C{\'e}bron}, {Le Bars}, {Le Gal}, {Moutou},
  {Leconte}  \& {Sauret}}{{C{\'e}bron} et~al.}{2013}]{CebronellipHJ}
{C{\'e}bron} D.,  {Le Bars} M.,  {Le Gal} P.,  {Moutou} C.,  {Leconte} J.,
  {Sauret} A.,  2013, \mn@doi [Icarus] {10.1016/j.icarus.2012.12.017}, \href
  {https://ui.adsabs.harvard.edu/abs/2013Icar..226.1642C} {226, 1642}

\bibitem[\protect\citeauthoryear{{Cox} \& {Giuli}}{{Cox} \&
  {Giuli}}{1968}]{Cox1968}
{Cox} J.~P.,  {Giuli} R.~T.,  1968, {Principles of stellar structure}.
{Gordon and Breach}

\bibitem[\protect\citeauthoryear{Cébron, Maubert  \& Le~Bars}{Cébron
  et~al.}{2010}]{Cebron2010}
Cébron D.,  Maubert P.,   Le~Bars M.,  2010, \mn@doi [Geophysical Journal
  International] {10.1111/j.1365-246X.2010.04712.x}, 182, 1311

\bibitem[\protect\citeauthoryear{Cébron, Le~Bars, Noir  \& Aurnou}{Cébron
  et~al.}{2012}]{librationellip}
Cébron D.,  Le~Bars M.,  Noir J.,   Aurnou J.~M.,  2012, \mn@doi [Physics of
  Fluids] {10.1063/1.4729296}, 24, 061703

\bibitem[\protect\citeauthoryear{{Damiani} \& {Mathis}}{{Damiani} \&
  {Mathis}}{2018}]{DamianiMathis2018}
{Damiani} C.,  {Mathis} S.,  2018, \mn@doi [\aap]
  {10.1051/0004-6361/201732538}, \href
  {https://ui.adsabs.harvard.edu/abs/2018A&A...618A..90D} {618, A90}

\bibitem[\protect\citeauthoryear{Dewberry}{Dewberry}{2023}]{Dewberry2023}
Dewberry J.~W.,  2023, \mn@doi [Monthly Notices of the Royal Astronomical
  Society] {10.1093/mnras/stad546}, 521, 5991

\bibitem[\protect\citeauthoryear{{Dewberry} \& {Lai}}{{Dewberry} \&
  {Lai}}{2022}]{Dewberry2022}
{Dewberry} J.~W.,  {Lai} D.,  2022, \mn@doi [The Astrophysical Journal]
  {10.3847/1538-4357/ac3ede}, 925, 124

\bibitem[\protect\citeauthoryear{{Dhouib}, {Baruteau}, {Mathis}, {Debras},
  {Astoul}  \& {Rieutord}}{{Dhouib} et~al.}{2024}]{Dhouib2024}
{Dhouib} H.,  {Baruteau} C.,  {Mathis} S.,  {Debras} F.,  {Astoul} A.,
  {Rieutord} M.,  2024, \mn@doi [Astronomy \& Astrophysics]
  {10.1051/0004-6361/202347703}, 682, A85

\bibitem[\protect\citeauthoryear{{Dobbs-Dixon}, {Lin}  \&
  {Mardling}}{{Dobbs-Dixon} et~al.}{2004}]{tidalcircularizationDobbs}
{Dobbs-Dixon} I.,  {Lin} D.~N.~C.,   {Mardling} R.~A.,  2004, \mn@doi [The
  Astrophysical Journal] {10.1086/421510}, \href
  {https://ui.adsabs.harvard.edu/abs/2004ApJ...610..464D} {610, 464}

\bibitem[\protect\citeauthoryear{{Duguid}, Barker  \& Jones}{{Duguid}
  et~al.}{2019}]{Craig2019effvisc}
{Duguid} C.~D.,  Barker A.~J.,   Jones C.~A.,  2019, \mn@doi [Monthly Notices
  of the Royal Astronomical Society] {10.1093/mnras/stz2899}, 491, 923

\bibitem[\protect\citeauthoryear{{Duguid}, Barker  \& Jones}{{Duguid}
  et~al.}{2020}]{Craig2020effvisc}
{Duguid} C.~D.,  Barker A.~J.,   Jones C.~A.,  2020, \mn@doi [Monthly Notices
  of the Royal Astronomical Society] {10.1093/mnras/staa2216}, 497, 3400

\bibitem[\protect\citeauthoryear{Durante et~al.,}{Durante
  et~al.}{2020}]{Durante2020}
Durante D.,  et~al., 2020, \mn@doi [Geophysical Research Letters]
  {10.1029/2019GL086572}, 47, e2019GL086572

\bibitem[\protect\citeauthoryear{Eggleton \& Kiseleva-Eggleton}{Eggleton \&
  Kiseleva-Eggleton}{2001}]{Eggleton2001}
Eggleton P.~P.,  Kiseleva-Eggleton L.,  2001, \mn@doi [The Astrophysical
  Journal] {10.1086/323843}, 562, 1012

\bibitem[\protect\citeauthoryear{Fischer, Lottes  \& Kerkemeier}{Fischer
  et~al.}{2008}]{nek5000-web-page}
Fischer P.~F.,  Lottes J.~W.,   Kerkemeier S.~G.,  2008, {Nek5000} {w}eb page,
  \url {http://nek5000.mcs.anl.gov}

\bibitem[\protect\citeauthoryear{{Fuller}, {Luan}  \& {Quataert}}{{Fuller}
  et~al.}{2016}]{F2016}
{Fuller} J.,  {Luan} J.,   {Quataert} E.,  2016, \mn@doi [Monthly Notices of
  the Royal Astronomical Society] {10.1093/mnras/stw609}, \href
  {https://ui.adsabs.harvard.edu/abs/2016MNRAS.458.3867F} {458, 3867}

\bibitem[\protect\citeauthoryear{{Goldreich} \& {Nicholson}}{{Goldreich} \&
  {Nicholson}}{1977}]{GoldreichNicholson1977}
{Goldreich} P.,  {Nicholson} P.~D.,  1977, \mn@doi [Icarus]
  {10.1016/0019-1035(77)90163-4}, \href
  {https://ui.adsabs.harvard.edu/abs/1977Icar...30..301G} {30, 301}

\bibitem[\protect\citeauthoryear{Goodman \& Oh}{Goodman \&
  Oh}{1997}]{Goodmaneffvisc}
Goodman J.,  Oh S.~P.,  1997, \mn@doi [The Astrophysical Journal]
  {10.1086/304505}, 486, 403

\bibitem[\protect\citeauthoryear{Greenspan}{Greenspan}{1968}]{greenspan1968}
Greenspan H.~P.,  1968, {The Theory of Rotating Fluids}.
Cambridge University Press

\bibitem[\protect\citeauthoryear{{Guervilly}, {Hughes}  \& {Jones}}{{Guervilly}
  et~al.}{2014}]{CelineLSV}
{Guervilly} C.,  {Hughes} D.~W.,   {Jones} C.~A.,  2014, \mn@doi [Journal of
  Fluid Mechanics] {10.1017/jfm.2014.542}, \href
  {https://ui.adsabs.harvard.edu/abs/2014JFM...758..407G} {758, 407}

\bibitem[\protect\citeauthoryear{{Guillot}, {Stevenson}, {Hubbard}  \&
  {Saumon}}{{Guillot} et~al.}{2004}]{Jupiterparams2004}
{Guillot} T.,  {Stevenson} D.~J.,  {Hubbard} W.~B.,   {Saumon} D.,  2004, in
  {Bagenal} F.,  {Dowling} T.~E.,   {McKinnon} W.~B.,  eds, , {Jupiter. The
  Planet, Satellites and Magnetosphere}.
Cambridge University Press, pp 35--57

\bibitem[\protect\citeauthoryear{Hollerbach \& Kerswell}{Hollerbach \&
  Kerswell}{1995}]{hollerbach_kerswell_1995}
Hollerbach R.,  Kerswell R.~R.,  1995, \mn@doi [Journal of Fluid Mechanics]
  {10.1017/S0022112095003338}, 298, 327–339

\bibitem[\protect\citeauthoryear{Jermyn et~al.,}{Jermyn
  et~al.}{2023}]{Jermyn2022}
Jermyn A.~S.,  et~al., 2023, \mn@doi [The Astrophysical Journal Supplement
  Series] {10.3847/1538-4365/acae8d}, 265, 15

\bibitem[\protect\citeauthoryear{Kerswell}{Kerswell}{1993}]{Kerswell1993}
Kerswell R.~R.,  1993, \mn@doi [Geophysical \& Astrophysical Fluid Dynamics]
  {10.1080/03091929308203609}, 72, 107

\bibitem[\protect\citeauthoryear{Kerswell}{Kerswell}{1996}]{Kerswell1996}
Kerswell R.~R.,  1996, \mn@doi [Journal of Fluid Mechanics]
  {10.1017/S0022112096007756}, 321, 335–370

\bibitem[\protect\citeauthoryear{Kerswell}{Kerswell}{2002}]{Ellipticalinstability}
Kerswell R.~R.,  2002, \mn@doi [Annual Review of Fluid Mechanics]
  {10.1146/annurev.fluid.34.081701.171829}, 34, 83

\bibitem[\protect\citeauthoryear{{Khlifi}, {Salhi}, {Nasraoui}, {Godeferd}  \&
  {Cambon}}{{Khlifi} et~al.}{2018}]{Khlifi2018}
{Khlifi} A.,  {Salhi} A.,  {Nasraoui} S.,  {Godeferd} F.,   {Cambon} C.,  2018,
  \mn@doi [\pre] {10.1103/PhysRevE.98.011102}, \href
  {https://ui.adsabs.harvard.edu/abs/2018PhRvE..98a1102K} {98, 011102}

\bibitem[\protect\citeauthoryear{{Kopal}}{{Kopal}}{1959}]{Kopal1959}
{Kopal} Z.,  1959, {Close binary systems}.
Wiley

\bibitem[\protect\citeauthoryear{Kumar, Pizzi, Mamatsashvili, Giesecke, Stefani
   \& Barker}{Kumar et~al.}{2024}]{kumar2023local}
Kumar V.,  Pizzi F.,  Mamatsashvili G.,  Giesecke A.,  Stefani F.,   Barker
  A.~J.,  2024, \mn@doi [Physical Review E] {10.1103/PhysRevE.109.065101}, 109,
  065101

\bibitem[\protect\citeauthoryear{{Lai}}{{Lai}}{2012}]{Lai2012}
{Lai} D.,  2012, \mn@doi [\mnras] {10.1111/j.1365-2966.2012.20893.x}, \href
  {https://ui.adsabs.harvard.edu/abs/2012MNRAS.423..486L} {423, 486}

\bibitem[\protect\citeauthoryear{{Lai}}{{Lai}}{2021}]{Lai2021}
{Lai} D.,  2021, \mn@doi [The Planetary Science Journal] {10.3847/PSJ/ac013b},
  2, 122

\bibitem[\protect\citeauthoryear{Lavorel \& Le~Bars}{Lavorel \&
  Le~Bars}{2010}]{Lavorelexperimentalellip}
Lavorel G.,  Le~Bars M.,  2010, \mn@doi [Physics of Fluids]
  {10.1063/1.3508946}, 22, 114101

\bibitem[\protect\citeauthoryear{Lazovik, Barker, de Vries  \& Astoul}{Lazovik
  et~al.}{2023}]{Lazovik2024}
Lazovik Y.~A.,  Barker A.~J.,  de Vries N.~B.,   Astoul A.,  2023, \mn@doi
  [Monthly Notices of the Royal Astronomical Society] {10.1093/mnras/stad3689},
  527, 8245

\bibitem[\protect\citeauthoryear{{Le Bars}, Lacaze, {Le Dizès}, {Le Gal}  \&
  Rieutord}{{Le Bars} et~al.}{2010}]{LE_BARS_elliptical}
{Le Bars} M.,  Lacaze L.,  {Le Dizès} S.,  {Le Gal} P.,   Rieutord M.,  2010,
  \mn@doi [Physics of the Earth and Planetary Interiors]
  {10.1016/j.pepi.2009.07.005}, 178, 48

\bibitem[\protect\citeauthoryear{Le~Bars, C\'{e}bron  \& Le~Gal}{Le~Bars
  et~al.}{2015}]{LeBars_prec_2015}
Le~Bars M.,  C\'{e}bron D.,   Le~Gal P.,  2015, \mn@doi [Annual Review of Fluid
  Mechanics] {10.1146/annurev-fluid-010814-014556}, 47, 163

\bibitem[\protect\citeauthoryear{Li}{Li}{2021}]{Li2021}
Li G.,  2021, \mn@doi [The Astrophysical Journal Letters]
  {10.3847/2041-8213/ac0620}, 915, L2

\bibitem[\protect\citeauthoryear{Li \& Lai}{Li \& Lai}{2020}]{Li2020}
Li J.,  Lai D.,  2020, \mn@doi [The Astrophysical Journal Letters]
  {10.3847/2041-8213/aba2c4}, 898, L20

\bibitem[\protect\citeauthoryear{{Lin}}{{Lin}}{2023}]{Lin2023}
{Lin} Y.,  2023, \mn@doi [Astronomy \& Astrophysics]
  {10.1051/0004-6361/202245112}, \href
  {https://ui.adsabs.harvard.edu/abs/2023A&A...671A..37L} {671, A37}

\bibitem[\protect\citeauthoryear{Lin \& Ogilvie}{Lin \&
  Ogilvie}{2017}]{LinOgilvie2017}
Lin Y.,  Ogilvie G.~I.,  2017, \mn@doi [Monthly Notices of the Royal
  Astronomical Society] {10.1093/mnras/stx540}, 468, 1387

\bibitem[\protect\citeauthoryear{Lorenzani \& Tilgner}{Lorenzani \&
  Tilgner}{2001}]{LORENZANI2001}
Lorenzani S.,  Tilgner A.,  2001, \mn@doi [Journal of Fluid Mechanics]
  {10.1017/S002211200100581X}, 447, 111–128

\bibitem[\protect\citeauthoryear{Lorenzani \& Tilgner}{Lorenzani \&
  Tilgner}{2003}]{LORENZANI2003}
Lorenzani S.,  Tilgner A.,  2003, \mn@doi [Journal of Fluid Mechanics]
  {10.1017/S002211200300572X}, 492, 363–379

\bibitem[\protect\citeauthoryear{Lurie et~al.,}{Lurie
  et~al.}{2017}]{Binarysync}
Lurie J.~C.,  et~al., 2017, \mn@doi [The Astronomical Journal]
  {10.3847/1538-3881/aa974d}, 154, 250

\bibitem[\protect\citeauthoryear{{Maciejewski} et~al.,}{{Maciejewski}
  et~al.}{2016}]{M2016}
{Maciejewski} G.,  et~al., 2016, \mn@doi [Astronomy \& Astrophysics]
  {10.1051/0004-6361/201628312}, \href
  {https://ui.adsabs.harvard.edu/abs/2016A&A...588L...6M} {588, L6}

\bibitem[\protect\citeauthoryear{Mak, Griffiths  \& Hughes}{Mak
  et~al.}{2017}]{Mak2017}
Mak J.,  Griffiths S.~D.,   Hughes D.~W.,  2017, \mn@doi [Physical Review
  Fluids] {10.1103/PhysRevFluids.2.113701}, 2, 113701

\bibitem[\protect\citeauthoryear{Malkus}{Malkus}{1968}]{Malkus1968}
Malkus W. V.~R.,  1968, \mn@doi [Science] {10.1126/science.160.3825.259}, 160,
  259

\bibitem[\protect\citeauthoryear{Martin \& Armitage}{Martin \&
  Armitage}{2021}]{Martin2021}
Martin R.~G.,  Armitage P.~J.,  2021, \mn@doi [The Astrophysical Journal
  Letters] {10.3847/2041-8213/abf736}, 912, L16

\bibitem[\protect\citeauthoryear{{Mason} \& {Kerswell}}{{Mason} \&
  {Kerswell}}{2002}]{masonkerswell}
{Mason} R.~M.,  {Kerswell} R.~R.,  2002, \mn@doi [Journal of Fluid Mechanics]
  {10.1017/S0022112002001994}, 471, 71–106

\bibitem[\protect\citeauthoryear{Millholland \& Batygin}{Millholland \&
  Batygin}{2019}]{Millholland2019}
Millholland S.,  Batygin K.,  2019, \mn@doi [The Astrophysical Journal]
  {10.3847/1538-4357/ab19be}, 876, 119

\bibitem[\protect\citeauthoryear{M\"uller \& Helled}{M\"uller \&
  Helled}{2023}]{Muller2023}
M\"uller S.,  Helled R.,  2023, \mn@doi [Astronomy \& Astrophysics]
  {10.1051/0004-6361/202244827}, 669, A24

\bibitem[\protect\citeauthoryear{M\"uller, Helled  \& Cumming}{M\"uller
  et~al.}{2020}]{Muller2020}
M\"uller S.,  Helled R.,   Cumming A.,  2020, \mn@doi [Astronomy \&
  Astrophysics] {10.1051/0004-6361/201937376}, 638, A121

\bibitem[\protect\citeauthoryear{Naing \& Fukumoto}{Naing \&
  Fukumoto}{2011}]{Naing2011}
Naing M.~M.,  Fukumoto Y.,  2011, \mn@doi [Fluid Dynamics Research]
  {10.1088/0169-5983/43/5/055502}, 43, 055502

\bibitem[\protect\citeauthoryear{{Nine}, {Milliman}, {Mathieu}, {Geller},
  {Leiner}, {Platais}  \& {Tofflemire}}{{Nine} et~al.}{2020}]{binarycirc}
{Nine} A.~C.,  {Milliman} K.~E.,  {Mathieu} R.~D.,  {Geller} A.~M.,  {Leiner}
  E.~M.,  {Platais} I.,   {Tofflemire} B.~M.,  2020, \mn@doi [The Astronomical
  Journal] {10.3847/1538-3881/abad3b}, \href
  {https://ui.adsabs.harvard.edu/abs/2020AJ....160..169N} {160, 169}

\bibitem[\protect\citeauthoryear{Noir, Brito, Aldridge  \& Cardin}{Noir
  et~al.}{2001}]{Noir2001}
Noir J.,  Brito D.,  Aldridge K.,   Cardin P.,  2001, \mn@doi [Geophysical
  Research Letters] {10.1029/2001GL012956}, 28, 3785

\bibitem[\protect\citeauthoryear{Ogilvie}{Ogilvie}{2013}]{OgilvieIW}
Ogilvie G.~I.,  2013, \mn@doi [Monthly Notices of the Royal Astronomical
  Society] {10.1093/mnras/sts362}, 429, 613

\bibitem[\protect\citeauthoryear{{Ogilvie}}{{Ogilvie}}{2014}]{Ogilvie2014}
{Ogilvie} G.~I.,  2014, \mn@doi [Annual Review of Astronomy and Astrophysics]
  {10.1146/annurev-astro-081913-035941}, \href
  {https://ui.adsabs.harvard.edu/abs/2014ARA&A..52..171O} {52, 171}

\bibitem[\protect\citeauthoryear{Ogilvie \& Lesur}{Ogilvie \&
  Lesur}{2012}]{Ogilvieeffvisc}
Ogilvie G.~I.,  Lesur G.,  2012, \mn@doi [Monthly Notices of the Royal
  Astronomical Society] {10.1111/j.1365-2966.2012.20630.x}, 422, 1975

\bibitem[\protect\citeauthoryear{{Ogilvie} \& {Lin}}{{Ogilvie} \&
  {Lin}}{2004}]{OL2004}
{Ogilvie} G.~I.,  {Lin} D.~N.~C.,  2004, \mn@doi [The Astrophysical Journal]
  {10.1086/421454}, \href
  {https://ui.adsabs.harvard.edu/abs/2004ApJ...610..477O} {610, 477}

\bibitem[\protect\citeauthoryear{Palma-Bifani et~al.,}{Palma-Bifani
  et~al.}{2023}]{Palma-Bifani2023}
Palma-Bifani P.,  et~al., 2023, \mn@doi [A&A] {10.1051/0004-6361/202244294},
  670, A90

\bibitem[\protect\citeauthoryear{Patra et~al.,}{Patra
  et~al.}{2020}]{Orbitaldecaysummary}
Patra K.~C.,  et~al., 2020, \mn@doi [The Astronomical Journal]
  {10.3847/1538-3881/ab7374}, 159, 150

\bibitem[\protect\citeauthoryear{{Paxton}, {Bildsten}, {Dotter}, {Herwig},
  {Lesaffre}  \& {Timmes}}{{Paxton} et~al.}{2011}]{Paxton2011}
{Paxton} B.,  {Bildsten} L.,  {Dotter} A.,  {Herwig} F.,  {Lesaffre} P.,
  {Timmes} F.,  2011, \mn@doi [The Astrophysical Journal Supplement Series]
  {10.1088/0067-0049/192/1/3}, \href
  {https://ui.adsabs.harvard.edu/abs/2011ApJS..192....3P} {192, 3}

\bibitem[\protect\citeauthoryear{{Paxton} et~al.,}{{Paxton}
  et~al.}{2013}]{Paxton2013}
{Paxton} B.,  et~al., 2013, \mn@doi [The Astrophysical Journal Supplement
  Series] {10.1088/0067-0049/208/1/4}, \href
  {https://ui.adsabs.harvard.edu/abs/2013ApJS..208....4P} {208, 4}

\bibitem[\protect\citeauthoryear{{Paxton} et~al.,}{{Paxton}
  et~al.}{2015}]{Paxton2015}
{Paxton} B.,  et~al., 2015, \mn@doi [The Astrophysical Journal Supplement
  Series] {10.1088/0067-0049/220/1/15}, \href
  {https://ui.adsabs.harvard.edu/abs/2015ApJS..220...15P} {220, 15}

\bibitem[\protect\citeauthoryear{{Paxton} et~al.,}{{Paxton}
  et~al.}{2018}]{Paxton2018}
{Paxton} B.,  et~al., 2018, \mn@doi [The Astrophysical Journal Supplement
  Series] {10.3847/1538-4365/aaa5a8}, \href
  {https://ui.adsabs.harvard.edu/abs/2018ApJS..234...34P} {234, 34}

\bibitem[\protect\citeauthoryear{{Paxton} et~al.,}{{Paxton}
  et~al.}{2019}]{Paxton2019}
{Paxton} B.,  et~al., 2019, \mn@doi [The Astrophysical Journal Supplement
  Series] {10.3847/1538-4365/ab2241}, \href
  {https://ui.adsabs.harvard.edu/abs/2019ApJS..243...10P} {243, 10}

\bibitem[\protect\citeauthoryear{{Penev}, Sasselov, Robinson  \&
  Demarque}{{Penev} et~al.}{2007}]{Penev2007}
{Penev} K.,  Sasselov D.,  Robinson F.,   Demarque P.,  2007, \mn@doi [The
  Astrophysical Journal] {10.1086/507937}, \href
  {https://ui.adsabs.harvard.edu/abs/2007ApJ...655.1166P} {655, 1166}

\bibitem[\protect\citeauthoryear{{Penev}, Sasselov, Robinson  \&
  Demarque}{{Penev} et~al.}{2009a}]{Penev2009a}
{Penev} K.,  Sasselov D.,  Robinson F.,   Demarque P.,  2009a, \mn@doi [The
  Astrophysical Journal] {10.1088/0004-637X/704/2/930}, \href
  {https://ui.adsabs.harvard.edu/abs/2009ApJ...704..930P} {704, 930}

\bibitem[\protect\citeauthoryear{Penev, Barranco  \& Sasselov}{Penev
  et~al.}{2009b}]{Penev2009b}
Penev K.,  Barranco J.,   Sasselov D.,  2009b, \mn@doi [The Astrophysical
  Journal] {10.1088/0004-637X/705/1/285}, \href
  {https://ui.adsabs.harvard.edu/abs/2009ApJ...705..285P} {705, 285}

\bibitem[\protect\citeauthoryear{Pizzi, Mamatsashvili, Barker, Giesecke  \&
  Stefani}{Pizzi et~al.}{2022}]{Pizzi2022}
Pizzi F.,  Mamatsashvili G.,  Barker A.~J.,  Giesecke A.,   Stefani F.,  2022,
  \mn@doi [Physics of Fluids] {10.1063/5.0131035}, 34, 125135

\bibitem[\protect\citeauthoryear{{Poincar{\'e}}}{{Poincar{\'e}}}{1910}]{Poincare1910}
{Poincar{\'e}} H.,  1910, Bulletin Astronomique, Serie I, \href
  {https://ui.adsabs.harvard.edu/abs/1910BuAsI..27..321P} {27, 321}

\bibitem[\protect\citeauthoryear{Pontin}{Pontin}{2022}]{Pontin_thesis}
Pontin C.~M.,  2022, Wave propagation and tidal dissipation in giant planets
  containing regions of stable stratification, Doctoral dissertation,
  University of Leeds

\bibitem[\protect\citeauthoryear{Pontin, Barker, Hollerbach, André  \&
  Mathis}{Pontin et~al.}{2020}]{Christina_2020}
Pontin C.~M.,  Barker A.~J.,  Hollerbach R.,  André Q.,   Mathis S.,  2020,
  \mn@doi [Monthly Notices of the Royal Astronomical Society]
  {10.1093/mnras/staa664}, 493, 5788

\bibitem[\protect\citeauthoryear{Pontin, Barker  \& Hollerbach}{Pontin
  et~al.}{2023}]{Pontin_2023}
Pontin C.~M.,  Barker A.~J.,   Hollerbach R.,  2023, \mn@doi [The Astrophysical
  Journal] {10.3847/1538-4357/accd67}, 950, 176

\bibitem[\protect\citeauthoryear{Pontin, Barker  \& Hollerbach}{Pontin
  et~al.}{2024}]{Pontin_2023b}
Pontin C.~M.,  Barker A.~J.,   Hollerbach R.,  2024, \mn@doi [The Astrophysical
  Journal] {10.3847/1538-4357/ad0a90}, 960, 32

\bibitem[\protect\citeauthoryear{Poon, Bryan, Rein, Morley, Mace, Zhou  \&
  Bowler}{Poon et~al.}{2024}]{Poon2024}
Poon M.,  Bryan M.~L.,  Rein H.,  Morley C.~V.,  Mace G.,  Zhou Y.,   Bowler
  B.~P.,  2024, \mn@doi [The Astronomical Journal] {10.3847/1538-3881/ad84e5},
  168, 270

\bibitem[\protect\citeauthoryear{{Salhi}, {Khlifi}  \& {Cambon}}{{Salhi}
  et~al.}{2019}]{Salhi2019}
{Salhi} A.,  {Khlifi} A.,   {Cambon} C.,  2019, \mn@doi [Atmosphere]
  {10.3390/atmos11010014}, \href
  {https://ui.adsabs.harvard.edu/abs/2019Atmos..11...14S} {11, 14}

\bibitem[\protect\citeauthoryear{Su \& Lai}{Su \& Lai}{2020}]{Su2020}
Su Y.,  Lai D.,  2020, \mn@doi [The Astrophysical Journal]
  {10.3847/1538-4357/abb6f3}, 903, 7

\bibitem[\protect\citeauthoryear{Tilgner}{Tilgner}{2005}]{Tilgner2005}
Tilgner A.,  2005, \mn@doi [Physics of Fluids] {10.1063/1.1852576}, 17, 034104

\bibitem[\protect\citeauthoryear{Tilgner \& Busse}{Tilgner \&
  Busse}{2001}]{TILGNER2001}
Tilgner A.,  Busse F.~H.,  2001, \mn@doi [Journal of Fluid Mechanics]
  {10.1017/S0022112000002536}, 426, 387–396

\bibitem[\protect\citeauthoryear{Townsend}{Townsend}{2022}]{richard_townsend_2022_7457723}
Townsend R.,  2022, {MESA SDK for Mac OS}, \mn@doi{10.5281/zenodo.7457723},
  \url {https://doi.org/10.5281/zenodo.7457723}

\bibitem[\protect\citeauthoryear{{Turner}, {Ridden-Harper}  \&
  {Jayawardhana}}{{Turner} et~al.}{2021}]{wasp12bdecay}
{Turner} J.~D.,  {Ridden-Harper} A.,   {Jayawardhana} R.,  2021, \mn@doi [The
  Astronomical Journal] {10.3847/1538-3881/abd178}, \href
  {https://ui.adsabs.harvard.edu/abs/2021AJ....161...72T} {161, 72}

\bibitem[\protect\citeauthoryear{\VRIES{Vries}{De}{de}~Vries, Barker  \&
  Hollerbach}{\VRIES{Vries}{De}{de}~Vries et~al.}{2023a}]{deVries2023}
\VRIES{Vries}{De}{de}~Vries N.~B.,  Barker A.~J.,   Hollerbach R.,  2023a,
  \mn@doi [Physics of Fluids] {10.1063/5.0135932}, 35, 024116

\bibitem[\protect\citeauthoryear{\VRIES{Vries}{De}{de}~Vries, Barker  \&
  Hollerbach}{\VRIES{Vries}{De}{de}~Vries et~al.}{2023b}]{deVries2023b}
\VRIES{Vries}{De}{de}~Vries N.~B.,  Barker A.~J.,   Hollerbach R.,  2023b,
  \mn@doi [Monthly Notices of the Royal Astronomical Society]
  {10.1093/mnras/stad1990}, 524, 2661

\bibitem[\protect\citeauthoryear{Vidal \& Barker}{Vidal \&
  Barker}{2020a}]{Vidal_Barker_2019}
Vidal J.,  Barker A.~J.,  2020a, \mn@doi [Monthly Notices of the Royal
  Astronomical Society] {10.1093/mnras/staa2239}, 497, 4472

\bibitem[\protect\citeauthoryear{Vidal \& Barker}{Vidal \&
  Barker}{2020b}]{Vidal_Barker_2020}
Vidal J.,  Barker A.~J.,  2020b, \mn@doi [The Astrophysical Journal Letters]
  {10.3847/2041-8213/ab6219}, 888, L31

\bibitem[\protect\citeauthoryear{{Wei} \& {Tilgner}}{{Wei} \&
  {Tilgner}}{2013}]{WeiTilgner2013}
{Wei} X.,  {Tilgner} A.,  2013, \mn@doi [Journal of Fluid Mechanics]
  {10.1017/jfm.2013.68}, \href
  {https://ui.adsabs.harvard.edu/abs/2013JFM...718R...2W} {718, R2}

\bibitem[\protect\citeauthoryear{Wu \& Roberts}{Wu \& Roberts}{2008}]{Wu2008}
Wu C.~C.,  Roberts P.~H.,  2008, \mn@doi [Geophysical \& Astrophysical Fluid
  Dynamics] {10.1080/03091920701450333}, 102, 1

\bibitem[\protect\citeauthoryear{{Zahn}}{{Zahn}}{1966}]{Zahn1966}
{Zahn} J.-P.,  1966, Annales d'Astrophysique, \href
  {https://ui.adsabs.harvard.edu/abs/1966AnAp...29..313Z} {29, 313}

\bibitem[\protect\citeauthoryear{{Zahn}}{{Zahn}}{1977}]{Zahn1977tidesplit}
{Zahn} J.-P.,  1977, Astronomy \& Astrophysics, \href
  {https://ui.adsabs.harvard.edu/abs/1977A&A....57..383Z} {57, 383}

\bibitem[\protect\citeauthoryear{{Zahn}}{{Zahn}}{1989}]{Zahn1989Turbvisceqtide}
{Zahn} J.-P.,  1989, Astronomy \& Astrophysics, \href
  {https://ui.adsabs.harvard.edu/abs/1989A&A...220..112Z} {220, 112}

\makeatother
\end{thebibliography}

%%%%%%%%%%%%%%%%%%%%%%%%%%%%%%%%%%%%%%%%%%%%%%%%%%

%%%%%%%%%%%%%%%%% APPENDICES %%%%%%%%%%%%%%%%%%%%%

\appendix

\section{Table of simulations}
\label{app:sim_tables}

The parameters of the simulations performed using \textsc{dedalus}, as well as the associated resolutions of those simulations, are given in Table~\ref{tab:resolutiontable_dedalus}. We have verified that the quantities computed here are independent of the resolution chosen such that for the reported quantities the simulations are all well-resolved. The resolutions reported show the number of grid points in the simulations. Because we utilise a real Fourier basis, however, we only have wavenumber modes up to $N_x/2,N_y/2$ in the $x$ and $y$ directions, respectively. The grid is expanded by a factor of $3/2$ for de-aliasing purposes when computing non-linear terms and those with spatially variable coefficients. Likewise, a table of the parameters for the simulations performed using \textsc{nek5000}, as well as the associated resolutions, is given in Table~\ref{tab:resolutiontable_nek5000}. We utilise polynomials of order $\mathcal{N}=9$ in all directions, inside $10$ elements in all directions, thus resulting in a resolution of $90^3$ for most simulations. For de-aliasing purposes we increase the polynomial order to $14$ when calculating the non-linear terms, satisfying the $3/2$ rule.

\subsection{Resolution of snapshots}
\label{app:snapshot_resolutions}

 To ensure that the simulations executed in \textsc{nek5000} are well-resolved, without resorting to interpolating the data to generate energy spectra, we verified that the following criterion:
\begin{equation}
    k_{\mathrm{max}} l_d > 1,
\end{equation}
\noindent with
\begin{equation}
    l_d=(\nu^3 / D_\nu )^{1/4},
\end{equation}
is satisfied. The quantity $l_d$ represents the Kolmogorov lengthscale, the scale at which viscosity dominates the flow, while $k_{\mathrm{max}}=(2\pi/L_x) N_x$ is the maximum wavenumber in one direction available in the simulation. This is a conservative estimate because the spectrum is steeper than the Kolmogorov spectrum, as shown in Fig.~\ref{fig:spectrum_Ra0_Po02} and Fig.~\ref{fig:spectrum_Ra4_Po01}, with less power in smaller-scale modes near the dissipation scale. Wavevectors are not employed in \textsc{nek5000}, and as such we have treated it is a regular grid for the purposes of this calculation, with $N_x=90$. We find for the simulations in Fig.~\ref{fig:Po0.2_2} and Fig.~\ref{fig:Po0.1_Ra6} that the ratios are $1.8$ and $1.6$ respectively, and the criterion is also satisfied for all other simulations. Thus we conclude that these simulations are indeed well-resolved, even though artefacts of the grid scale appear in the figures. The well-resolvedness of the simulations executed in \textsc{dedalus}, and the agreement between the simulations in \textsc{nek5000} and \textsc{dedalus}, further demonstrates the validity of our \textsc{nek5000} time series results.

\begin{table}
\caption{Table of parameters used in the \textsc{dedalus} simulations to study the precessional instability. The Poincar\'{e}  numbers of all simulations at each employed combination of Rayleigh and Ekman numbers are shown. All simulations in this table were executed using a resolution $N_x=N_y=N_z=96$ and with box size $L_x=L_y=L_z=1$. These resolutions are the resolutions prior to expanding the grid with a factor $3/2$ for de-aliasing.}
\centering
\begin{tabular}{|l|l|}
\hline
\textsc{dedalus}                               & Po                       \\ \hline
$\mathrm{Ek} = 2.5\cdot10^{-5}, R =0$ & 0.15,0.17,0.18,0.20,0.25 \\ \hline
$\mathrm{Ek} = 2.5\cdot10^{-5}, R =2$ & 0.25                     \\ \hline
$\mathrm{Ek} = 2.5\cdot10^{-5}, R =4$ & 0.0,0.10,0.15,0.25,0.40  \\ \hline
$\mathrm{Ek} = 2.5\cdot10^{-5}, R =6$ & 0.25                     \\ \hline
\end{tabular}
\label{tab:resolutiontable_dedalus}
\end{table}

\begin{table}
\caption{Same as Table~\ref{tab:resolutiontable_dedalus} for the \textsc{nek5000} simulations. All simulations in this table were executed using a resolution  of $N_x=N_y=N_z=90$, obtained by using polynomials of order $9$ in all directions in a cube of $10$ elements in all directions. We expand the polynomials to order $14$ for de-aliasing purposes.}
\centering
\begin{tabular}{|l|l|l|}
\hline
\textsc{nek5000} & $\mathrm{Po}$\\ \hline 

\hspace{-1mm} $\mathrm{Ek}=5\cdot10^{-5}$, $R=0$ &  $0.12,0.14,0.15,0.16,0.18,0.20,0.22,0.24,$ \\ 
& \! $0.25,0.26,0.28,0.30$ \\ \hline

\hspace{-1mm}$\mathrm{Ek}=2.5\cdot10^{-5}$, $R=0$                                   & $0.06,0.08,0.10,0.12,0.14,0.15,0.16,0.17,$ \\ 
& $0.18,0.20,0.25,0.30,0.35,0.40,0.45,0.50,$ \\ 
& $0.60,0.80,1.00,1.10$ \\ \hline

\hspace{-1mm}$\mathrm{Ek}=10^{-5}$, $R=0$
& $0.03,0.04,0.05,0.06,0.08,0.10,0.11,0.12,$\\ 
& $0.13,0.14,0.15,0.16,0.18,0.20,0.25,0.30$ \\ \hline

\hspace{-1.75mm} $\mathrm{Ek}=2.5\cdot10^{-5}$, $R=1.5$ \hspace{-3mm}  & $0.04,0.10,0.25$ \\ \hline

\hspace{-1mm}$\mathrm{Ek}=2.5\cdot10^{-5}$, $R=2$ & $0.01,0.02,0.03,0.04,0.06,0.08,0.10,0.12,$ \\
& $0.14,0.15,0.16,0.18,0.20,0.25,0.30,0.35,$\\
& $0.40,0.45,0.50$\\ \hline

\hspace{-1mm}$\mathrm{Ek}=2.5\cdot10^{-5}$, $R=3$  & $0.04,0.10,0.25$ \\ \hline

\hspace{-1mm}$\mathrm{Ek}=2.5\cdot10^{-5}$, $R=4$ & $0.01,0.02,0.03,0.04,0.06,0.08,0.10,0.15,$ \\ 
& $0.20,0.25,0.30,0.35,0.40,0.45,0.50$\\ \hline

\hspace{-1mm}$\mathrm{Ek}=2.5\cdot10^{-5}$, $R=5$  & $0.04,0.10,0.25$ \\ \hline

\hspace{-1mm}$\mathrm{Ek}=2.5\cdot10^{-5}$, $R=6$  & $0.01,0.02,0.03,0.04,0.06,0.08,0.10,0.15,$ \\ 
& $0.20,0.25,0.30,0.35,0.40,0.45,0.50$\\ \hline

\hspace{-1mm}$\mathrm{Ek}=2.5\cdot10^{-5}$, $R=8$  & $0.04,0.10,0.25$ \\ \hline

\hspace{-1mm}$\mathrm{Ek}=2.5\cdot10^{-5}$, $R=10$  & $0.04,0.10,0.25$ \\ \hline

\hspace{-1mm}$\mathrm{Ek}=2.5\cdot10^{-5}$, $R=15$  & $0.04,0.10,0.25$ \\ \hline

\end{tabular}
\label{tab:resolutiontable_nek5000}
\end{table}

\section{Examination of the 2D energy}

\label{app:Dedalus_2D}

\begin{figure*}
\subfloat[$\mathrm{Po}=0.1$, $\mathrm{Ra}=0$. \label{fig:Ded_Po0.1}]{
         \includegraphics[width=0.45\textwidth]{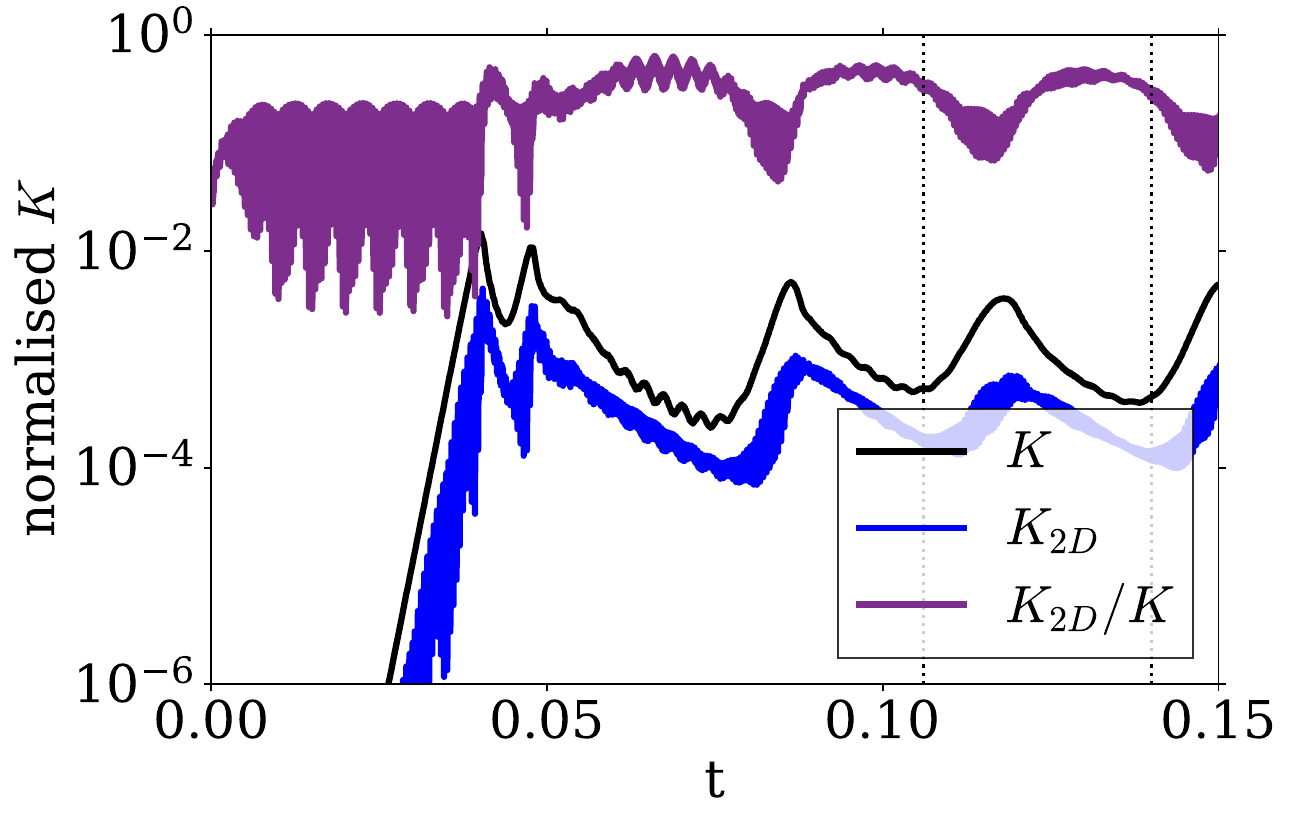}}
\hfill
\subfloat[$\mathrm{Po}=0.17$, $\mathrm{Ra}=0$. \label{fig:Ded_Po0.17}]{ \includegraphics[width=0.45\textwidth]{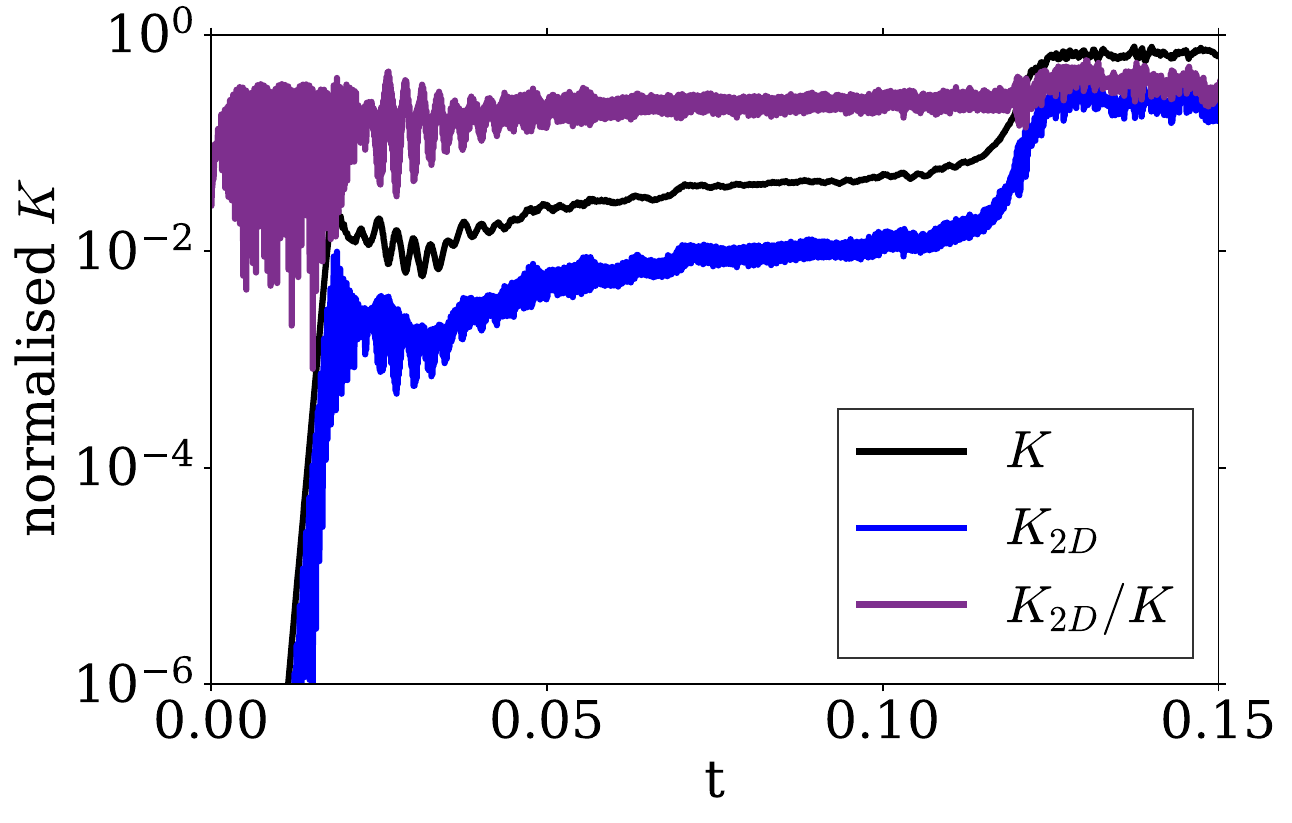}}\\

\subfloat[$\mathrm{Po}=0.18$, $\mathrm{Ra}=0$. \label{fig:Ded_Po0.18}]{\includegraphics[width=0.45\textwidth]{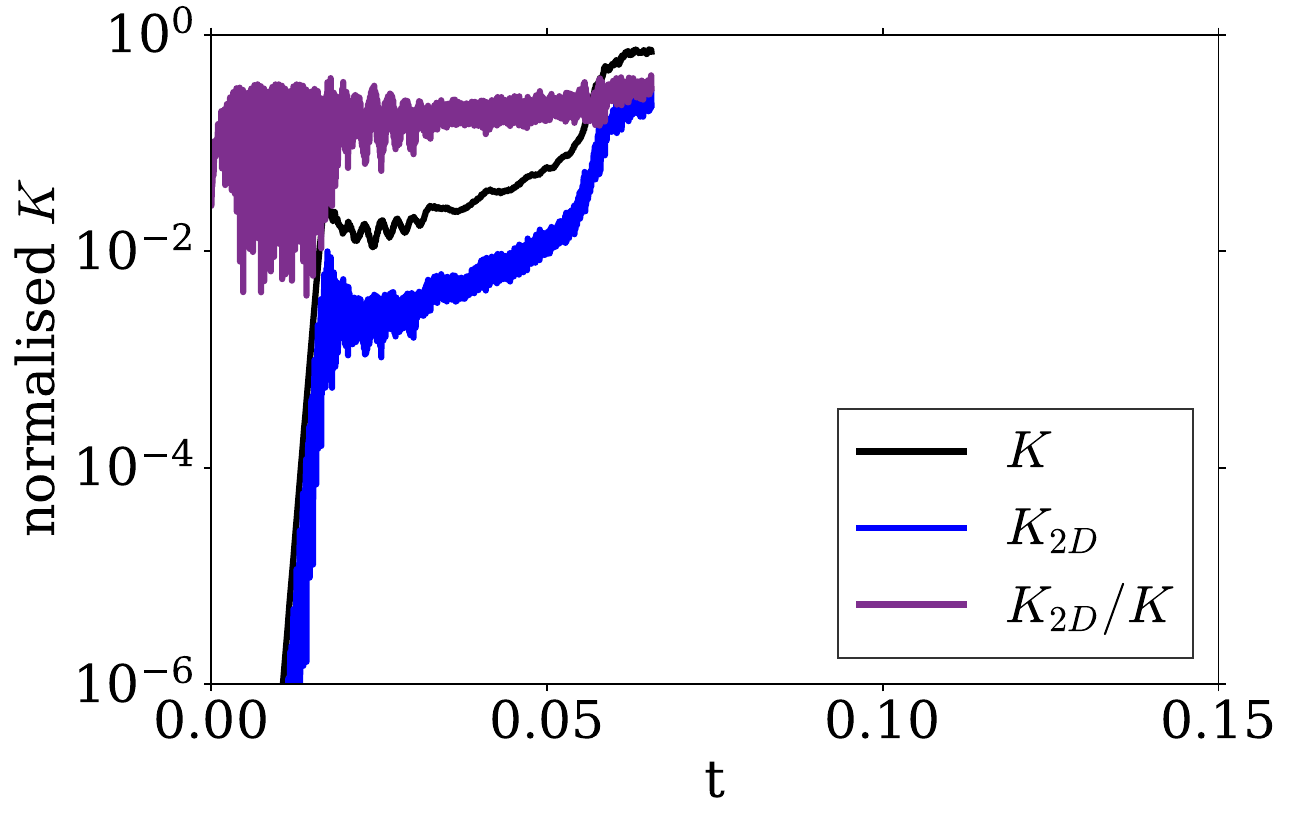}}
\hfill
\subfloat[$\mathrm{Po}=0.2$, $\mathrm{Ra}=0$. \label{fig:Ded_Po0.2}]{\includegraphics[width=0.45\textwidth]{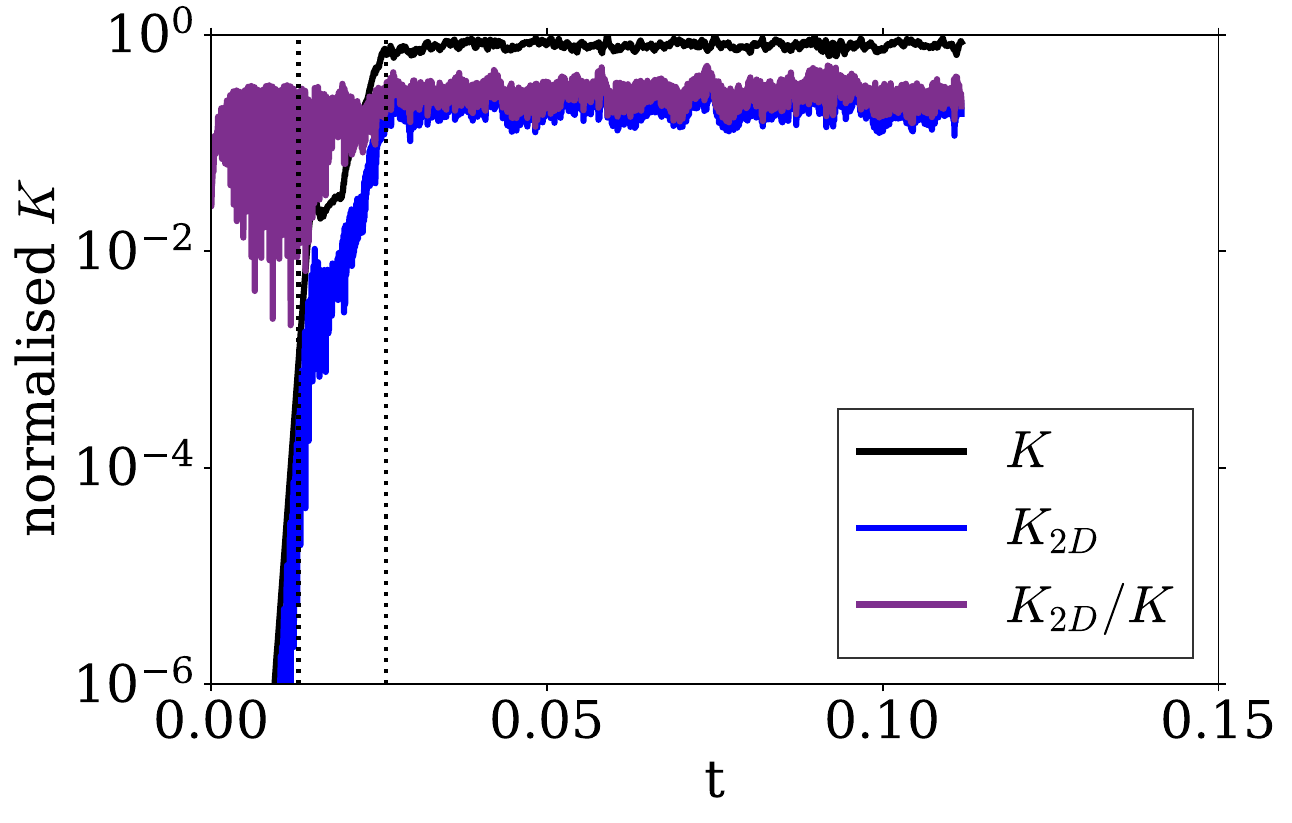}}\\

\subfloat[$\mathrm{Po}=0$, $\mathrm{Ra}=4\mathrm{Ra}_c$. \label{fig:Ded_Ra4}]{\includegraphics[width=0.45\textwidth]{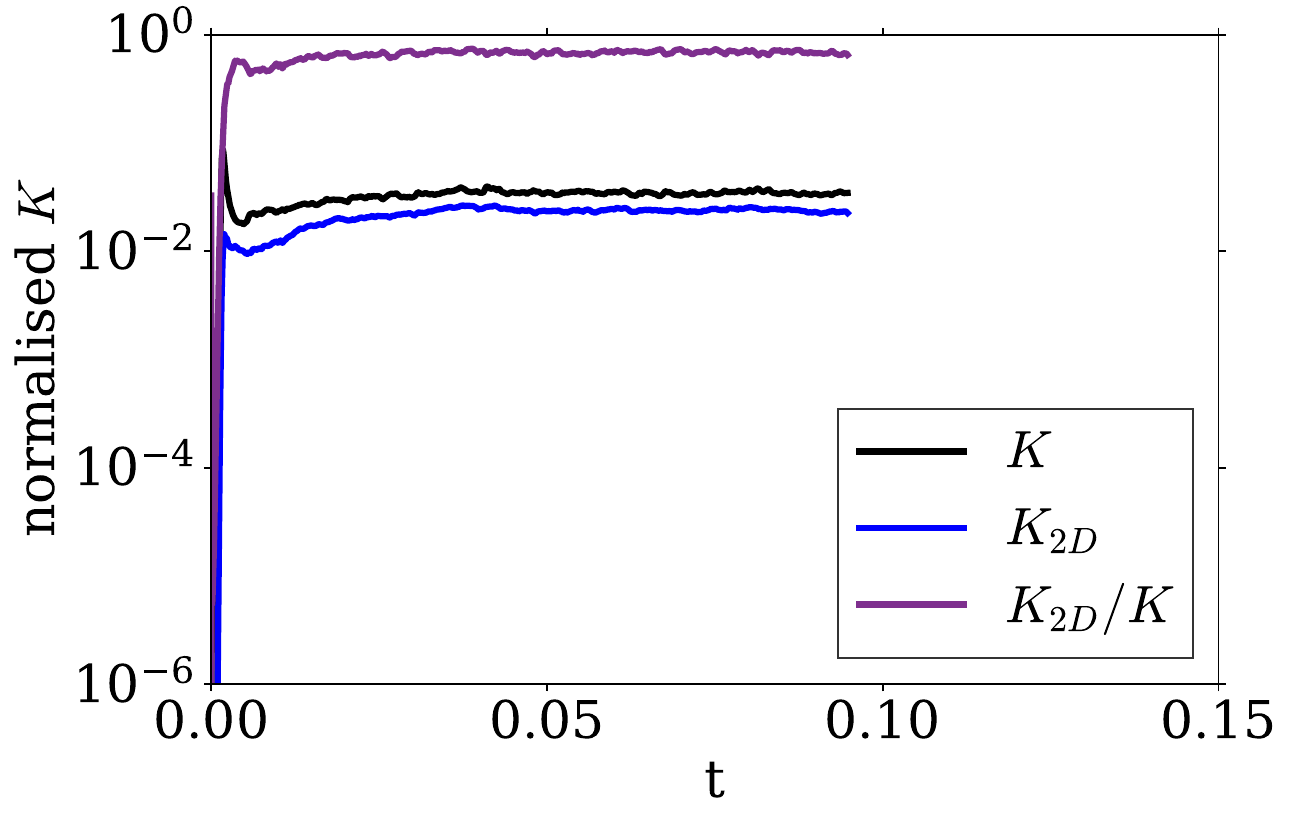}}
\hfill
\subfloat[$\mathrm{Po}=0.1$, $\mathrm{Ra}=4\mathrm{Ra}_c$. \label{fig:Ded_Po0.1_Ra4}]{\includegraphics[width=0.45\textwidth]{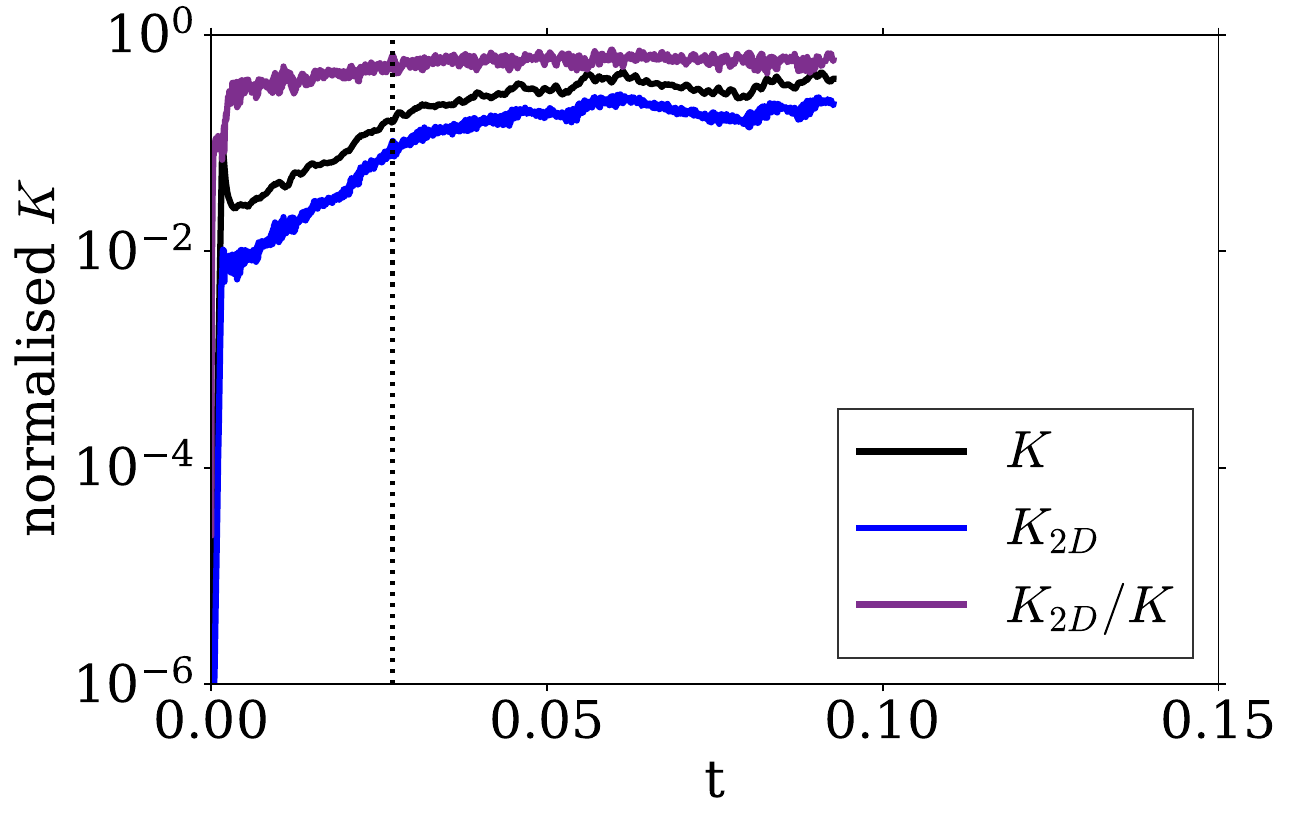}}\\

     \caption[Normalised energy time series of the precessional instability and convection with fixed Ekman number, executed using \textsc{dedalus}.]{Normalised energy time series of the precessional instability and convection with $\mathrm{Ek}=2.5\cdot10^{-5}$, executed using \textsc{dedalus}. Vertical dotted lines correspond to the ends of the intervals of the spectra shown in Figs.~\ref{fig:spectrum_Ra0_Po01}, \ref{fig:spectrum_Ra0_Po02} and \ref{fig:spectrum_Ra4_Po01}. The energies are normalised by the largest energy attained in this group of six simulations, which occurs in the simulation in the middle right panel. The expected bursty behaviour is observed in the top left panel, as well as dominance of $K_{2D}$ in the absence of the bursts of the precessional instability. The top right panel features an absence of bursty behaviour and a very gradual increase in energy until the secondary transition is reached. The middle left panel, with stronger precessional driving than the top right panel, attains the secondary transition faster, and the deciding factor when the secondary transition occurs appears to be the value of $K_{2D}$. The precessional driving in the simulation in the middle right panel is even stronger, such that the secondary transition is achieved very rapidly. In the bottom left panel, a purely convective simulation is shown, which achieves rapid saturation with weaker fluctuations compared to the precessional cases, and a clear dominance of the $K_{2D}$ energy. In the convective and precessional simulation in the bottom right panel, the energy far exceeds that of the convective simulation and the precessional instability in isolation with $\mathrm{Po}=0.1$; thus the combination indeed allows for larger energies to be achieved.}
    \label{fig:dedalus_E2D_ratios}
\end{figure*}

To shed more light on whether and, if so, how, the convective instability allows the precessional instability to achieve the continuously turbulent regime at lower values of the Poincar\'{e} number we examined $K_{2D}$ obtained using \textsc{dedalus} simulations. We again note that the vortices in the snapshots in Fig.~\ref{fig:Nek_snapshots} do not appear to be $z$-invariant, and thus $K_{2D}$ does not fully capture the energy in these vortices. To be able to better compare and study the values of $K_{2D}$ and its importance in these simulations relative to $K$, we plot normalised values of the total kinetic energy $K$, the 2D energy $K_{2D}$, as well as their ratio in Fig.~\ref{fig:dedalus_E2D_ratios}. We have normalised all these simulations by the same value, namely the largest value of the total kinetic energy reached in this set of six simulations, which is found in the simulation with $\mathrm{Po}=0.2,\ \mathrm{Ra}=0$.

In Fig.~\ref{fig:Ded_Po0.1} the simulation with $\mathrm{Po}=0.1,\ \mathrm{Ra}=0$ is plotted. The expected bursty behaviour is evident in the values of $K$, and the behaviour closely resembles that of the simulation executed using \textsc{nek5000} in Fig.~\ref{fig:NEK_Po0.1}. The 2D energy follows closely behind the burst in total kinetic energy, and is maintained more strongly after the burst than the 3D energy, such that after the burst the ratio of the total to the 2D energy attains the largest values, while during the bursts this ratio dips. The 2D energy appears to peak at normalised values of $\mathcal{O}(10^{-3})$ in this simulation. 
In Fig.~\ref{fig:Ded_Po0.17} the simulation with $\mathrm{Po}=0.17,\ \mathrm{Ra}=0$ is plotted; this simulation was found, when examining the \textsc{nek5000} simulations (not shown), to transition into the continuously turbulent regime after many rotation times in the lower energy turbulent state and we observe the same here. Prior to the secondary transition, the bursty behaviour has disappeared. It has been replaced by the flow containing a vortex, like the one in Fig.~\ref{fig:Po0.2_1}. $K_{2D}$ keeps growing in this simulation, and the secondary transition starts once $K_{2D}$ has attained a normalised value of $\approx10^{-2}$. In Fig.~\ref{fig:Ded_Po0.18} we plot the simulation with $\mathrm{Po}=0.18,\ \mathrm{Ra}=0$. This simulation goes through the secondary transition faster, because the $K_{2D}$ energy increases faster and more steadily than in the $\mathrm{Po}=0.17,\ \mathrm{Ra}=0$ case. The transition again occurs roughly when the normalised value of $K_{2D}$ exceeds $10^{-2}$. Next, in Fig.~\ref{fig:Ded_Po0.2} we plot the simulation with $\mathrm{Po}=0.2,\ \mathrm{Ra}=0$; the $K_{2D}$ energies attain large values more rapidly than the simulations at lower values of the Poincar\'{e} number, and again the transition happens faster than at lower values of $\mathrm{Po}$. We also note that the final saturation energies increase with increasing Poincar\'{e} number, because the energy injection increases with $\mathrm{Po}$.

Next we examine these same quantities in the presence of convection in the bottom panels of Fig.~\ref{fig:dedalus_E2D_ratios}. In Fig.~\ref{fig:Ded_Ra4} the simulation with $\mathrm{Po}=0$, $\mathrm{Ra}=4\mathrm{Ra}_c$ is plotted; we can see that the energies saturate rapidly at values around the required transition value, with $K_{2D}$ dominating the flow from the start as we would expect. The flow saturates and attains its final statistically steady state very rapidly. Furthermore, it displays much fewer high frequency fluctuations than the cases where the precession is present. Finally, we examine Fig.~\ref{fig:Ded_Po0.1_Ra4} in which we have plotted the simulation with $\mathrm{Po}=0.1$, $\mathrm{Ra}=4\mathrm{Ra}_c$. A much higher energy state is achieved in this simulation compared to both simulations with precession and convection in isolation for these same parameters. The transition to the high energy state again appears more gradual than the secondary transitions of the precessional instability in isolation. This gradual transition arises roughly when the normalised value of $K_{2D}$ exceeds $10^{-2}$. The ratio of $K_{2D}$ to $K$ is maintained at a larger value compared to the purely precessional simulations that have gone through a turbulent transition, although the value is smaller than the one of the purely convective case. Therefore, if the convection reaches a certain energy, and the precessional instability is able to operate, then the convection assists the precessional instability in achieving the continuously turbulent regime.

%%%%%%%%%%%%%%%%%%%%%%%%%%%%%%%%%%%%%%%%%%%%%%%%%%

% Don't change these lines
\bsp	% typesetting comment
\label{lastpage}
\end{document}